\documentclass[11pt,a4paper]{article}
\usepackage{float}
\usepackage[T1]{fontenc}
\usepackage[utf8]{inputenc}
\usepackage[greek,english]{babel}
\usepackage{jcappub}
\usepackage{bm}
\usepackage{lmodern}
\usepackage{booktabs}
\usepackage{array}
\usepackage{multirow}
\usepackage[table]{xcolor}
\usepackage{relsize}
\usepackage{mathrsfs}
\usepackage{subcaption}

\RequirePackage{ifpdf}
\ifpdf
\else 
\fi

\newcommand{\renormalizedtag}{\text{\textsc{r}}}
\newcommand{\renormalized}[1]{{#1}^{\renormalizedtag}}

\newcommand{\kdamp}{k_{\text{damp}}}

\newcommand{\sigmav}{\sigma_{\text{v}}}

\newcommand{\Legendre}[2]{\mathscr{P}_{#1}(#2)}
\newcommand{\FabrikantThree}[3]{\mathscr{J}_{{#1}{#2}{#3}}}

\newcommand{\FabrikantOne}[1]{\mathscr{J}_{#1}}

\newcommand{\semibold}[1]{{\fontseries{b}\selectfont{#1}}}
\newcommand{\sansbold}[1]{{\sffamily\fontseries{sbc}\selectfont{#1}}}

\newcommand{\para}[1]{\par\vspace{2mm}\noindent\semibold{{#1.}---}\ignorespaces}

\newcommand{\modellabel}[1]{\textsf{\textsl{{#1}}}}
\newcommand{\KaiserTree}{\modellabel{KaiserTree}}
\newcommand{\KaiserHalo}{\modellabel{KaiserHalo}}
\newcommand{\EFT}{\modellabel{EFT}}
\newcommand{\SPT}{\modellabel{SPT}}
\newcommand{\Linear}{\modellabel{Linear}}
\newcommand{\MRoy}{\modellabel{M{\&}R}}
\newcommand{\Coevo}{\modellabel{Coevo}}
\newcommand{\Advective}{\modellabel{Advective}}

\newcommand{\deltahalo}{\delta^h}
\newcommand{\deltaadvect}{\delta_{\text{Adv}}}
\newcommand{\deltaadvects}{\delta_{\text{Adv,s}}}

\renewcommand{\geq}{\geqslant}

\newcommand{\llangle}{\langle\kern-2\nulldelimiterspace\langle}
\newcommand{\rrangle}{\rangle\kern-2\nulldelimiterspace \rangle}

\renewcommand{\d}{\mathrm{d}}
\newcommand{\D}{\mathrm{D}}
\newcommand{\im}{\mathrm{i}}
\newcommand{\e}[1]{\mathrm{e}^{{#1}}}

\newcommand{\Mp}{M_{\mathrm{P}}}

\newcommand{\vect}[1]{\bm{\mathrm{{#1}}}}

\newcommand{\Mpc}{\text{Mpc}}

\newcommand{\kmax}{k_{\text{max}}}
\newcommand{\kmin}{k_{\text{min}}}
\newcommand{\xfl}{\vect{x}_{\text{fl}}}

\newcommand{\bnl}{b_{\text{3nl}}}

\newcommand{\Iint}[1]{\mathscr{I}^{({#1})}}

\newcommand{\DA}{D_A}
\newcommand{\DB}{D_B}
\newcommand{\DD}{D_D}
\newcommand{\DE}{D_E}
\newcommand{\DF}{D_F}
\newcommand{\DG}{D_G}
\renewcommand{\DJ}{D_J}

\newcommand{\DK}{D_K}
\newcommand{\DL}{D_L}

\newcommand{\fA}{f_A}
\newcommand{\fB}{f_B}

\DeclareMathOperator{\Or}{\mathrm{O}}
\newcommand{\Operator}{\mathcal{O}}

\renewcommand{\geq}{\geqslant}

\newcommand{\WizCOLA}{\sansbold{WizCOLA}}
\newcommand{\Halofit}{\sansbold{Halofit}}

\newcommand{\GitRevision}[2]{\sansbold{\href{#2}{{#1}}}}

\newcommand\CC{C\nolinebreak\hspace{-.05em}\raisebox{.4ex}{\relsize{-3}{\textbf{+}}}\nolinebreak\hspace{-.10em}\raisebox{.4ex}{\relsize{-3}{\textbf{+}}}}

\newcolumntype{Q}{>{$\displaystyle}r<{$}}
\newcolumntype{R}{>{$\displaystyle}c<{$}}
\newcolumntype{S}{>{$\displaystyle}l<{$}}

\graphicspath{{./}}

\usepackage{rotating}

\usepackage{natbib}

\begin{document}

\title{\Huge\fontseries{sbc} Impact of bias and redshift-space
modelling for the halo power spectrum
\selectfont  \\[2mm]
\LARGE\fontseries{s}\selectfont Testing the effective field theory of
large-scale structure\vspace{5mm}\hrule}
\author{\fontseries{s}\selectfont\large Luc\'{\i}a Fonseca de la Bella$^{1,2}$, Donough Regan$^1$, David Seery$^1$ and David Parkinson$^3$}

\affiliation{\vspace{2mm}\small
$^1$ Astronomy Centre, University of Sussex, Falmer, Brighton BN1 9QH, UK \\
$^2$ Department of Physics and Astronomy, University of Manchester, Oxford Road, Manchester M13 9PL,UK \\
$^3$  Korea Astronomy and Space Science Institute 776, Daedeokdae-ro, Yuseong-gu, Daejeon, Republic of Korea (34055)
}

\emailAdd{lucia.fonsecadelabella@manchester.ac.uk}
\emailAdd{donough.regan@gmail.com}
\emailAdd{D.Seery@sussex.ac.uk}
\emailAdd{davidparkinson@kasi.re.kr}

\abstract{We study the impact of different bias and redshift-space models
on the halo power spectrum, quantifying their effect
by comparing the fit to a subset of realizations taken from the
{\WizCOLA} suite.
These provide simulated power spectrum measurements between
$\kmin = 0.03 h/\Mpc$ and $\kmax = 0.29 h/\Mpc$, constructed using the
comoving Lagrangian acceleration method.
For the bias prescription we include (i) simple linear bias;
(ii) the McDonald \& Roy model and (iii) its coevolution
variant introduced by Saito et al.;
and (iv) a very general model including all terms up to one-loop
and corrections from advection.
For the redshift-space modelling we include the Kaiser formula
with exponential damping and
the power spectrum provided by
(i) tree-level perturbation theory and (ii) the {\Halofit} prescription;
(iii) one-loop perturbation theory,
also with exponential damping;
and (iv) an effective field theory description, also at one-loop,
with damping represented by the EFT subtractions.
We quantify the improvement from each layer of modelling by measuring the typical
improvement in $\chi^2$
when fitting to a member of the simulation suite.
We attempt to detect overfitting by testing for compatibility between the best-fit
power spectrum
per realization and the best-fit over the entire {\WizCOLA} suite.
For both bias and the redshift-space map we find that increasingly permissive models
yield improvements in $\chi^2$ but with diminishing returns.
The most permissive models show modest evidence for overfitting.
Accounting for model complexity using the Bayesian Information Criterion, we argue
that standard perturbation theory up to one-loop,
or a related model such as that of Taruya, Nishimichi \& Saito,
coupled to the Saito et al. coevolution bias model, is likely to provide a good
compromise for near-future galaxy surveys operating with comparable $\kmax$.
}

\maketitle
\newpage

\section{Introduction}
\label{sec:introduction}

An era of large galaxy surveys will soon inaugurate a fertile period
for the study of large-scale structure in the Universe.
Current or planned surveys include the Dark Energy Survey (`DES')~\cite{Abbott:2018jhe},
Euclid~\cite{Laureijs:2011gra}, the Dark Energy Spectroscopic Instrument (`DESI'),
the Large Synoptic Survey Telescope (`LSST'), the Square Kilometre Array (`SKA')
and the 4-metre Multi Object Spectroscopic Telescope
(`4MOST')~\cite{4most}.
Each survey measures the distribution of \emph{tracers} of the
underlying dark matter distribution. Examples include dark matter
haloes and galaxy clusters, but also quasars, the Lyman-$\alpha$ forest,
or radiation from the 21-centimetre hyperfine transition of hydrogen.
Unfortunately it is not trivial to predict the statistical distribution
of these tracers, even if the distribution of dark matter is known
accurately, because it involves poorly-understood astrophysical
processes such as galaxy formation.

Whatever statistical relation
exists between tracers and dark matter,
it can be expanded in perturbation theory
on scales much larger than the characteristic scale of the tracers.
On these scales the density contrast is small and it is reasonable to expect
that an adequate description can be found by retaining only a few low-order terms
in the perturbation expansion; for a recent review, see, eg., Ref.~\cite{Desjacques:2016bnm}.
The unknown astrophysical processes that characterize the relationship (or `bias')
are encoded in the coefficients of this expansion.
They are very difficult to predict from first principles, but
if not too many
are required then we can simply \emph{measure} them. The disadvantage of this
approach is that we must expend some data in constraining the unknown bias
coefficients. This inevitably increases the uncertainty with which
remaining physical
quantities can be measured.

 The number of terms in the perturbation expansion relates directly to the range of predictions that can be made. A simple model will lead to a smaller range---but, while the resulting statistical measurements of the parameters may be more precise, any recovered values are likely to be inaccurate because the model fails to the relationship between matter and galaxies correctly. We describe this situation as \emph{underfitting the data}. Conversely, allowing a more general relationship will expand the range of predictions that can be made, but some statistical power will be wasted in constraining unnecessary coefficients. It also exposes us to the risk of \emph{overfitting}---that is, misinterpreting the random variation between realizations as a meaningful signal. To make best use of the data as it arrives we must find a compromise, balancing simplicity of description against the minimum complexity needed to match the sophistication of next-generation surveys.

 What is the appropriate level of complexity of modelling for a present-day or near-future survey? A typical previous-generation survey would have used a simple linear truncation. But, as the sample of galaxies available for large-scale analysis increases, an opportunity exists for extracting cosmological information on smaller and smaller scales where more complex biasing schemes have been found necessary to extract unbiased estimates (in the statistical sense) of the underlying cosmological parameters, or to obtain consistent results between different $n$-point functions. For this reason more recent surveys have begun to adopt prescriptions that include terms at quadratic order or higher, representing the strength of the tidal gravitational field or related quantities~\cite{Krause:2017ekm,delaTorre:2016rxm,Slepian:2016kfz,Beutler:2016arn,Gil-Marin:2016wya}.

\para{Bias modelling}
In this paper we quantitatively address the question of the appropriate
level of modelling sophistication required for analysis of a
present-day or near-future galaxy survey. In `Standard Perturbation Theory'
(or `SPT') we organize each contribution to a correlation function according to
its order in the `loop expansion'. In this scheme each loop corresponds to
an unrestricted average over wavenumbers~\cite{1981MNRAS.197..931J,1983MNRAS.203..345V,
Fry:1983cj,Goroff:1986ep,Suto:1990wf,Makino:1991rp,Scoccimarro:1995if,Scoccimarro:1996se}.

Up to one-loop, McDonald \& Roy wrote down the most general bias prescription
that respects the equivalence principle and is local in the sense of depending
on the fields and their spatial derivatives at a single time~\cite{McDonald:2009dh}.
The McDonald \& Roy scheme therefore bears the same relationship to the linear
truncation as one-loop SPT does to tree-level.
Their discussion was refined by Chan et al., who phrased their analysis
in terms of a slightly different basis of local operators~\cite{Chan:2012jj}.
We give the mapping between the McDonald \& Roy and Chan et al. expansions in
Appendix~\ref{app: Dictionary Perko-Assassi}.

Generalizations of McDonald \& Roy's prescription are possible.
Haloes are assembled over cosmological timescales, so it could happen that
the tracer density at time $t$ depends on the advective history of the
dark matter fluid at earlier
times~\cite{Wang:2012fr,Senatore:2014eva,Mirbabayi:2014zca}. This makes the
bias expansion more complex.
But it could also happen that there are dynamical reasons for different bias coefficients
to be related to each other, making the expansion simpler~\cite{Saito:2014qha}.

\para{Redshift-space modelling}
Bias is not the only effect that must be modelled carefully.
A galaxy survey does not measure actual spatial configurations, because the
radial distance to each galaxy must be estimated from its redshift.
This determination is confused by peculiar motions, giving rise to
so-called `redshift-space distortions'~\cite{Jackson:2008yv,Kaiser:1987qv}.
The large-scale angular distortion due to coherent infall
towards local overdensities
is known to all orders, assuming non-relativistic motion
of the sources~\cite{Scoccimarro:2004tg}.
We may account for its effects to as many
loop orders as we wish, but these all depend on modelling of local velocities.

Tree-level contributions to the two-point function are
well-understood.
The analogue of tree-level perturbation theory is the
Kaiser formula, which is the basis for most existing
treatments~\cite{Kaiser:1987qv,1989MNRAS.236..851L,1990MNRAS.242..428M,Cole:1993kh}.
Its reach in $k$ is known to be limited because it does not account
for the `fingers-of-God'
suppression at quasi-linear scales to be described below.
The extension to one-loop contributions was performed in SPT by
Matsubara~\cite{Matsubara:2007wj}.
Inclusion of time dependence beyond the Einstein-de Sitter approximation
at one-loop
was discussed in Ref.~\cite{Fasiello:2016qpn},
and in redshift space in Ref.~\cite{delaBella:2017qjy}.
A phenomenologically-successful hybrid formula, including some elements of the one-loop
SPT result together with other elements of a different origin,
was given by Taruya, Nishimichi \& Saito~\cite{Taruya:2010mx}.
This is commonly known as the `TNS model'.

\para{Fingers-of-God and Effective Field Theory}
Coherent infall is not the only source of distortion.
$N$-body simulations consistently exhibit strong suppression of power at
quasi-linear wavenumbers and above, often ascribed to erasure
of correlations due to virialization on subhalo scales.
This is the `fingers-of-God' effect~\cite{Jackson:2008yv,Taruya:2007xy}.
It is a short-wavelength phenomenon that is not captured at low orders in
perturbation theory
and must be accounted for in some other way.
One option is to introduce an ad hoc phenomenological
suppression~\cite{Peacock1992,Peacock:1993xg}, usually by guessing a
suitable functional form.
Alternatively, the methods of effective field theory furnish a systematic
parametrization of the influence of short-wavelength modes on long-wavelength
physics.

This procedure has been elaborated by a number of authors as the
`Effective Field Theory of Large-Scale
Structure'~\cite{Baumann:2010tm,Carrasco:2012cv,
Carrasco:2013mua,Porto:2013qua,Senatore:2014via,Senatore:2014vja,
Vlah:2015sea,Vlah:2015zda}.
In this framework, as in any application of effective field theory,
one introduces counterterms to supply the ultraviolet parts of loop
integrals that cannot be estimated reliably on the basis of low-energy
perturbation theory.
Counterterms due to loops from the redshift-space map should be regarded as
the natural means to describe subtraction of power due to short-scale motions.
As we shall see, they give a rather more flexible description of the subtraction
than the most widely-used phenomenological parametrizations.

Estimates for the power spectrum
of a tracer population including some or all of these effects have appeared
in the literature.
The one-loop bias
corrections to the matter--tracer correlation function and tracer autocorrelation function
were computed by McDonald \& Roy~\cite{McDonald:2009dh}.
Formulae for the same correlation functions including all one-loop effects in clustering
and bias (this time accounting for advective contributions from the fluid history)
were given by Angulo et al.~\cite{Angulo:2015eqa},%
    \footnote{The expressions in this reference were later corrected by
    Fujita et al.~\cite{Fujita:2016dne}.}
and after including all one-loop effects in clustering, bias and redshift-space
by Perko et al.~\cite{Perko:2016puo}.
These authors worked within the effective field theory framework and determined
the counterterms necessary to parametrize unknown short-wavelength effects.
They both applied a resummation technique to account
for damping of the baryon acoustic oscillation
due to large-scale coherent motions.

The modelling burden due to accounting for all these
different effects is significant.
Taking one-loop effective field theory as an example,
we must obtain analytic expressions for the one-loop integrals
due to clustering, bias, and redshift-space effects,
and use these to deduce the pattern of counterterms.
The integrals themselves must
be evaluated numerically, usually by Monte Carlo methods,
requiring non-negligible compute time.
Further integrations are typically required to resum displacements,
if this step is performed.
Finally, the free counterterms must be determined, either in a 2-step process
in which some data is sacrificed for the purpose of obtaining a fit, or by
marginalizing over them as nuisance parameters in a larger Markov chain.
All this requires a significant investment in analytic calculations,
software pipelines, and compute time for parameter fits---and
generalization to nonstandard scenarios,
such as modified gravities, requires further investment of a similar scale.

\para{Outline}
In this paper we study whether this modelling burden is justified.
For each combination of bias and redshft-space model,
we determine the goodness-of-fit to an ensemble of
COLA-accelerated $N$-body simulations~\cite{Tassev:2013pn}.
Our realizations are drawn from the {\WizCOLA} simulation suite, which
was generated to provide accurate covariance matrices for the WiggleZ galaxy redshift
survey~\cite{Kazin:2014qga,Johnson:2015aaa}.
Each mock halo catalogue was used to simulate the power spectrum reconstructed
from real WiggleZ measurements, accounting for effects of measurement error,
survey geometry, incompleteness, and uncertainties due to the use of photometric redshifts.
Therefore our performance measurements will account for
a typical selection of effects
relevant to a modern galaxy redshift survey, rather than representing the
performance of each layer under idealized conditions.
We regard this as a significant virtue of using the {\WizCOLA} suite.

The {\WizCOLA} realizations give measurements for the
$\ell=0$, $\ell=2$ and $\ell=4$ Legendre modes of the redshift-space power spectrum
up to $\kmax = 0.29 h/\Mpc$.
With this relatively low value of $\kmax$,
the WiggleZ team found that they could recover unbiased estimates of the underlying
cosmological parameters using a linear bias model
and the tree-level Kaiser formula for the redshift-space map.
The linear power spectrum was estimated using a {\Halofit} prescription.
We take this simple model as the baseline for all our comparisons.

This paper is organized as follows.
In {\S}\ref{sec:modelling} we define and review the different models for bias and
redshift-space effects that will be used in our performance analysis.
In {\S}\ref{sec:Analysis} we present our results for fitting to a subsample
of ten realizations drawn from the {\WizCOLA} suite.
We contrast these with fits to the ensemble average over the full set of
600 {\WizCOLA} realizations as a means to test for overfitting.
Our main conclusions are summmarized in {\S}\ref{sec:Conclusions}.

\para{Code availability}
The calculations needed for the complete one-loop power spectrum, including
clustering, bias and redshift-space loops, are very lengthy. To validate our results
we compare outputs between independent implementations of the entire pipeline.

The first implementation is based on traditional hand-calculation of the loop integrals.
These are translated into Mathematica, also by hand.
The second implementation uses semi-custom computer algebra methods to automate
the computation of all loop integrals directly from an SPT expression for the
overdensity $\delta$.
Appropriate {\CC}
code to perform numerical integration is generated automatically to avoid
errors due to typos, omissions, or other accidents of translation.
We find agreement between these pipelines up to the expected variance
due to Monte Carlo implementation.

To assist those wishing to replicate our results we have made the code
available for our most developed pipeline.
There are three components.
\begin{enumerate}
    \item The main integration engine (`\sansbold{LSSEFT}').
    This can be downloaded from GitHub. The
    calculations reported in this paper were performed using
    revision
    \GitRevision{134a6bcb}{https://github.com/ds283/LSSEFT/commit/134a6bcbaf0d1658f3d134573b25fd1c2f931d1b}.

    \item The analytic support tool (`\sansbold{LSSEFT-analytic}'), used to
    autogenerate code to perform the 1-loop integrals. The calculations
    reported in this paper were performed using revision
    \GitRevision{82c0465f}{https://github.com/ds283/LSSEFT-analytic/commit/82c0465f2d44682d347a7db6c970fdef0d0fbe7b}.

    \item The parameter-fitting pipeline (`\sansbold{LSSEFT-haloeft}'),
    which is implemented
    using the \sansbold{CosmoSIS} platform~\cite{Zuntz:2014csq}.
    The calculations reported in this paper were performed using revision
    \GitRevision{6c82cc08}{https://github.com/ds283/LSSEFT-haloeft/commit/6c82cc08c6c6639ab898970a589204699e3fb977}.
\end{enumerate}

\para{Data availability}
The data products used in this paper are
a subsample of the {\WizCOLA} power spectrum suite~\cite{Koda:2015mca}.
We use only the processed power spectra, covariance matrices and
convolution matrices produced by the WiggleZ team.
The simulations and their derived products are curated by the WiggleZ
team and are not yet available from a public repository.

\para{Notation}
Throughout this paper we use units in which $c = \hbar = 1$.
We define the reduced
Planck mass to be $\Mp \equiv (8\pi G)^{-1/2}$.
Our Fourier convention
is
\begin{align}
f(\vect{x}) & = \int \d^3 k \, (2\pi)^{-3} \,f(\vect{k}) \e{\im \vect{k}\cdot\vect{x}}, \\
[f(\vect{x})]_{\vect{k}}\equiv f(\vect{k}) & = \int \d^3 x \, f(\vect{x}) \e{- \im \vect{k}\cdot\vect{x}}.
\end{align}
Latin indices $a$, $b$, \ldots, from the beginning of the alphabet
range over space-time coordinates $(t, x, y, z)$ or $(0, 1, 2, 3)$.
Latin indices $i$, $j$, \ldots, from the middle of the alphabet
range over spatial indices only.
Repeated space-time indices are taken to be contracted with the
metric $g_{ab}$. Repeated spatial indices in the `up' position are contracted with the
three-dimensional Euclidean metric $\delta_{ij}$, so that (for example)
$v^2 = v^i v^i = \delta_{ij} v^i v^j = \sum_{i} (v^i)^2$. Finally,
we use subscripts to identify different variants of the same field:
$s$ refers to a redshift-space quantity;
$\delta$ refers to dark matter;
$h$ refers to halo quantities;
$\theta$ refers to the velocity divergence.
A superscript $\renormalizedtag$
denotes a renormalised operator,
and
a superscript
$(n)$ refers to the $n^{\text{th}}$ order in perturbation theory.

\section{Modelling: bias prescriptions and the redshift-space map}
\label{sec:modelling}
In this section we review a very general model for perturbative bias
up to one-loop, which we call the `advective bias model'.
It is constructed from local operators in the fields and their
spatial derivatives, integrated over the advective past of the
dark matter fluid,
and includes stochastic contributions.
It is therefore weakly%
    \footnote{By `weakly nonlocal' at a position $\vect{x}$
    we mean that the expansion
    includes information from the neighbourhood of $\vect{x}$
    carried by derivatives of the fields, but does not explicitly
    depend on field values at positions $\vect{x}' \neq \vect{x}$.}
nonlocal in space and nonlocal in time.

\subsection{Perturbative bias}
\label{sec:bias}

\subsubsection{Non-local Eulerian bias}
\label{sec:nonlocal-bias}
\para{Expansion in rotational invariants}
We first review the model,
building on the discussion by Assassi et al.~\cite{Assassi:2014fva}.
The halo overdensity at a given location
is not restricted to depend only on the underlying dark
matter distribution; it may
also depend on rotational invariants formed from
the gravitational potential and the matter velocity field.
If we require the expansion to satisfy the equivalence principle,
we may use arbitrary rotational invariants formed from the
gravitational tidal tensor $\partial_i \partial_j \Phi_g$
and the fluid expansion tensor $\partial_i \partial_j \Phi_v
= \partial_i v_j$~\cite{McDonald:2009dh,Chan:2012jj,Assassi:2014fva,Senatore:2014eva},
or their derivatives.
Here, $\Phi_g$ is the Newtonian gravitational potential
and $\partial_i \Phi_v = v_j$,
where $\vect{v}$ is the comoving fluid peculiar velocity.

\begin{figure}
    \centering
        \includegraphics[scale=0.4]{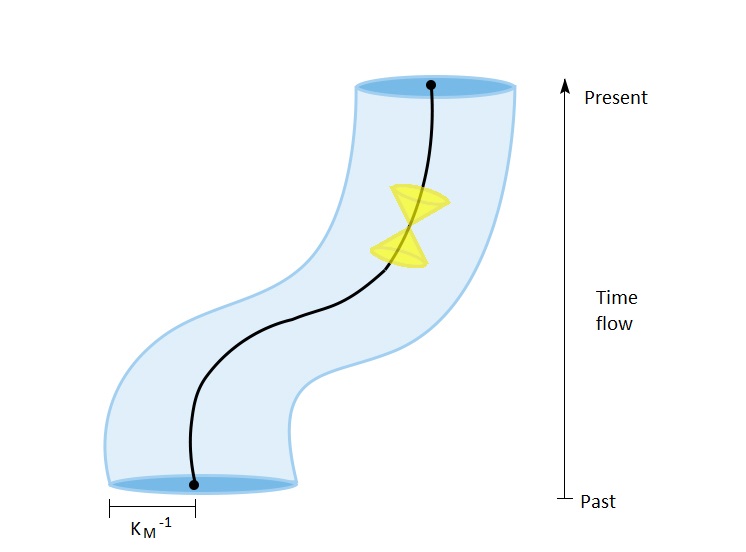}
    \caption[Halo time and space non-local]{Spacetime history of a collapsed object
    from the past to today along its worldline, showing the halo mass scale,
    $k_M \sim 2\pi / R_M$,
    and the light cone (in yellow). \\
    {\footnotesize \sansbold{CREDIT} M. de la Bella.}}
    \label{figure:cuchy}
\end{figure}

At lowest order, the expansion begins with terms containing only two spatial derivatives
for each factor of $\Phi_g$ or $\Phi_v$.
Higher-derivative terms would typically be suppressed by a dimensionful scale of order
$1/R_M$, where $\rho R_M^3 \sim M$.
Here, $\rho$ is the mean energy density of matter and $M$ is the halo mass.
Terms of this kind
represent a sensitivity to the environment of the halo during formation,
defined as the approximate region from which matter is swept up to form
the overdensity; see Fig.~\ref{figure:cuchy}.
In addition there may be `stochastic' contributions
that capture injection of energy from the bath of
modes above the cutoff of the theory.
We write these generically as $\epsilon$.

Table~\ref{tab:operators}
lists the eligible operators---in general, denoted by $\Operator (x)$---that satisfy both rotational invariance and the
equivalence principle, up to third order in perturbation theory and with
two spatial derivatives per field.
\begin{table}
	\setlength{\tabcolsep}{10pt} 
	\renewcommand{\arraystretch}{1.5} 
	\rowcolors{2}{gray!25}{white}
	\centering
	\begin{tabular}{ll}
		\toprule
		Order & Operators
		\\
		\midrule
		Linear & $\delta = \, \partial ^2 \Phi _g$ \\
		Quadratic &
		\begin{tabular}[t]{@{}S@{}}
			\delta ^2 = (\partial ^2 \Phi _g)^2,
			\quad \mathcal{G}_2(\Phi _g) \equiv (\partial _i \partial _j \Phi _g)^2 - (\partial ^2 \Phi _g)^2
		\end{tabular}
		\\
		Cubic &
		\begin{tabular}[t]{@{}S@{}}
			\delta ^3,
			\quad \mathcal{G}_2(\Phi _g)\delta ,
			\\
			\mathcal{G}_3(\Phi _g) \equiv  - \frac{1}{2} \left[
				2 (\partial _i \partial _j \Phi _g)(\partial ^i \partial _k \Phi _g)(\partial ^k \partial ^j \Phi _g)
				+ (\partial ^2 \Phi _g)^3
				- 3(\partial _i \partial _j \Phi _g)^2 (\partial ^2 \Phi _g)
			\right], \\
			\Gamma _3 (\Phi _g, \Phi _v) \equiv \mathcal{G}_2(\Phi _g) - \mathcal{G}_2(\Phi _v)
		\end{tabular}
		\\
		\bottomrule
	\end{tabular}
	\caption[Building blocks of local Eulerian bias]{Independent operators contributing
	to local Eulerian bias up to third order in perturbation theory~\cite{Chan:2012jj}.}
	\label{tab:operators}
\end{table}
It follows that the leading contribution to the bias expansion, up to one-loop order,
can be written~\cite{Desjacques:2016bnm}
\begin{equation}\label{eq:bias-integral}
\deltahalo(x) =
	\int_{\mathcal{I}^{-}(x)}
	\d^4 x' \sum_{\Operator} \kappa_{\Operator}(x,x')	b_{\Operator} \Operator(x')
\end{equation}
where $x$ and $x'$ are spacetime coordinates,
$\mathcal{I}^-(x)$ is the interior of the past lightcone at $x$,
and $\kappa_{\Operator}(x,x')$ is a `memory kernel'
that describes the influence of fluctuations at location $x'$ on
the halo density field at location $x$. In this form the bias expansion is nonlocal in both time and space. Although not required,
it is usually assumed that
the bias parameters
$b_{\Operator} = \{ b_1, b_2/2!, b_3/3!, b_{\mathcal{G}_2}, b_{\mathcal{G}_2\delta}, b_{\mathcal{G}_3}, b_{\Gamma_3} \}$
are functions of time but not space, so Eq.~\eqref{eq:bias-integral}
implicitly requires a choice of slicing. From now on, we work in
spacetime coordinates adapted to this slicing.

\para{Halo formation history}
Consider the worldline of a nascent halo.
This satisfies
$\D \vect{\xfl} / \D t = 0$,
where $\D/\D t = \partial / \partial t + \vect{v} \cdot \nabla$
is the advective derivative along the halo velocity field $\vect{v}$.
Although this velocity might not coincide with the dark matter
velocity, they agree at zero wavenumber and
therefore differ only by terms suppressed by at least one derivative.
In this paper we assume they can be regarded as equal
on sufficiently large scales.
The worldline with boundary condition $\xfl = \vect{x}$
at time $t$ has position
\begin{equation}
    \label{eq:xfl}
    \xfl(t,t')
    =
    \vect{x}
    - \int^t_{t'} \d t'' \; \vect{v} [ \xfl(t,t'') ]
\end{equation}
at earlier times $t' < t$, and
if the displacement $\xfl - \vect{x}$ is not large we may expand
perturbatively in $\vect{v}$, giving
\begin{equation}
    \label{eq:xfl1}
    \xfl(t,t')
    =
    \vect{x}
    - \int_{t'}^t \d t'' \; \vect{v} (\vect{x},t'')
    + \int_{t'}^t \d t'' \int _{t''}^{t'} \d t'''
        \; \vect{v} (\vect{x},t''') \cdot \nabla \vect{v} (\vect{x},t'')
    + \Or(\vect{v}^3).
\end{equation}

We expect $\kappa_{\Operator}(x,x')$ to be strongly peaked within a radius $\sim R_M$ of
the worldline $\xfl(t,t')$.
Therefore~\eqref{eq:bias-integral}
will be dominated by an integral along the worldline, accompanied by
higher-derivative corrections suppressed by the localization scale
$k_M \sim 1/R_M$. These can be neglected on scales much larger than
$R_M$, which is sufficient for the description of large-scale clustering.%
    \footnote{This discussion shows that
    higher-derivative terms in the bias expansion
    are likely to be more relevant for very massive tracers.
    For the most massive objects it is possible that higher-derivative
    terms should be taken into account.}
Therefore
\begin{equation}
    \deltahalo(\vect{x},t)
    =
    \int^t_{-\infty} \d t' \sum_{\Operator} \kappa_{\Operator}(t,t')	b_{\Operator} \Operator(\xfl(t,t'))
    \, + \Or(\partial^2/k_M^2)
\label{eq:worldline-bias-integral}
\end{equation}
The lower limit of integration $t=-\infty$ is schematic. In reality, the
memory kernel $\kappa(t,t')$ will cut off contributions at times much earlier than $t$.
Provided the displacement $\xfl-\vect{x}$ does not grow too large before
this suppression becomes effective we can use
Eq.\eqref{eq:xfl1}
to determine $\delta\big( \xfl[t,t'] \big)$ in terms of
$\delta(\vect{x})$ and its derivatives,
and likewise for all other operators.
This recovers weak nonlocality in space
while retaining explicit nonlocality in time,
at the cost of introducing higher-order
derivatives contracted with $\vect{v}$.

\para{Einstein--de Sitter approximation}
In general this is as far as we can go, because the integral
in~\eqref{eq:worldline-bias-integral} cannot be performed analytically.
But there is a simplification
in the special case that $\delta$ describes the
overdensity of cold dark matter in the matter era.
In this case $\delta$ factorizes into a sum
of products of
time- and space-dependent functions
at each order in perturbation theory.
If we write
\begin{equation}
	\delta(t, \vect{x})
	=
	\sum_{n \geq 1} \delta^{(n)}(t, \vect{x}) ,
	\label{eq:deltan-def}
\end{equation}
where $\delta^{(n)}(t,\vect{x})$ contains terms of exactly $n^{\mathrm{th}}$ order
in perturbation theory,
then
\begin{equation}
	\delta^{(n)}(t,\vect{x})
	= \sum_j D_j^{(n)}(t) F_j^{(n)}(\vect{x}) .
	\label{eq:deltan-decompose}
\end{equation}
We describe the functions
$D_j^{(n)}(t)$ as `growth functions'.
At linear order the sum contains just one term,
$D^{(1)}(t) F^{(1)}(\vect{x})$.
The function $D^{(1)}(t)$ is the linear growth function,
often abbreviated simply $D(t)$,
and $F^{(1)}(\vect{x}) = \delta^\ast(\vect{x})$.
The superscript `$\ast$' denotes evaluation at some
initial time $t^\ast$ early in the matter era,
at which time
$D(t)$ satisfies the boundary
condition $D(t^\ast) = 1$.
A similar factorization applies to $\theta(t,\vect{x})$.

In much of the literature Eqs.~\eqref{eq:deltan-def}--\eqref{eq:deltan-decompose}
are simplified by adopting
the \emph{Einstein--de Sitter approximation}.
In an EdS universe each $D_j^{(n)}(t)$ can be written as a multiple of $D(t)^n$,
and therefore
$\delta^{(n)}$ can be written
$\delta^{(n)}(t,\vect{x}) = D(t)^n F_n(\vect{x})$.
For non-EdS cosmologies
the relation
$D_j^{(n)}(t) \sim D(t)^n$
is still approximately true,
with $D(t)$ the appropriate linear growth function
for the cosmology in question.%
	\footnote{Note that the Einstein--de Sitter approximation consists in
	assuming the relationship
	$D_j^{(n)} \sim D(t)^n$, \emph{not} in assuming that
	$D(t)$ is the linear growth function in an EdS cosmology.
	While the Einstein--de Sitter approximation is accurate to a few percent,
	$D(t)$ usually differs much more significantly from its EdS counterpart.}
Fasiello \& Vlah~\cite{Fasiello:2016qpn}
and Ref.~\cite{delaBella:2017qjy}
estimated the impact of this approximation to be of order a few percent
for the multipole power spectra.
Since this may be insufficient for very accurate calculations
we generally retain the full time dependence
for each $\delta^{(n)}$ and $\theta^{(n)}$.

\para{Quasi-local expansion}
However, as we now explain, we find it necessary for
practical reasons to
resort to the Einstein--de Sitter approximation in the bias prescription.
Temporarily
assuming the approximation, a term such
as $b_1(t) \delta(t, \vect{x})$
in the integrand of Eq.~\eqref{eq:worldline-bias-integral}
would yield~\cite{Senatore:2014eva}
\begin{equation}
\begin{split}
	\int_{-\infty}^t \d t' \,
	& \kappa_1(t,t') b_1(t')
		\Big[
			D(t') F_1(\vect{x})
			+ D(t')^2 F_2(\vect{x})
			+ D(t')^3 F_3(\vect{x})
			+ \cdots
		\Big]
	\\
	&
	\mbox{}
	=
	b_1^{(1)}(t) \delta^{(1)}(t,\vect{x})
	+
	b_1^{(2)}(t) \delta^{(2)}(t,\vect{x})
	+
	b_1^{(3)}(t) \delta^{(3)}(t,\vect{x})
	+
	\cdots .
\end{split}
\label{eq:quasi-local-b1}
\end{equation}
Following Senatore,
we have defined the time-dependent functions
$b_1^{(1)}$, $b_1^{(2)}$ and $b_1^{(3)}$ to satisfy~\cite{Senatore:2014eva}
\begin{subequations}
\begin{align}
	\label{eq:b1-1}
	\int_{-\infty}^t \d t' \, \kappa_1(t,t') b_1(t') D(t') \,  & \equiv b_1^{(1)}(t) D(t) \\
	\label{eq:b1-2}
	\int_{-\infty}^t \d t' \, \kappa_1(t,t') b_1(t') D(t')^2 \,  & \equiv b_1^{(2)}(t) D(t)^2 \\
	\label{eq:b1-3}
	\int_{-\infty}^t \d t' \, \kappa_1(t,t') b_1(t') D(t')^3 \,  & \equiv b_1^{(3)}(t) D(t)^3
\end{align}
\end{subequations}
Eq.~\eqref{eq:quasi-local-b1}
appears to be local in time,
but it splits the
perturbative orders $\delta^{(n)}(t,\vect{x})$ in a way
that cannot occur in a truly local expression:
the different orders of perturbation theory
$\delta^{(n)}$ have no meaning by themselves,
so Eq.~\eqref{eq:quasi-local-b1}
has meaning only as a convenient representation for the nonlocal time
integral in~\eqref{eq:worldline-bias-integral}.
We describe the expansion in~\eqref{eq:quasi-local-b1} as `quasi-local'.
The net effect is that, instead of a \emph{single} bias parameter $b_1(t)$
at linear order, we have three such parameters.
The degree of splitting between these parameters can be regarded as a
measure of the importance of halo histories in describing their statistical properties.
A similar discussion can be given for each operator appearing in the
integrand of~\eqref{eq:worldline-bias-integral}.

If we were to drop the Einstein--de Sitter approximation, then
a similar analysis would apply; except that we would
obtain an independent bias parameter for each growth function
$D_j^{(n)}(t)$.
The resulting bias expansion would be extremely complex
and contain an impractically large number of bias
coefficients.
To prevent the statistical power of our datasets becoming
diluted by this proliferation of bias parameters we
assume that~\eqref{eq:quasi-local-b1}
will approximately apply, even though we retain the full
time dependence of $\delta^{(n)}(t,\vect{x})$.
More specifically, this can be regarded as a sufficient approximation,
since all the bias parameters are determined at a fixed time. Additionally, further non-degenerate terms only appear at fourth order, as discussed in App. B of Desjacques \textit{et al.}~\cite{Desjacques:2016bnm}.
Effectively, this means that we use the Einstein--de Sitter
approximation in the \emph{bias expansion} even though we
do not use it for the description of \emph{dark matter clustering}.

\subsubsection{Advective terms}
\label{sec:Adective Terms}
Neglecting the velocity-dependent
corrections from Eq.~\eqref{eq:xfl1},
which we describe as `advective terms',
the quasi-local representation of the
bias prescription~\eqref{eq:worldline-bias-integral}
can be written in Fourier space as
\begin{equation}
\label{eq:quasi-local-simple}
\begin{split}
\deltahalo_{\vect{k}}
	\supseteq
	\mbox{} &
    b_1^{(1)} \delta^{(1)}_{\vect{k}}
    \\
	&
	+
	b_1^{(2)} \delta ^{(2)}_{\vect{k}}
	+ \frac{b_{2}^{(2)}}{2!} [ \delta ^{(1)\,2} ]_{\vect{k}}
	+ b_{\mathcal{G}_2}^{(2)} [\mathcal{G}_2^{(2)}]_{\vect{k}}
	\\
	&
	+
	b_1^{(3)} \delta^{(3)}_{\vect{k}}
	+ 2 \frac{b_{2}^{(3)}}{2!}
	   [ \delta ^{(1)} \delta ^{(2)} ]_{\vect{k}}
	+ b_{\mathcal{G}_2}^{(3)} [\mathcal{G}_2^{(3)}]_{\vect{k}}
	+ \frac{b_{3}}{3!} [ \delta ^{(1)\,3} ]_{\vect{k}}
	+ b_{1 \mathcal{G}_2} [\delta \mathcal{G}_2]_{\vect{k}}
	+ b_{\mathcal{G}_3} [\mathcal{G}_3]_{\vect{k}}
	+ b_{\Gamma _3} [ \Gamma_3] _{\vect{k}} ,
\end{split}
\end{equation}
where the symbol
`$\supseteq$' means that $\deltahalo$ contains these terms
among others that we have not written explicitly.

Eq.~\eqref{eq:quasi-local-simple} is corrected by
a number of advective terms.
To compute them
we apply the Einstein--de Sitter approximation
to~\eqref{eq:xfl1}
and obtain another quasi-local
expansion,
\begin{equation}\label{eq:xfl2}
\begin{split}
	\xfl(t,t') =
	    \vect{x}
	    - \mbox{}
	    &
	    \bigg(
	        1
	        - \frac{D(t')}{D(t)}
	    \bigg)
	    \frac{\vect{v}^{(1)}}{fH}
	    -
	    \frac{1}{2}
	    \bigg(
	        1
	        - \frac{D(t')^2}{D(t)^2}
	    \bigg)
	    \frac{\vect{v}^{(2)}}{fH}
	    \\ & \mbox{}
	        -
	        \Bigg[
	            \frac{D(t')}{D(t)}
	            - \frac{1}{2}
	            \bigg( 1 - \frac{D(t')^2}{D(t)^2} \bigg)
	        \Bigg]
	        \bigg(
	            \frac{\vect{v}^{(1)}}{fH} \cdot \nabla
	        \bigg) \frac{\vect{v}^{(1)}}{fH}
	    + \Or(\vect{v}^3)
	,
\end{split}
\end{equation}
where $\vect{v}^{(1)}$ and $\vect{v}^{(2)}$
are the terms in $\vect{v}$ of first and second order
in perturbation theory;
the linear growth factor is
$f= (\d \ln D /\d t)/H$;
and $D(t)$ is the linear growth
function~\cite{1981MNRAS.197..931J}.

Using~\eqref{eq:xfl2}
to expand around the fixed spatial coordinate $\vect{x}$,
as explained above,
together with Eqs.~\eqref{eq:b1-1}--\eqref{eq:b1-3},
the extra contributions
to $\deltahalo$ from use of
$b_1 \delta(\xfl)$ rather than
$b_1 \delta(\vect{x})$
in the integrand of~\eqref{eq:worldline-bias-integral}
can be written
up to third order in perturbation theory as
\begin{equation}
\label{eq:delta1-adv}
\begin{split}
	\deltahalo_{\vect{k}}
	\supseteq \mbox{}
	& -
	\Big(
        b_1^{(1)}
        - b_1^{(2)}
    \Big)
    \Big[
    	\frac{\vect{v}^{(1)}}{fH} \cdot \nabla \delta ^{(1)}
    \Big]_{\vect{k}}
    -
    \frac{1}{2}
    \Big(
        b_1^{(1)} - b_1^{(3)}
    \Big)
    \Big[
    	\frac{\vect{v}^{(2)}}{fH} \cdot \nabla \delta ^{(1)}
    \Big]_{\vect{k}}
    -
    \Big(
        b_1^{(2)} - b_1^{(3)}
    \Big)
    \Big[
	    \frac{\vect{v}^{(1)}}{fH} \cdot \nabla \delta ^{(2)}
	\Big]_{\vect{k}}
    \\
    &
    +
    \bigg[
        \frac{1}{2} \Big( b_1^{(1)} - b_1^{(3)} \Big) - b_1^{(2)}
    \bigg]
    \Bigg[
        \bigg(
            \frac{\vect{v}^{(1)}}{fH} \cdot \nabla
            \frac{\vect{v}^{(1)}}{fH}
        \bigg)
        \cdot
        \nabla \delta ^{(1)}(t,\vect{x})
        +
        \frac{v^{(1)\,i}}{fH} \frac{v^{(1)\,j}}{fH}
        \partial_i \partial_j \delta ^{(1)}
	\Bigg]_{\vect{k}} ,
\end{split}
\end{equation}
where all fields are evaluated at the same time and space coordinates
$(t,\vect{x})$.
Observe that in the limit that all $b_1^{(n)}$ are equal,
making the halo formation history irrelevant,
the corrections in~\eqref{eq:delta1-adv} disappear.

Structurally similar terms arise from the second-order
pieces $[\delta^2]^{(2)}(\xfl)$
and $\mathcal{G}_2^{(2)}(\xfl)$,
giving
\begin{equation}
	\label{eq:thirdorder-adv}
	\deltahalo_{\vect{k}}
	\supseteq
    -
    2
    \Big(
        b_2^{(2)}
        - b_2^{(3)}
    \Big)
    \Big[
	    \delta ^{(1)}
	    \frac{\vect{v}^{(1)}}{fH}
	    \cdot
	    \nabla \delta^{(1)}
	\Big]_{\vect{k}}
    -
    \Big(
        b_{\mathcal{G}_2}^{(2)}
        - b_{\mathcal{G}_2}^{(3)}
    \Big)
    \Big[
	    \frac{\vect{v}^{(1)}}{fH} \cdot \nabla \mathcal{G}_2^{(2)}
	\Big]_{\vect{k}}
    ,
\end{equation}
with the same understanding that all terms are to be
evaluated at $(t,\vect{x})$.
There are no advective corrections to third-order
terms in the expansion because these would enter at
fourth-order. It follows that these terms do not
contribute to the power spectrum at one-loop.

Notice here that we choose to neglect the difference between the dark matter and the halo velocity fields, $\vect{v}$ and $\vect{v}_h$, up to one loop, in the advective terms---equations \eqref{eq:delta1-adv}  and \eqref{eq:thirdorder-adv} above. That is mainly because the advective terms begin at second order in perturbation theory, but the counterterms are only supposed to appear at tree level. That means the difference can be neglected, at least for the power spectrum, in second-order terms. Therefore, working with the dark matter velocity field, $\vect{v}$, suffices.

The full advection-corrected bias expansion
comprises the sum of
Eq.~\eqref{eq:quasi-local-simple},
giving the `non-advective' pieces,
and
Eqs.~\eqref{eq:delta1-adv}--\eqref{eq:thirdorder-adv}
which comprise the advective terms.
In what follows we sometimes collect
Eqs.~\eqref{eq:delta1-adv}--\eqref{eq:thirdorder-adv}
into an advective piece $\deltaadvect$.
This contains
second- and third-order contributions
$\deltaadvect^{(2)}$
and
$\deltaadvect^{(3)}$,
ie. $\deltaadvect = \deltaadvect^{(2)} + \deltaadvect^{(3)}$.

\subsection{Redshift-space distortions}
\label{sec:Redshift-space distortions}
In addition to bias, we must model the effect of systematic
mis-estimation of radial distances due to the
peculiar velocities of whatever population we use as a tracer~\cite{Kaiser:1987qv,1989MNRAS.236..851L,
1990MNRAS.242..428M,Cole:1993kh, delaBella:2017qjy}.
As explained in~\S\ref{sec:introduction},
these are `redshift-space distortions'~\cite{Jackson:2008yv,Kaiser:1987qv}.
The one-loop calculations needed to match the one-loop bias
prescription developed in~\S\ref{sec:bias}
were carried out by Matsubara~\cite{Matsubara:2004fr}
and later replicated by Senatore \& Zaldarriaga~\cite{Senatore:2014vja}.
The calculation was extensively reviewed by
de la Bella et al.~\cite{delaBella:2017qjy},
whose terminology and notation we adopt.
Generalizations to non-Einstein cosmologies were studied by
Fasiello \& Vlah~\cite{Fasiello:2016qpn}.

Galaxy surveys estimate radial position by measuring
recession velocities from redshifts.
A survey operating in this way will mis-locate
a galaxy at real-space position $\vect{r}$
with peculiar velocity $\vect{v}$
to a displaced position
$\vect{s}$,
\begin{equation}
    \vect{s} = \vect{r} + \frac{\vect{v} \cdot \hat{\vect{r}}}{H} \hat{\vect{r}} .
    \label{eq:rsd-map}
\end{equation}
Eq.~\eqref{eq:rsd-map} applies in physical or comoving coordinates.
It neglects relativistic corrections of order $\sim v/c$,
but allows a significant displacement whenever the peculiar velocity $v$
is comparable to the Hubble flow $\sim Hr$ at the location of the tracer. Note the reader that to be rigorous, one could write the relation between the halo (tracer) velocity and the dark matter velocity as $v^i_h = v^i  + \beta(t) \partial^i \delta$, where $\beta(t)$ is an arbitrary time-dependent bias function---see equation (2.3) in Perko \textit{et al}~\cite{Perko:2016puo}. However, working with the tracer velocity field appears to be unnecessary, because its effect will be degenerate with some renormalization parameters, as we will show on section \ref{sec: Renormalisation}.

The mapping from $\vect{r}$ to $\vect{s}$
changes the appearance of matter
lumps but does not alter their concentration,
and therefore the total mass contained in a small coordinate region
is unchanged,
\begin{equation}
    \rho_s(\vect{s}) \, \d^3 s = \rho(\vect{r}) \, \d^3 r ,
    \label{eq:mass-conservation}
\end{equation}
where $\rho$ and $\rho_s$ are the matter density in real- and redshift-space,
respectively. Using Eqs.~\eqref{eq:rsd-map} and~\eqref{eq:mass-conservation},
Scoccimarro showed that
the overdensity fields
$\delta_s$ and $\delta$
for a fixed tracer,
measured respectively in the $\vect{s}$ and $\vect{r}$ coordinates,
are related by
\begin{equation}
    \delta_s(\vect{k})
    =
    \delta(\vect{k})
    +
    \int \d^3 r \; \e{-\im \vect{k}\cdot\vect{r}}
    \Bigg[
        \exp
        \Big(
            {-\frac{\im}{H}}
            (\vect{k}\cdot\hat{\vect{r}})
            (\vect{v}\cdot\hat{\vect{r}})
        \Big)
        - 1
    \Bigg]
    \Big[
        1 + \delta(\vect{r})
    \Big] .
    \label{eq:deltah-s}
\end{equation}

Eq.~\eqref{eq:deltah-s} applies
for any tracer and therefore can be used with an arbitrary bias
prescription.
Because it follows from~\eqref{eq:rsd-map}
it applies to all orders in $v/Hr$ but  only to leading order in $v/c$.
To use it for computation of correlation functions we should treat it
perturbatively in $\vect{v}$
and expand to whatever order is required.
Up to third order, appropriate for a one-loop calculation,
this yields~\cite{Heavens:1998es,Scoccimarro:1999ed,Matsubara:2007wj,Senatore:2014vja}
\begin{equation}
\begin{split}
    [\delta_s]_{\vect{k}}
    =
    [\delta]_{\vect{k}}
    &
    \mbox{}
    - \frac{\im}{H} k \mu
        [ \hat{\vect{r}}\cdot\vect{v} ]_{\vect{k}}
    - \frac{\im}{H} k \mu
        [ \hat{\vect{r}}\cdot\vect{v} \delta]_{\vect{k}}
    - \frac{1}{2!H^2} (k \mu)^2
        [ (\hat{\vect{r}}\cdot\vect{v})^2 ]_{\vect{k}}
    \\
    &
    \mbox{}
    - \frac{1}{2!H^2} (k \mu)^2
        [ (\hat{\vect{r}}\cdot\vect{v})^2\delta]_{\vect{k}}
    - \frac{\im}{3!H^3} (k \mu)^3
        [ (\hat{\vect{r}}\cdot\vect{v})^3 ]_{\vect{k}}
    +
    \Or(\delta^{4})
    .
\end{split}
\label{eq:delta-s-oneloop}
\end{equation}
where $\mu = \hat{\vect{k}}\cdot\hat{\vect{r}}$ is the cosine of the
angle between the line-of-sight from Earth $\hat{\vect{r}}$
and the Fourier mode $\hat{\vect{k}}$.

The full expression for the linear, quadratic and cubic part of the
redshift-space density contrast---using the advective bias model---can
be found in Appendix~\ref{app:p_calculations}.
The restriction of~\eqref{eq:delta-s-oneloop} to linear terms
gives the Kaiser formula,
\begin{equation}
	[\delta_s]_{\vect{k}}
	=
	[\delta]_{\vect{k}}
	-
	\frac{\im}{H} k \mu [ \hat{\vect{r}}\cdot\vect{v}]_{\vect{k}} .
	\label{eq:Kaiser-formula}
\end{equation}

As for~\eqref{eq:deltah-s},
the one-loop transformation~\eqref{eq:delta-s-oneloop}
applies for any bias prescription or none;
it is simply a transformation that maps a given overdensity field from
real-space to redshift-space.
We describe it as the `redshift-space map'.
These two layers of modelling---the bias and the redshift-space
power spectrum---are therefore entirely independent
and can be varied separately.

\para{Advective terms}
The term $[\hat{\vect{r}}\cdot\vect{v}\delta]_{\vect{k}}$
combines with the lowest-order advective term
$\deltaadvect^{(2)}$ to produce a new third-order
term that we also describe as advective.
This term is
\begin{equation}
\begin{split}
	\label{eq:adv_s}
	\deltaadvects^{(3)} & =
		- \im \frac{\mu}{H} \Big[ (\hat{\vect{r}} \cdot \vect{v}^{(1)}) \deltaadvect^{(2)} \Big]_{\vect{k}}
	\\
	& =
	\left( b_1^{(1)} - b_1^{(2)} \right) D^3 f \mu k \int_{\vect{q},\,\vect{r}}^{\vect{k}} \frac{\mu_{\vect{q}}}{q}
	\int_{\vect{s},\,\vect{p}}^{\vect{r}} \frac{\vect{s}\cdot\vect{p}}{s^2}\,
	\delta_{\vect{q}}^{*} \delta_{\vect{s}}^{*}\delta_{\vect{p}}^{*}
\end{split}
\end{equation}
where we have used the notation of Appendix~\ref{app:p_calculations}.

\subsection{Renormalization and effective field theory}
\label{sec: Renormalisation}
Up to this point we have developed the bias prescription and redshift-space
map in standard perturbation theory (`SPT'),
which
expresses how a set of initial Fourier modes
are transformed over time
by the clustering action of fluid flow under gravity.
Given the statistical properties of these initial Fourier modes
we can estimate the corresponding properties of their late-time
counterparts.
We classify each contribution by its order in the `loop expansion', with each
loop corresponding to one unconstrained momentum integration.
The one-loop terms in the power spectrum arise from
correlating two second-order terms, giving `22-type' contributions,
or correlating a first-order term with a third-order term, giving `13-type' contributions.

The loop integrals can be regarded as weighted averages
measuring the backreaction on a Fourier mode $\vect{k}$
from fluctuations at other scales.
Because the momenta contributing to these averages are
unrestricted, they include effects
from wavenumbers where SPT
does not provide even an approximately correct description.
This might happen because some physical process cannot be described at
fixed order in perturbation theory,
or simply because the relevant physics
was not included in the approximate description
with which we started;
typical examples of the latter
include virialization or stream-crossing on subhalo scales.
Therefore, by themselves, these averages are not trustworthy.
The solution is borrowed from quantum field theory, where
the correct description of physics at wavenumbers about a few hundred GeV
is not currently known
even in principle.
Rather than trying to predict the needed averages \emph{ab initio},
we parametrize their values in regions where our
description is inadequate and extract the necessary parameters from
measurement.
This is the programme of effective field theory.

\subsubsection{Renormalized operators}
Attempts have been made to apply this logic to structure
formation~\cite{Baumann:2010tm,Carrasco:2012cv,
Carrasco:2013mua,Porto:2013qua,Senatore:2014via,Senatore:2014vja,
Vlah:2015sea,Vlah:2015zda}.
A brief review with references to the original literature
is given in Ref.~\cite{delaBella:2017qjy}.
The procedure of parametrizing the unknown high-energy part of
each loop integral is equivalent to replacing
fields such as $\delta$
with a renormalized counterpart $\renormalized{\delta}$; eg. (see Ref.~\cite{delaBella:2017qjy})
\begin{equation}
	[\renormalized{\delta}]_{\vect{k}}
	=
	(1 - c_0)[\delta]_{\vect{k}}
	-
	c_2[\partial^2 \delta]_{\vect{k}}
	-
	c_4[\partial^4 \delta]_{\vect{k}}
	+
	\cdots
	-
	d_0 [\epsilon]_{\vect{k}}
	-
	d_2 [\partial^2 \epsilon]_{\vect{k}}
	+
	\cdots
	-
	\tilde{d}_1 [\epsilon \delta]_{\vect{k}}
	+
	\cdots
	.
\end{equation}
The renormalized field $\renormalized{\delta}$ may differ from the bare
field $\delta$
by the multiplicative renormalization $(1-c_0)$,
which represents an amplitude adjustment,
or by the additive renormalizations proportional to
$c_2$, $c_4$, {\ldots}, $d_0$, $d_2$, {\ldots}, $\tilde{d}_1$, {\ldots}
which represent mixing with other operators.
We treat the $c_i$ and products $d_i d_j$ of the $d_i$
as one-loop terms.
Each of the coefficients $c_n$, $d_n$, $\tilde{d}_n$ (and so on) may be time dependent,
and some or all of them may be zero,
but they are spatially independent.
For dark matter, conservation of mass and momentum force
$c_0 = d_0 = d_2 = 0$ when applied to the overdensity field~\cite{1980lssu.book.....P}.
For a tracer population whose
comoving number density is not conserved---for example, due to
formation or mergers of haloes or galaxies---these constraints are removed.

Mixing with operators such as $\partial^2 \delta$ describes backreaction
correlated with a field $\delta$
belonging to the low-energy effective description.
In constrast the field $\epsilon$ is stochastically independent of
all fields in the effective description.
Baumann et al. labelled such fields \emph{stochastic counterterms}~\cite{Baumann:2010tm}.
They describe random injection of energy from the bath of short-scale modes
to which the effective long-wavelength description is
coupled~\cite{Feynman:1963fq,Caldeira:1981rx,Caldeira:1982uj,Calzetta:1996sy}.
The inclusion of such `noise' terms
to describe uncorrelated short-scale effects
in the bias prescription has a long
history; see eg. Refs.~\cite{McDonald:2009dh,Desjacques:2016bnm}.
At one-loop in the power spectrum,
the $c_n$ counterterms renormalize 13-type integrals
and the $d_n$ counterterms renormalize 22-type integrals.

\para{Renormalization and bias}
The overdensity field for a biased tracer will typically not be associated with a
conserved comoving number density, and therefore we can expect
both additive and multiplicative renormalization at $\partial^0 \sim k^0$
and $\partial^2 \sim k^2$.
This corresponds to nonzero values for $c_0$, $d_0$ and $d_2$.

However, in an expansion such as~\eqref{eq:quasi-local-simple}
there is no need to include multiplicative counterterms.
When renormalizing an operator of the form
\begin{equation}
	[\deltahalo]_{\vect{k}}
	= b_1 [\delta]_{\vect{k}} + b_2 [\delta^2]_{\vect{k}} + b_3 [\delta^3]_{\vect{k}} + \cdots
\end{equation}
such counterterms would correspond to a
redefinition of the bias coefficients $b_1$, $b_2$, $b_3$, \ldots, and so on,
which are already arbitrary.
Also, because at each order in perturbation
theory we expand into a complete set of rotational invariants satisfying
the equivalence principle,
renormalization cannot generate mixing with new operators, but only higher derivatives
of operators that are already present~\cite{Assassi:2014fva}.
Further, it is plausible that the functional form of
the bias parameters
appearing in
Eqs.~\eqref{eq:delta1-adv} and~\eqref{eq:thirdorder-adv}
may be preserved by renormalization,
since the limit in which the order-by-order parameters $b_n^{(i)}$
are equal should still correspond to decoupling of the halo formation history.
On the other hand, if we artificially cut down the bias expansion
by omitting some operators
(as we will do for some models in {\S}\ref{sec:Analysis}),
then the resulting model need not be stable under renormalization, even if it
is phenomenologically acceptable.

\subsubsection{Renormalization of the tracer overdensity}
In practice we wish to renormalize
the overdensity for some biased tracer
$\deltahalo$
up to one loop.
The discussion above shows that we should allow the mixings
\begin{equation}
	\renormalized{[\deltahalo]}_{\vect{k}}
	=
	[\deltahalo]_{\vect{k}}
	-
	c_2 [\partial^2 \delta]_{\vect{k}}
	-
	d_0 [\epsilon]_{\vect{k}}
	-
	d_2 [\partial^2 \epsilon]_{\vect{k}}
	+
	\Or(\partial^4) .
	\label{eq:renormalization-prescription}
\end{equation}
At one loop we need only consider linear operators;
at two or more loops we should have to consider higher-order
operators built from products of $\delta$ and $\epsilon$.
The big-O
term $\Or(\partial^4)$ denotes operators of the form
$[\partial^4 \delta]_{\vect{k}}$
or
$[\partial^4 \epsilon]_{\vect{k}}$,
or their higher-derivative counterparts,
that we have not written. In principle we could consider them
(even at one loop), and they may assist in matching the power
spectrum at increasing $k$.
In practice, however, it is often not possible to obtain a good description
for such wavenumbers without also including nonanalytic terms from the
two- or higher-loop contribution.
Unfortunately there is no theoretical criterion that can be used to decide
ahead of time which order in $\partial^2$ matches which order in the loop expansion.
This is a question to be decided by comparison to data, even for the low-order terms
$[\partial^2 \delta]_{\vect{k}}$
and
$[\partial^2 \epsilon]_{\vect{k}}$.
In this paper we assume it is consistent to combine the one-loop power spectrum
with counterterms up to $\partial^2$, as in~\eqref{eq:renormalization-prescription}.%
    \footnote{Ref.~\cite{delaBella:2017qjy} validated this supposition for the
    dark matter power spectrum by showing that there is a range of $k$ in which
    the predicted and measured power spectrum differ by a term $\sim k^2 P(k)$.
    In principle the same could be done here, although the test becomes more complex
    as we introduce further counterterms.}

\para{Redshift space}
In redshift space the same considerations apply, except that the counterterms
$c_2$, $d_0$ and $d_2$ become functions of $\mu = \hat{\vect{k}} \cdot \hat{\vect{r}}$
in addition to time $t$.
As explained in Ref.~\cite{delaBella:2017qjy} they admit a series expansion
into powers of $\mu^2$ even though $\deltahalo$ itself does not.
Odd powers of $\mu$ are forbidden due to statistical isotropy.

The counterterms at $\mu^0$ are inherited from~\eqref{eq:renormalization-prescription}.
At higher powers of $\mu$ they can be regarded as corrections to the
redshift map due to small-scale motion.
It is known empirically that
the leading effect of these corrections
is to damp the power on quasilinear scales, perhaps because
randomized virial velocities erase coherent infall.
The presence of multiple counterterms means that the power
spectrum can be damped differently as $\mu$ varies.
In the multipole formalism to be described in~\S\ref{sec:Analysis}
this will allow the damping scale for the lowest few multipoles to be adjusted
independently.

This capacity for multiple adjustment is finite, however.
Whatever their origin,
the coefficients of the counterterm
series must be compatible with the condition
that $\renormalized{[\deltahalo]}_{\vect{k}}$ is renormalized
by local, rotationally-covariant operators.
As a consequence, the coefficients obey constraints imposed by
the operator product expansion.
Specifically, at lowest order in $k$
the composite operators appearing in
Eq.~\eqref{eq:delta-s-oneloop}
satisfy the operator product
expansions~\cite{Senatore:2014vja,Lewandowski:2015ziq,Perko:2016puo}
\begin{subequations}
\begin{align}
    \renormalized{[v_i v_j]}_{\vect{k}} & =
        [v_i v_j]_{\vect{k}}
        +
        \frac{H^2}{\kdamp^2}
        \Big(
            Z_1 \delta_{ij} + Z_2 \hat{k}_i \hat{k}_j
        \Big)
        [ \delta ]_{\vect{k}}
        + \epsilon_{ij}
    \label{eq:ope-vv}
    \\
    \renormalized{[v_i v_j \delta]}_{\vect{k}} & =
        [v_i v_j \delta]_{\vect{k}}
        +
        \frac{H^2}{\kdamp^2} Z_3 \delta_{ij}
        [ \delta ]_{\vect{k}}
        + \epsilon'_{ij}
    \label{eq:ope-vvd}
    \\
    \renormalized{[v_i v_j v_k]}_{\vect{k}} & =
        [v_i v_j v_k]_{\vect{k}}
        +
        \frac{H^2}{\kdamp^2}
        Z_4
        \Big(
            \delta_{ij} [v_k]_{\vect{k}} + \text{cyclic}
        \Big)
        +
        \text{stochastic} ,
    \label{eq:ope-vvv}
\end{align}
\end{subequations}
where $Z_1$, $Z_2$, $Z_3$, $Z_4$ are constants;
$\kdamp$ is a typical scale associated with damping of
power, as discussed above;
and $\epsilon_{ij}$ and $\epsilon'_{ij}$
are independent stochastic fields transforming as 2-index tensors
under rotations.
We have not written an explicit stochastic field for
$\renormalized{[v_i v_j v_k]}_{\vect{k}}$
because, as will be seen, it would not contribute to
$\renormalized{[\deltahalo]}_{\vect{k}}$
at $\Or(k^2)$.

In principle,
rotational covariance would allow
the OPEs for
$\renormalized{[v_i v_j \delta]}_{\vect{k}}$
and
$\renormalized{[v_i v_j v_k]}_{\vect{k}}$
(given in Eqs.~\eqref{eq:ope-vvd} and~\eqref{eq:ope-vvv})
to contain terms involving the tensor factor
$\hat{k}_i \hat{k}_j$ in addition to $\delta_{ij}$.
However a short calculation shows that they are absent at one loop,
and therefore~\eqref{eq:ope-vvv} cannot generate a counterterm at
$\mu^6$ in $\renormalized{[\deltahalo]}_{\vect{k}}$.

\para{Power spectrum renormalization}
So far we have discussed only renormalization of operators,
but similar arguments apply to correlation functions.
As we will see, the property that $\renormalized{[\deltahalo]}_{\vect{k}}$
has no renormalization at $\mu^6$ or above
will be critical in determining the structure
of the counterterms for the two-point function.

The renormalized redshift-space halo--halo
two-point function
is defined by
\begin{equation}
    \langle \renormalized{[\deltahalo_s]}_{\vect{k}}
    \renormalized{[\deltahalo_s]}_{\vect{k}'} \rangle
    = (2\pi)^3 \delta(\vect{k} + \vect{k}') \renormalized{[P^{hh}_s]}(k) ,
    \label{eq:def-Phhs}
\end{equation}
with a similar definition for its bare
counterpart
$\langle [\deltahalo_s]_{\vect{k}} [\deltahalo_s]_{\vect{k}'} \rangle$
and bare power spectrum $P^{hh}_s(k)$.
Renormalization can be performed using an expansion similar to
Eq.~\eqref{eq:renormalization-prescription}.
Specifically, at one loop we find
\begin{equation}
    \label{eq:renormalized-phh}
    \renormalized{[P^{hh}_s]}(k)
    = P^{hh}_{s}(k)
    + C_2(t,\mu) k^2 P_{\delta\delta}^{\text{tree}}(k)
    + D_0(t,\mu)
    + D_2(t,\mu) k^2 ,
\end{equation}
where $P^{hh}_s(k)$ is also computed to one loop,
and $P_{\delta\delta}^{\text{tree}}(k)$ is the matter
power spectrum computed to tree-level.
Concretely,
for $C_2(t,\mu)$ we find the missing
$\hat{k}_i \hat{k}_j$ terms in the OPE~\eqref{eq:ope-vvd}--\eqref{eq:ope-vvv}
imply
\begin{equation}\label{eq:C_2}
    C_2(t,\mu) = C_{2|0}(t) + C_{2|2}(t) \mu^2 + C_{2|4}(t) \mu^4 + C_{2|6}(t) \mu^6 ,
\end{equation}
subject to the constraint~\cite{delaBella:2017qjy}
\begin{equation}
    C_{2|6}(t) = f C_{2|4}(t) - f^2 C_{2|2}(t) + f^3 C_{2|0}(t) ,
    \label{eq:ope-c-constraint}
\end{equation}
where $f$ is the usual linear growth factor
$f = H^{-1} \d \ln D / \d t$.
There is no renormalization at $\mu^8$.

A clarification note on the difference between the dark matter and the halo (tracer) velocity fields. The only term in which we need to remember the difference between the tracer velocity and the dark matter velocity is the term linear in $\vect{v}$ in the lowest-order Kaiser formula---second term in equation~\eqref{eq:delta-s-oneloop}. By considering the tracer velocity, a new term of the form $k^2 \mu^2 \beta(t) \delta$ would rise, \textit{ie.} an unknown time-dependent term at order $k^2 \mu^2$ in the expansion. However, this is degenerate with the $\Or(\mu^2)$ term coming from the OPE renormalization of the composite operators in the redshift-space map---that is, the term $C_{2|2}$ in equation~\eqref{eq:C_2}.

\para{Stochastic terms}
The OPE also determines the stochastic counterterms.
The lowest-order stochastic counterterm in $\vect{v}$ will
be $\sim \nabla \epsilon$~\cite{Perko:2016puo},
and therefore any counterterms involving $\vect{v}$
will appear at $\Or(k^2)$ or above.
Rotational covariance forces expectation values such as
$\langle \epsilon \epsilon_{ij} \rangle$
or
$\langle \epsilon \epsilon'_{ij} \rangle$
to be proportional to either of the tensor factors
$\delta_{ij}$ or $k_i k_j$.
(Note that, unlike the case of $\vect{v}$,
there are no factors of $1/k$ available that could
convert $k_i k_j$ to $\hat{k}_i \hat{k}_j$.
Therefore going up a power of $\mu^2$ from the
same operator also costs a power of $k^2$.)
Assuming these expansions,
after translation to the two-point function,
we find the OPE implies $P^{hh}_s$ can be renormalized
by~\cite{Perko:2016puo}
\begin{subequations}
\begin{align}
    D_0(t,\mu) & = D_{0|0}(t) + D_{0|2}(t) \mu^2 \\
    D_2(t,\mu) & = D_{2|0}(t) \mu^2 .
\end{align}
\end{subequations}
Only the terms proportional to $\delta_{ij}$ are required, because
$k_i k_j$ produces contributions at $\Or(k^4)$.
The same is true for any stochastic term accompanying
$\renormalized{[v_i v_j v_k]}_{\vect{k}}$.

\subsection{Bias and redshift-space models}
\label{sec:Comparison with other models}

We are now in a position to build our different models for the bias
prescription and the redshift-space map.
The main raw material is the redshift-space halo--halo
power spectrum defined in~\eqref{eq:def-Phhs},
either renormalized or in its bare form.
The bare power spectrum can be computed up to one-loop using using
\begin{equation}
\label{eq:phh}
    [P_s^{hh}]^{\text{1-loop}}(k,z) = P^{hh}_{s,11}(k,z) + P^{hh}_{s,13}(k,z)  + P^{hh}_{s,22}(k,z) .
\end{equation}
We have switched the time variable from cosmic time $t$
to redshift $z$, defined as usual by $1+z(t) = a(t) / a_0$
where $a_0 = a(t_0)$ is the value of the scale factor today.
The required components are
\begin{subequations}
	\begin{align}
		\langle [\deltahalo_s]_{\vect{k}}^{(1)} [\deltahalo_s]_{\vect{k}'}^{(1)} \rangle
	    & =
		(2 \pi)^3 \delta(\vect{k} + \vect{k}')P^{hh}_{s,11}(k) , \\
		\langle [\deltahalo_s]_{\vect{k}}^{(1)} [\deltahalo_s]_{\vect{k}'}^{(3)} \rangle
		+
		\langle [\deltahalo_s]_{\vect{k}}^{(3)} [\deltahalo_s]_{\vect{k}'}^{(1)} \rangle
		& =
		(2 \pi)^3 \delta(\vect{k} + \vect{k}')P^{hh}_{s,13}(k) , \\
		\langle [\deltahalo_s]_{\vect{k}}^{(2)} [\deltahalo_s]_{\vect{k}'}^{(2)} \rangle
		& =
		(2 \pi)^3 \delta(\vect{k} + \vect{k}')P^{hh}_{s,22}(k)
		.
	\end{align}
\end{subequations}
For some models we also require the advective contributions, which likewise
break up into `13' and `22' components,
\begin{equation}
\label{eq:padv}
    P^{\text{Adv}}_{s}(k) = P^{\text{Adv}}_{s,\,13}(k) + P^{\text{Adv}}_{s,\,22}(k) .
\end{equation}
These are defined by
\begin{subequations}
	\begin{align}
		2 \langle [\deltahalo_s]^{(1)}_{\vect{k}} [\deltaadvect]^{(3)}_{\vect{k}'} \rangle
		+
		2 \langle [\deltahalo_s]^{(1)}_{\vect{k}} [\deltaadvects]^{(3)}_{\vect{k}'} \rangle
		& =
		(2 \pi)^3 \delta(\vect{k} + \vect{k}')P^{\text{Adv}}_{s,\,13}(k)
		,
		\label{eq:Padv-13}
		\\
		\langle [\deltahalo_s]^{(2)}_{\vect{k}} [\deltaadvect]^{(2)}_{\vect{k}'} \rangle
		+
		\langle [\deltaadvect]^{(2)}_{\vect{k}} [\deltahalo_s]^{(2)}_{\vect{k}'} \rangle
		+
		\langle [\deltaadvect]^{(2)}_{\vect{k}} [\deltaadvect]^{(2)}_{\vect{k}'} \rangle
		& =
		(2 \pi)^3 \delta(\vect{k} + \vect{k}')P^{\text{Adv}}_{s,\,22}(k)
		.
		\label{eq:Padv-22}
	\end{align}
\end{subequations}
where, as above,
$\deltaadvect$ is the sum of Eqs.~\eqref{eq:delta1-adv} and\eqref{eq:thirdorder-adv},
and
$\deltaadvects$ was defined in Eq.~\eqref{eq:adv_s}.
The `22' contributions contain cross-correlations between the nonlinear advective, bias and
clustering models.
Full details of all these calculations can be found in Appendix~\ref{app:p_calculations}.

\subsubsection{Power spectrum models}
We use the following models to describe the redshift-space power spectrum:
\begin{itemize}
\item \semibold{Kaiser formula in redshift-space} \cite{Kaiser:1987qv}
\begin{equation}\label{eq:kaisert}
    P^{hh}_s(k,z)
    = P^{hh}_{s,11}(k,z)
    = P_{\delta\delta}(k,z) + 2 f \mu ^2 P_{\delta \theta}(k,z) + f^2 \mu ^4 P_{\theta \theta}(k,z).
\end{equation}
There are no counterterms.
We consider two variants of this model.
In both cases we use the tree-level Kaiser formula~\eqref{eq:Kaiser-formula},
but the underlying matter power spectra can vary.
A simple choice is to use the linear power spectrum, eg. as computed by CAMB.
This is the {\KaiserTree} model.
Alternatively we can use a parametrized nonlinear power spectrum such as
{\Halofit};
we use the most recent implementation
due to Takahashi et al.~\cite{Takahashi:2012em}.
This produces a hybrid model that is tree-level in the redshift-space map,
but incorporates nonlinear corrections in the description of clustering.
We label this the {\KaiserHalo} model.

\item \semibold{One-loop standard perturbation theory in redshift space} \cite{Bernardeau:2001qr}
\begin{equation}\label{eq:sptt}
    P^{hh}_{s}(k,z) = [P^{hh}_s]^{\text{1-loop}}(k,z) .
\end{equation}
This is the {\SPT} model. It is given by Eq.~\eqref{eq:phh} but does not include any
counterterms.

\item \semibold{One-loop effective field theory in redshift
space} \cite{Perko:2016puo,Angulo:2015eqa,delaBella:2017qjy}
\begin{equation}\label{eq:eftt}
    P^{hh}_s(k,z) = \renormalized{[P^{hh}_s]}(k,z) .
\end{equation}
This is the {\EFT} model, using the renormalized
power spectrum~\eqref{eq:renormalized-phh}.
It differs from the {\SPT} model only through the
addition of the counterterms
$C_2$, $D_0$ and $D_2$. As we have seen, their
form is relatively restricted due to constraints imposed by the OPE.
\end{itemize}

\para{Fingers-of-God damping}
As we have explained,
to produce an acceptable power spectrum using any of these models
we must account for damping on quasilinear scales.
The effective field-theory framework already
accounts for such effects through its counterterms.
On the other hand, the {\KaiserTree}, {\KaiserHalo} and {\SPT} models by themselves
contain no mechanism to describe this damping.
Various phenomenological prescriptions exist to describe its effect, but a simple
model is suppress the power spectrum by an exponential factor
$\e{- \mu ^2 f^2  k^2 \sigmav^2}$~\cite{Peacock1992,Peacock:1993xg}.
This approximately describes damping on small scales due to the velocity dispersion $\sigmav$.

Once this exponential suppression factor has been included, the {\SPT} model is very similar to an alternative model
suggested by Taruya, Nishimichi \& Saito~\cite{Taruya:2010mx}. This `TNS' model has become increasingly
popular. Although we do not include it explicitly, for the value of ${\kmax}$
appropriate for {\WizCOLA} we expect the
{\SPT} model to approximate its predictions quite closely.

The key feature of the effective field theory model in redshift space is that
the independently adjustable counterterms associated with different powers of
$\mu$ allow \emph{independent suppression} of the $P_0$, $P_2$ and $P_4$
multipole power spectra
(see Eq.~\eqref{eq:def-Pell} below).
Using the exponential factor $\mathrm{e}^{- \mu ^2 f^2  k^2 \sigmav^2}$ there is
just a \emph{single} parameter $\sigmav$ that describes a common degree of
suppression in all $P_\ell$. The usefulness of the {\EFT} description compared
to the other models will
therefore largely depend on whether the data need to make use of this feature,
or whether a single suppression scale is sufficient.

\subsubsection{Bias models}
The bias models we use are:
\begin{itemize}
\item \semibold{Linear bias model} \\
This is the simplest bias model with which one could work.
The statistical relation between haloes and the underlying
dark matter perturbations in real space \cite{Davis:1985} becomes linear
\begin{equation}\label{eq:linearb}
    [\deltahalo]_{\vect{k}}  = b_1 [\delta]_{\vect{k}}.
\end{equation}
We label this model {\Linear}.

\item\semibold{Local McDonald \& Roy model} \cite{McDonald:2009dh} \\
McDonald \& Roy constructed the most general bias expansion up to one-loop using
purely local,
rotationally covariant functions of the fields and their spatial derivatives.
Their model is therefore a subset of the bias expansion
described in~\S\ref{sec:bias}.
When applied to the power spectrum their result can be written in the form
\begin{equation}\label{eq:M&Rb}
    [\deltahalo]_{\vect{k}} =
    b_1 [\delta]_{\vect{k}}
    + \frac{b_2}{2!} [\delta ^2]_{\vect{k}}
    + \frac{b_{s^2}}{2!} [s ^2]_{\vect{k}}
    + \bnl[\sigma_3^2 \delta]_{\vect{k}},
\end{equation}
where $\sigma_3^2(k) = \int \d(\ln r) \, \Delta ^2(k r) I_R(r)$;
for further details see Ref.~\cite{McDonald:2009dh}.
In our language
$[\sigma_3^2 \delta]_{\vect{k}} \equiv 105 \Big( s_{ij} t^{ij} + 8 \delta ^3 / 189\Big) / 32$
and, according to the dictionary in Appendix \ref{app: Dictionary Perko-Assassi},
the McDonald \& Roy model reduces to
\begin{equation}
    [\deltahalo]_{\vect{k}} =
    b_1 \delta _{\vect{k}}
    + \frac{b_{2}}{2!} \delta ^2_{\vect{k}}
    + b_{\mathcal{G}_2}[\mathcal{G}_2^{(2)}]_{\vect{k}}
	+ \frac{b_{3}}{3!} \delta ^{3}_{\vect{k}}
	+  b_{\mathcal{G}_2\delta} [\mathcal{G}_2 \delta]_{\vect{k}}
	+ b_{\Gamma _3} [ \Gamma _3] _{\vect{k}}
\end{equation}
with the following assignments
\begin{equation}
\begin{aligned}
    \frac{b_{2}}{2!} & \equiv \frac{b_{2}}{2!} + \frac{2}{3} \frac{b_{s^2}}{2!},
    &
    b_{\mathcal{G}_2} & \equiv \frac{b_{s^2}}{2!},
    \\
    \frac{b_{3}}{3!} & \equiv \frac{5}{6} \bnl,
    &
    b_{\mathcal{G}_2\delta} & \equiv - \frac{5}{8} \bnl,
    &
    \text{and} \hspace{3mm} b_{\Gamma _3} & \equiv \frac{105}{64} \bnl.
\end{aligned}
\end{equation}
This is properly local in time, rather than quasi-local.
The model therefore does not depend on halo formation history and does not
include advective contributions.
We label this model {\MRoy}.

\item\semibold{Co-evolution model} \cite{Saito:2014qha} \\
Saito et al. compared the McDonald \& Roy expansion to time-evolution
models using Lagrangian perturbation theory, and also the
results from $N$-body simulations.
On this basis they suggested a more constrained version of the
bias prescription that retains only $b_1$ and $b_2$ as free parameters.
The remaining coefficients $b_{s^2}$ and $\bnl$
satisfy
\begin{equation}\label{eq:coevb}
    \frac{b_{s^2}}{2!} \equiv -\frac{4}{7} \Big( b_1 -1 \Big)
    \quad
    \text{and}
    \quad
    \bnl \equiv \frac{32}{315} \Big( b_1 -1 \Big) .
\end{equation}
This is the {\Coevo} model.
As with {\MRoy} it is properly local in time and has no
advective terms.

\item\semibold{Advective model} \cite{Desjacques:2016bnm,Senatore:2014eva} \\
This is our most general model and retains all the features described in~\S\ref{sec:bias},
including advective contributions~\eqref{eq:Padv-13}--\eqref{eq:Padv-22}
and quasi-locality in time.

The full model contains a large number of parameters,
but at one-loop in the power spectrum not all of them are
independent~\cite{McDonald:2009dh,Perko:2016puo}.
We find it is possible to remove the operators
$[\delta^{(1)}\delta^{(2)}]$,
$[\delta^{(1) \, 3}]$,
$[\mathcal{G}_2]$,
$[\mathcal{G}_2 \delta]$
and $[\Gamma_3]$
by making the redefinitions
\begin{subequations}
\label{eq:deg}
\begin{align}
    b_1^{(1)} & \longrightarrow
    \begin{aligned}[t]
    b_1^{(1)}
        +
        \bigg[
            &
            \bigg( 1+ \frac{6 \DA + 8 \DB }{3D^2}\bigg) b_2^{(3)}
            + \frac{1}{2} b_3
            - \frac{4}{3} b_{\mathcal{G}_2\delta}
        \\ & \mbox{}
            + \frac{4}{3}
                \frac{\DA + \DB - f(\fA \DA + \fB \DB) + f^2 D^2}
                {\DA + \DB}
                b_{\Gamma_3}
        \bigg]
        \sigma^2
    \end{aligned}
    \\
     b_{\mathcal{G}_2}^{(3)} & \longrightarrow  b_{\mathcal{G}_2}^{(3)}  + \frac{\DA + \DB - f(\fA \DA + \fB \DB) + f^2 D^2}{\DA + \DB} b_{\Gamma _3}
\end{align}
\end{subequations}
where the variance $\sigma^2$ is defined by
\begin{equation}
    \sigma^2 = D(z)^2 \int ^{\Lambda} \frac{\d q}{4 \pi ^2} q^2 P(q) .
\end{equation}
In the Einstein--de Sitter approximation these redefinitions become
\begin{subequations}
\begin{align}
    b_1^{(1)} & \longrightarrow
    b_1^{(1)}
    +
    \bigg[
        \frac{55}{21} b_2^{(3)}
        + \frac{1}{2} b_3
        - \frac{4}{3} b_{\mathcal{G}_2\delta}
        + \frac{4}{3} \Big( 1 - \frac{3}{5} f^2 \Big) b_{\Gamma _3}
    \bigg] \sigma^2
    \\
    b_{\mathcal{G}_2}^{(3)} & \longrightarrow
    b_{\mathcal{G}_2}^{(3)}
    + \Big( 1 - \frac{3}{5} f^2 \Big) b_{\Gamma _3} .
\end{align}
\end{subequations}
After these redefinitions the number of \emph{independent} parameters needed to
describe the bias model is reduced to six,
\begin{equation*}
\{ b_1^{(1)}, b_1^{(2)}, b_1^{(3)}, b_2^{(2)}, b_{\mathcal{G}_2}^{(2)},  b_{\mathcal{G}_2}^{(3)} \}.
\end{equation*}
We describe this as the {\Advective} model.

Notice that the degeneracies we have identified
are accidental properties of the one-loop power spectrum.
Were we to compute higher-order correlation functions,
or to carry the calculation to further loops, we would eventually find that
all terms in the model become non-degenerate.
\end{itemize}

As a clarification note, we would like to remark the difference between the {\Advective} model and the {\MRoy} prescription. The latter started from a completely general basis of operators, so their model is supposed to be the unique local bias expansion up to one-loop in the power spectrum. Conversely, for the {\Advective} model, we  allow the expansion to be built up in terms of velocities, as well as including third-order matter fields, and derivatives\footnote{We prove that McDonald $\&$ Roy and Chan \textit{et al.}~\cite{Chan:2012jj} bases are completely equivalent -- see App.~\ref{Dictionary Perko-Assassi}. We derive a dictionary for up to third-order matter density fields that allows the reader to translate {\MRoy} notation into Chan et al. notation -- used in the ``advective'' model. }. Therefore, at the level of $\delta$ and derivatives, the {\Advective} model is degenerate with {\MRoy}. However, the products with the velocities are different, making both prescriptions fundamentally different.

Our model grid is constructed by taking all possible combinations of
bias and RSD model.
We summarize the possibilities in Table~\ref{tab:RSD+BIAS}.

\begin{table}
\centering
\begin{turn}{90}
\rowcolors{2}{gray!25}{white}
\begin{tabular}{cllll}
\toprule
& \multicolumn{1}{c}{\Linear}
& \multicolumn{1}{c}{\MRoy}
& \multicolumn{1}{c}{\Coevo}
& \multicolumn{1}{c}{\Advective} \\ \midrule
\KaiserTree & \begin{tabular}[c]{@{}l@{}}Eq. \eqref{eq:linearb} and\\ \eqref{eq:kaisert} $\times{\mathrm{e}} ^{- \mu ^2 f^2 k^2 \sigmav^2}$\\ and considering $P^{\text{CAMB}}_{\delta \delta}$\end{tabular} & \begin{tabular}[c]{@{}l@{}}Eq. \eqref{eq:M&Rb} \\ with \eqref{eq:kaisert} $\times{\mathrm{e}} ^{- \mu ^2 f^2k^2 \sigmav^2}$\\  and $P^{\text{CAMB}}_{\delta \delta}$\end{tabular}   & \begin{tabular}[c]{@{}l@{}}Eqs. \eqref{eq:M&Rb} \& \eqref{eq:coevb}\\ with \eqref{eq:kaisert}$\times{\mathrm{e}} ^{- \mu ^2 f^2 k^2 \sigmav^2}$\\ and $P^{\text{CAMB}}_{\delta \delta}$\end{tabular}    & \begin{tabular}[c]{@{}l@{}}Eq. \eqref{eq:delta-s-oneloop}\\ and \eqref{eq:adv_s}\\ with \eqref{eq:kaisert}$\times{\mathrm{e}} ^{- \mu ^2 f^2k^2 \sigmav^2}$\\ and $P^{\text{CAMB}}_{\delta \delta}$\end{tabular} \\
\KaiserHalo & \begin{tabular}[c]{@{}l@{}}Eq. \eqref{eq:linearb} \\  with \eqref{eq:kaisert} $\times{\mathrm{e}} ^{- \mu ^2 f^2k^2 \sigmav^2}$\\ and $P^{\text{Halofit}}_{\delta \delta}$\end{tabular}                      & \begin{tabular}[c]{@{}l@{}}Eq. \eqref{eq:M&Rb}\\ with \eqref{eq:kaisert} $\times{\mathrm{e}} ^{- \mu ^2 f^2 k^2 \sigmav^2}$\\ and $P^{\text{Halofit}}_{\delta \delta}$\end{tabular} & \begin{tabular}[c]{@{}l@{}}Eq. \eqref{eq:coevb}\\ with \eqref{eq:kaisert}$\times{\mathrm{e}} ^{- \mu ^2 f^2 k^2 \sigmav^2}$\\ and $P^{\text{Halofit}}_{\delta \delta}$\end{tabular} & \begin{tabular}[c]{@{}l@{}}Eq. \eqref{eq:delta-s-oneloop}\\ and \eqref{eq:adv_s}\\ with \eqref{eq:kaisert}$\times{\mathrm{e}} ^{- \mu ^2 f^2 k^2 \sigmav^2}$\\  and $P^{\text{Halofit}}_{\delta \delta}$\end{tabular} \\
\SPT & \begin{tabular}[c]{@{}l@{}}Eq. \eqref{eq:linearb}\\ with \eqref{eq:sptt}$\times{\mathrm{e}} ^{- \mu ^2 f^2 k^2 \sigmav^2}$\end{tabular}                                                                        & \begin{tabular}[c]{@{}l@{}}Eq. \eqref{eq:M&Rb}\\ with \eqref{eq:sptt} $\times{\mathrm{e}} ^{- \mu ^2 f^2 k^2 \sigmav^2}$\end{tabular}                                                 & \begin{tabular}[c]{@{}l@{}}Eq. \eqref{eq:coevb}\\ with \eqref{eq:sptt} $\times{\mathrm{e}} ^{- \mu ^2 f^2k^2 \sigmav^2}$\end{tabular}                                                 & \begin{tabular}[c]{@{}l@{}}Eq. \eqref{eq:delta-s-oneloop}\\ and \eqref{eq:adv_s}\\ with \eqref{eq:sptt}$\times{\mathrm{e}} ^{- \mu ^2 f^2k^2 \sigmav^2}$\end{tabular} \\
\EFT & \begin{tabular}[c]{@{}l@{}}Eq. \eqref{eq:linearb} with \eqref{eq:eftt}\end{tabular}                                                                                                                      & \begin{tabular}[c]{@{}l@{}}Eq. \eqref{eq:M&Rb} with \eqref{eq:eftt}\end{tabular}                                                                                                 & Eq. \eqref{eq:coevb} with \eqref{eq:eftt}                                                                                                                                          & \begin{tabular}[c]{@{}l@{}}Eq. \eqref{eq:renormalized-phh}\\ (this work)\end{tabular}                                                                                                                                      \\
\bottomrule
\end{tabular}
\end{turn}
\caption[RSD+BIAS]{Grid of different RSD and bias models considered in this analysis. The first row refers to the different bias models, whereas the first column represents the RSD frameworks.}
\label{tab:RSD+BIAS}
\end{table}

\section{Analysis}
\label{sec:Analysis}

The main goal of this section is to analyse a broad combination of bias and redshift-space models.
A similar analysis in real space, focusing on the EFT model
and using a sample of 20 COLA simulations,
was discussed by Bose et al.~\cite{Bose:2018orj}.
Since the {\WizCOLA} realizations report their results as measurements of the Legendre modes
of the redshift-space power spectrum
we must first apply this decomposition to Eq.~\eqref{eq:renormalized-phh},
together with Eq.~\eqref{eq:padv} for the
advective contribution if required.

As explained in the introduction, we validate our analysis using two different pipelines: one developed
using manual calculations and coding, and the other using semi-custom computer algebra methods
to automate the computation of the 1-loop integrals and their translation
into {\CC}.
The manual pipeline does not perform infrared resummation of
displacements to damp the baryon acoustic oscillation
feature, whereas the automated pipeline uses the
damping prescription of Vlah et al.~\cite{Vlah:2015zda},
which was translated to redshift space in Ref.~\cite{delaBella:2017qjy}.
The inclusion of resummation does not significantly affect our results,
and we find consistent $\chi^2$ values whether or not it is used.
The pipelines show excellent agreement up to the expected variance in nondeterministic
integration.

\subsection{The {\WizCOLA} simulation suite}
\label{sec: WizCola simulations}

To analyse the performance of each model we use a subset of $N$-body realizations drawn from
the {\WizCOLA} suite~\cite{Koda:2015mca}.
{\WizCOLA} is a set of COLA-accelerated simulations~\cite{Tassev:2013pn}
performed by the WiggleZ team and designed to resemble
observational data from the WiggleZ Dark Energy Survey.%
    \footnote{The WiggleZ Dark Energy survey was originally designed to detect the scale of the baryon acoustic oscillations at high redshift~\cite{2012PhRvD..86j3518P}.
        The survey was carried out at the Australian Astronomical Observatory
        over 276 nights and obtained redshifts for
        225,415 galaxy spectra~\cite{2010MNRAS.401.1429D,Blake:2011en}.}
The power spectra derived from each realization use
the same angular mask
and the same redshift distribution of observed galaxies
as the full survey.
The suite was used to obtain accurate non-Gaussian estimates
of covariances.

The full suite is a set of 3,600 COLA
simulations with different initial conditions.
These are used to generate 600 independent
mock galaxy catalogues covering the six independent WiggleZ
survey regions.
The survey volume covers roughly 1,000 square
degrees up to redshift 1, broken into three redshift bins.
Of these we use only the lowest bin centred on $z = 0.44$,
where nonlinearities are expected to be most pronounced.
To cover this volume the simulation uses a box of side
$600 h^{-1} \, \Mpc$
containing $1,296^3$ particles
of mass $7.5 \times 10^9 h^{-1} M_{\odot}$.
The dark matter haloes hosting WiggleZ emission-line
galaxies, which are the mass tracer whose bias we are
fitting, are inferred to have typical mass
$\sim 10^{12} h^{-1} M_{\odot}$.
This means that haloes of the necessary size are resolved
with more than $10^2$ particles per halo in the simulations.

The {\WizCOLA} cosmology is chosen to match WMAP5~\cite{Komatsu:2008hk}
with parameters
$\Omega _m = 0.273$,
$\Omega_{\Lambda} = 0.727$,
$\Omega _b= 0.0456$,
$h = 0.705$,
$\sigma _8 = 0.812$
and
$n_s = 0.961$.
The full suite comprises of 600 mock catalogues,
of which we use a subsample of 10 because of limits on
computational time.
We hope to discuss a larger subsample in future work.

\para{Multipole power spectra}
Observations are typically reported in terms of
`multipole power spectra' rather than
values of
$P^s_g(k,\mu)$
on a grid of wavenumber $k$ and cosine $\mu$.
Cole et al. defined the multipoles $P_\ell$ to satisfy~\cite{Cole:1993kh}
\begin{equation}
\label{eq:def-Pell}
    P^s_g(k,\mu)=\sum_{\ell=0}^{\infty} P_\ell(k) \Legendre{\ell}{\mu} .
\end{equation}
Each multipole may be computed by Legendre decomposition of $P^s_g(k,\mu)$,
\begin{equation}
    P_\ell(k) = \frac{2\ell+1}{2} \int_{-1}^1 \d\mu \; P^s_g(k,\mu)
    \Legendre{\ell}{\mu} ,
\end{equation}
where $\Legendre{\ell}{\mu}$ is the Legendre polynomial in $\mu$ of order $\ell$.

\para{Mask, selection function and finite volume effects}
A realistic galaxy survey is forced to deal with
irregularly-shaped regions in order to mask the galaxy or
other bright objects.
Together with the varying redshift distribution of tracers,
and their restriction to our past lightcone,
this means that the Fourier modes measured by the
survey are a biased sample of the Fourier modes in infinite volume.
(We do not take into account relativistic projection effects
that arise because we observe on the past lightcone, rather than
inside a Euclidean box.)
The power spectrum measured by averaging over these Fourier modes
will therefore be a biased version of the infinite-volume power spectrum
predicted by theory.
Fortunately, the necessary corrections can be determined by using
the simulation suite to characterize how each Fourier mode is mis-weighted
in the average.
For the WizCOLA suite,
the correction can be approximately written
\begin{equation}
    P^{\text{obs}}_i = \sum_{j} D_{ij} P^{\text{theory}}_j ,
\end{equation}
where the vectors $P^{\text{obs}}$, $P^{\text{theory}}$ consist of measurements
for $P_0$, $P_2$, $P_4$ at a set of $k$-modes,
which for WiggleZ consist of 25 sample points between
$k = 0.01 h/\Mpc$ and $k = 0.49 h/\Mpc$ inclusive, in steps of $\Delta k = 0.02 h/\Mpc$.
The convolution matrices $D_{ij}$ are supplied as part of the {\WizCOLA} data products.

\para{Likelihood}
The final likelihood $\mathscr{L}$ can be
written~\cite{stw447, blakew}
\begin{equation}
    \chi^2
    =
    -2 \ln \mathscr{L}^{\text{\WizCOLA}}
    =
    \sum_{r}
    \sum_{ij}
    \Delta_{i}^r [C^r]^{-1}_{ij} \Delta_j^r ,
\end{equation}
where we recall that just one redshift bin is under discussion,
$r$ sums over regions,
$C^r_{ij}$ is the covariance matrix in region $r$,
and
$\Delta^r_{i}$ is the difference between measured and predicted values
for the multipole power spectra in region $r$,
\begin{equation}
    \Delta^r_{i} = P^{r,\text{\WizCOLA}}_i - P^{r,\text{obs}}_i .
\end{equation}
To minimize uncertainties with measurements on very large or very small
scales
only 14 $k$-modes are used in the fit, between
$k = 0.03 h/\Mpc$ and $k = 0.29 h/\Mpc$ inclusive in steps
of $\Delta k = 0.02 h/\Mpc$.
We use
the \sansbold{emcee} sampler~\cite{2013PASP..125..306F}
to explore the parameter space and extract best-fit
values after the chain has converged.

\subsection{Bayesian information criterion}
\label{app:bic}

Before presenting our results
we should consider the issue of
fits to models containing
a large number of parameters.
As the number of parameters increases, so does the risk of `over-fitting'---that is,
misinterpreting realization variance as signal.

There are various empirical prescriptions
to penalize models that introduce a large number of unnecessary parameters.
Ultimately, the measure in which we are most interested is the degree to which
marginalization over these unnecessary parameters would artificially inflate
the error bars for other quantities of interest.
To determine the degree of inflation would require a ensemble of Monte Carlo
fits for the cosmological parameters in addition to bias and redshift-space
parameters, which is computationally challenging due to the time cost
of performing the various one-loop integrals.
We hope to return to this interesting question in the future.
In the interim, experience has shown that a useful empirical
penalty is the \emph{Bayesian information criterion} or `BIC'~\cite{Schwarz:1978}.

The BIC is defined as the raw $\chi^2$ adjusted by an offset
$\varpi = N_p \ln N_d$,
where $N_p$ is the number of model parameters and $N_d$ is to the number of data points.
Specifically, we define
\begin{equation}\label{eq:bic}
    BIC = \chi ^2 + \varpi .
\end{equation}
Smaller values of the BIC are preferred.
Large values indicate either that the model is a poor fit, or includes an
unjustified number of extra parameters.
Tables \ref{tab:param} and \ref{tab:bic} exhibit the number of parameters
and
corresponding penalty factor for each model we consider.
\begin{table}
\centering
\begin{tabular}{cccc}
\toprule
\multicolumn{4}{c}{\semibold{Bias model}} \\
\Linear & \Coevo & \MRoy & \Advective \\ \midrule
1 & 2 & 4 & 6 \\
\bottomrule
\end{tabular}
\hspace{5mm}
\begin{tabular}{cccc}
\toprule
\multicolumn{4}{c}{\semibold{Redshift-space model}} \\
\KaiserTree & \KaiserHalo & \SPT & \EFT \\ \midrule
1 & $1^+$ & 1 & 6 \\
\bottomrule
\end{tabular}
\caption[Models]{Number of parameters associated to each RSD and bias parameter. The $+$ sign for Kaiser {\Halofit} indicates that the power spectrum generated by CAMB {\Halofit} uses some parameters to calibrate and match data.}
\label{tab:param}
\end{table}
\begin{table}
\centering
\rowcolors{2}{gray!25}{white}
\begin{tabular}{clrlrlrlr}
\toprule
& \multicolumn{2}{c}{\Linear} & \multicolumn{2}{c}{\Coevo} & \multicolumn{2}{c}{\MRoy} & \multicolumn{2}{c}{\Advective} \\ \midrule
\KaiserTree & 2 & \textcolor{blue}{11.1} & 3 & \textcolor{blue}{16.6} & 5  & \textcolor{blue}{27.6} & 7  & \textcolor{blue}{38.7} \\
\KaiserHalo & 2 & \textcolor{blue}{11.1} & 3 & \textcolor{blue}{16.6} & 5  & \textcolor{blue}{27.6} & 7  & \textcolor{blue}{38.7} \\
\SPT        & 2 & \textcolor{blue}{11.1} & 3 & \textcolor{blue}{16.6} & 5  & \textcolor{blue}{27.6} & 7  & \textcolor{blue}{38.7} \\
\EFT        & 7 & \textcolor{blue}{38.7} & 8 & \textcolor{blue}{44.2} & 10 & \textcolor{blue}{55.3} & 12 & \textcolor{blue}{66.4} \\
\bottomrule
\end{tabular}
\caption[BIC]{Number of parameters (black) and penalty factor (blue) of every model combination, RSD+BIAS.}
\label{tab:bic}
\end{table}

\subsection{Results}
\label{Results}

We now summarize the outcome of our analysis.
First, we study the relative performance of each combination of bias and
redshift-space model, quantified in units of $\chi^2$.
Second,
we use the
Bayesian Information Criterion to compensate
for the number of free parameters involved in each
model.
We will use a ranking by improvement in BIC to identify which modelling choices
represent the best compromise between
flexibility to capture physically meaningful adjustments
and rigidity to prevent overfitting.
For our purposes, `overfitting'
means that the fit per-realization is sufficiently permissive that it can
adjust to match realization variance.
This is an unwanted effect: it is likely to bias attempts to recover
the underlying cosmological parameters by fitting the power
spectrum.
Finally, we validate our conclusions by
comparing the fit per-realization to the fit to the ensemble average.
Where the model is too permissive we should expect these to differ
significantly because of adaptation to realization-specific features.

\subsubsection{Improvement in $\chi^2$}
\label{sec:improvement}
For each realization, we define the $\chi^2$-improvement
relative to the WiggleZ
baseline model
{\KaiserHalo}+{\Linear}.
This model gives an overall $\chi^2$ of $296.4 \pm 25.5$
with $\chi^2/\text{dof} \sim 1.176$.
(There are $14 \times 3 \times 6 = 252$ degrees of freedom in the dataset.)
While this is a reasonable fit, there is scope for improved modelling to reduce
the mean $\chi^2$ by up to $\sim 50$ to obtain $\chi^2/\text{dof} \sim 1.0$.
A significant portion of the variability in fit is attributable
to variability in the baryon acoustic oscillation (`BAO') feature.
For a survey of size comparable to WiggleZ the BAO feature is not very well resolved
in a typical realization,
and can appear with large phase shifts or even be absent entirely.
This is a purely statistical effect due to the number
of available modes. It has no dependence on the underlying
cosmological model.
For larger surveys such as DESI or LSST the BAO feature is expected to be
defined much more clearly.
However, should these surveys elect to subdivide their volume then similar
variability could reappear.

\begin{figure}
    \centering
        \includegraphics[width=.7\linewidth]{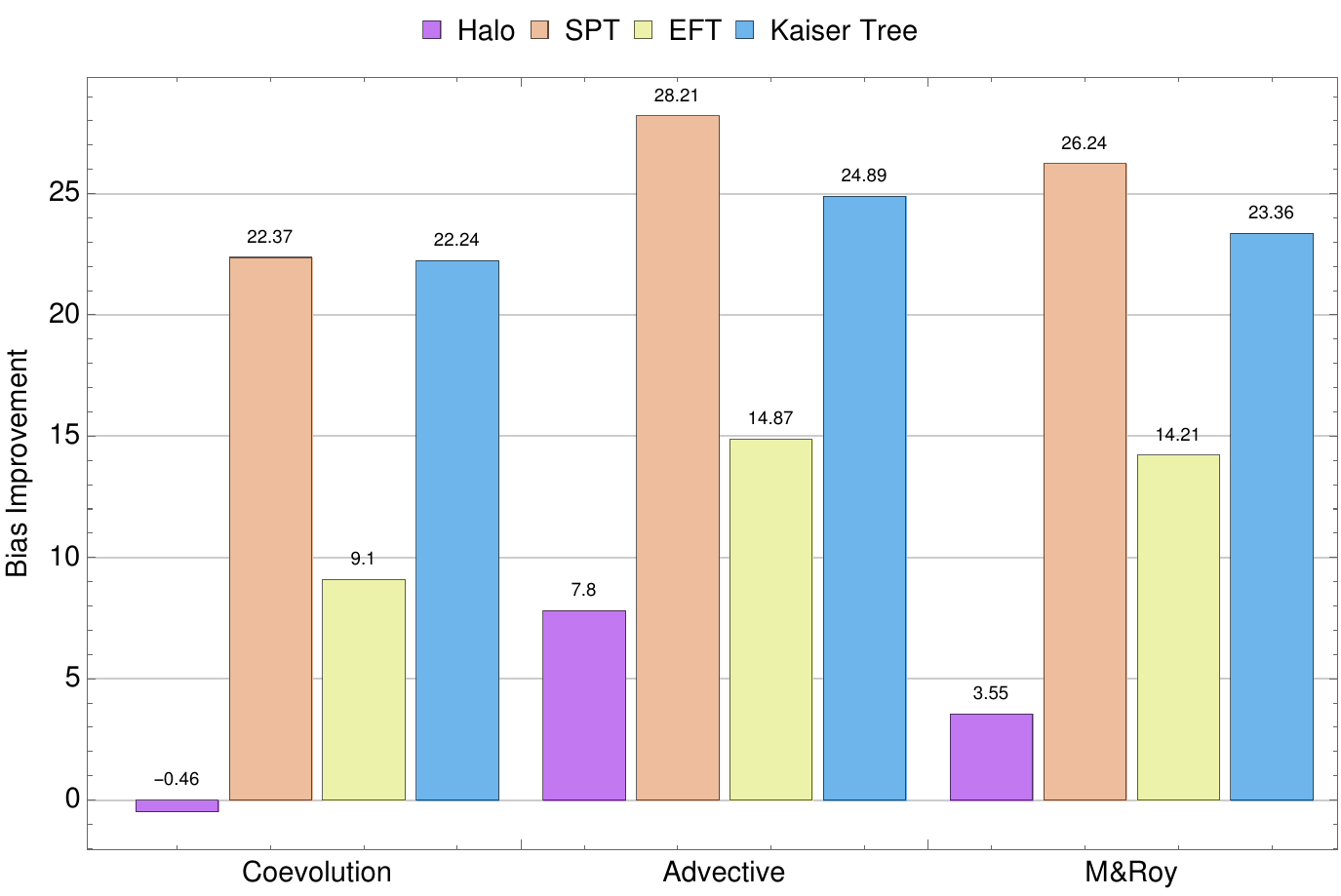}
\caption[Bias improvement]{Improvement in $\chi^2$ for each bias model, measured
relative to the WiggleZ $\Linear$ baseline with a fixed redshift-space model.
Negative values mean that the model performs more poorly than {\Linear}.}
\label{figure:bias-improvement}
\end{figure}

\begin{figure}
    \centering
        \includegraphics[width=.7\linewidth]{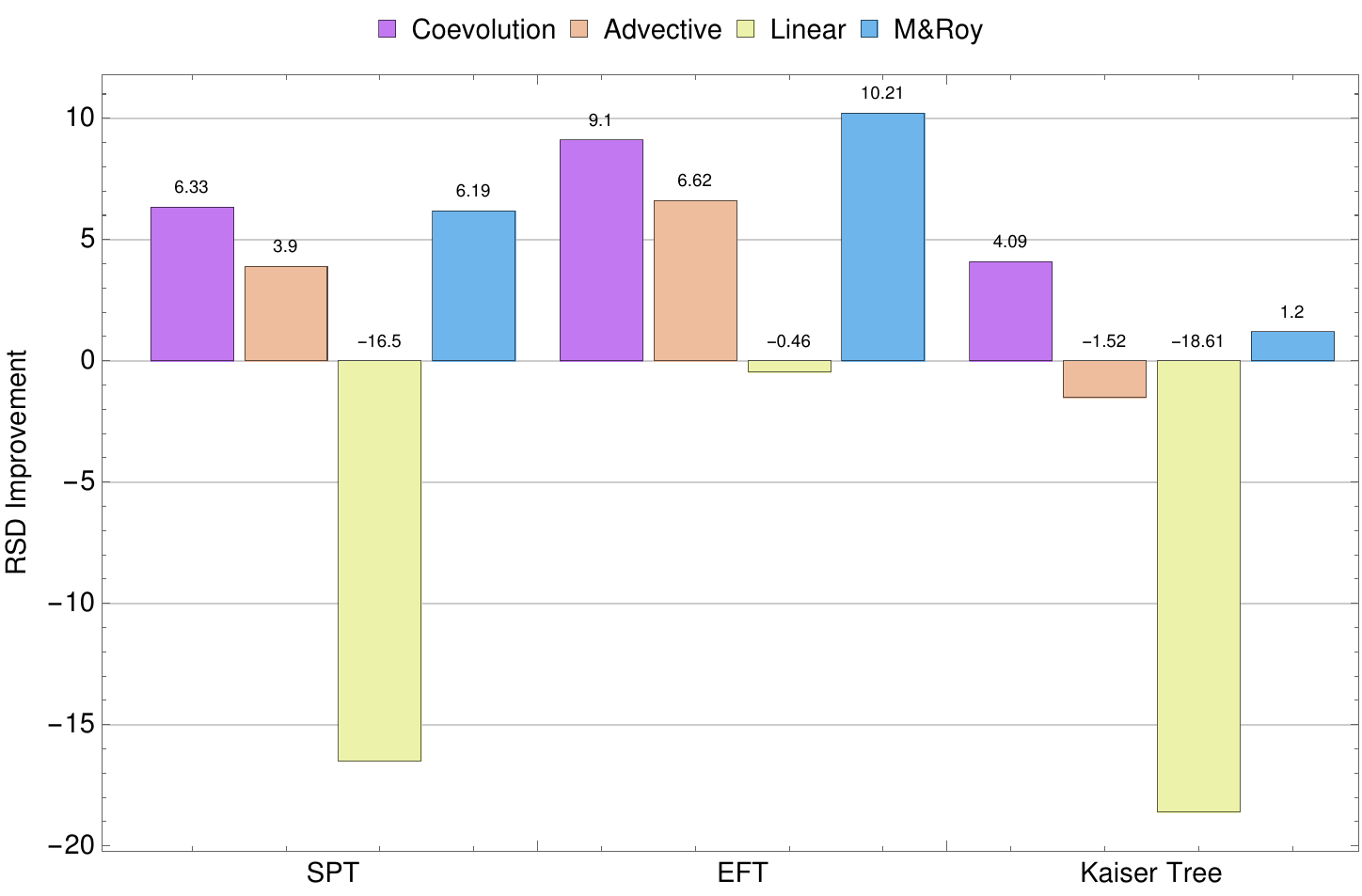}
    \caption{Improvement in $\chi^2$ for each redshift-space model,
    measured relative to the WiggleZ baseline of {\KaiserHalo}
    with a fixed bias model.
    Negative values mean that the model performs more poorly than {\KaiserHalo}.}
    \label{figure:rsd-improvement}
\end{figure}

For each realization, the improvement is
\begin{equation}
    \Delta \chi^2 \equiv \chi^2_\text{fit} - \chi^2_\text{base} .
\end{equation}
Here, $\chi^2_\text{base}$ is the $\chi^2$ achieved by the baseline model
and $\chi^2_\text{fit}$ is the $\chi^2$ achieved when fitting
whatever combination of bias
and redshift-space models is under discussion.
Occasionally we will refer to the $\chi^2$ for a specific realization,
but generally we quote the mean improvement over all ten realizations.
The results are given in Table~\ref{table:results}
and summarized in Figs.~\ref{figure:bias-improvement}
and~\ref{figure:rsd-improvement}, which show the improvement due to changes in
bias model (with fixed redshift-space model)
and redshift-space model (with fixed bias model), respectively.
\begin{table}
    \centering
    \rowcolors{2}{gray!25}{white}
    \begin{tabular}{lQQQQ}
        \toprule
        Model
            & \multicolumn{1}{R}{\text{mean} \pm \sigma}
            & \multicolumn{1}{R}{\chi^2/\text{dof}}
            & \multicolumn{1}{R}{\min \chi^2}
            & \multicolumn{1}{R}{\max \chi^2}
        \\ \midrule
        {\Linear}+{\KaiserTree} & \mathbf{315.0 \pm 28.7} & 1.250 & 260.2 & 347.2 \\
        {\Linear}+{\SPT} & \mathbf{312.9 \pm 25.3} & 1.242 & 256.6 & 338.0 \\
        {\Linear}+{\KaiserHalo} & \mathbf{296.4 \pm 25.5} & 1.176 & 240.6 & 322.7 \\
        {\Linear}+{\EFT} & \mathbf{296.9 \pm 25.2} & 1.178 & 245.1 & 322.1 \\
        {\Coevo}+{\KaiserTree} & \mathbf{292.7 \pm 26.0} & 1.162 & 236.1 & 321.2 \\
        {\Coevo}+{\SPT} & \mathbf{290.5 \pm 24.7} & 1.153 & 240.0 & 316.3 \\
        {\Coevo}+{\KaiserHalo} & \mathbf{296.8 \pm 25.4} & 1.178 & 241.2 & 323.0 \\
        {\Coevo}+{\EFT} & \mathbf{287.7 \pm 24.9} & 1.142 & 238.9 & 315.5 \\
        {\MRoy}+{\KaiserTree} & \mathbf{291.6 \pm 25.9} & 1.157 & 235.2 & 320.1 \\
        {\MRoy}+{\SPT} & \mathbf{286.6 \pm 23.5} & 1.137 & 237.3 & 309.4 \\
        {\MRoy}+{\KaiserHalo} & \mathbf{292.8 \pm 25.9} & 1.162 & 236.7 & 320.2 \\
        {\MRoy}+{\EFT} & \mathbf{282.6 \pm 24.0} & 1.121 & 233.5 & 310.7 \\
        {\Advective}+{\KaiserTree} & \mathbf{290.1 \pm 25.6} & 1.151 & 235.3 & 319.3 \\
        {\Advective}+{\SPT} & \mathbf{284.7 \pm 23.6} & 1.130 & 236.3 & 308.7 \\
        {\Advective}+{\KaiserHalo} & \mathbf{288.6 \pm 25.4} & 1.145 & 234.4 & 317.2 \\
        {\Advective}+{\EFT} & \mathbf{281.9 \pm 23.8} & 1.119 & 232.9 & 309.6 \\
        \bottomrule
    \end{tabular}
    \caption{Summary statistics for fit to subsample of ten realizations from
    the {\WizCOLA} suite.\label{table:results}}
\end{table}

As we would expect, the most permissive combination
$\Advective+\EFT$ yields the best overall fit, giving
a mean
$\chi^2 = 281.9 \pm 23.8$.
But given the complexity of the modelling,
this improvement is strikingly modest---just $\sim 14$ units of $\chi^2$
compared to the baseline model.
Meanwhile, we
note that the variability of fit over the subsample barely changes, no matter
which model is in use.

\para{Improvements due to bias model}
To break these results down in detail, consider first
Fig.~\ref{figure:bias-improvement} which represents the improvement due to
modifying the bias model to be more flexible than the simple linear truncation.
The models are superclasses of each other in the order
{\Advective} $\supseteq$ {\MRoy} $\supseteq$ {\Coevo}, and therefore we have
a strict ordering of improvements:
{\Advective} $\geq$ {\MRoy} $\geq$ {\Coevo}
for all redshift-space models.
Because the advective model is most general, it automatically shows the largest
improvement.
However the performance of all the bias models other than {\Linear}
is similar, and there is not much to choose between them.

The breakdown by redshift-space model is more variable,
but the structure is similar in each case.
The {\KaiserHalo} model barely benefits from addition of flexibility in the bias
prescription, while the {\SPT} model
and {\KaiserTree} models show very significant improvements that are nearly
independent of the model actually chosen.
We will comment on these features in more detail below.

\para{Improvements due to redshift-space model}
Second, consider Fig.~\ref{figure:rsd-improvement}.
This time no one model is a strict superclass of any other,
except that the {\EFT} model can be regarded as a superclass of {\SPT}
if terms of order $\Or(k^4)$ and higher in the phenomenological
fingers-of-God damping term are not relevant.
Neither the {\EFT} or {\SPT} model has a simple relationship to the
{\KaiserTree} model since there is no continuous parameter that can be
varied to connect them.

\begin{figure}
    \centering
    \includegraphics[scale=0.58]{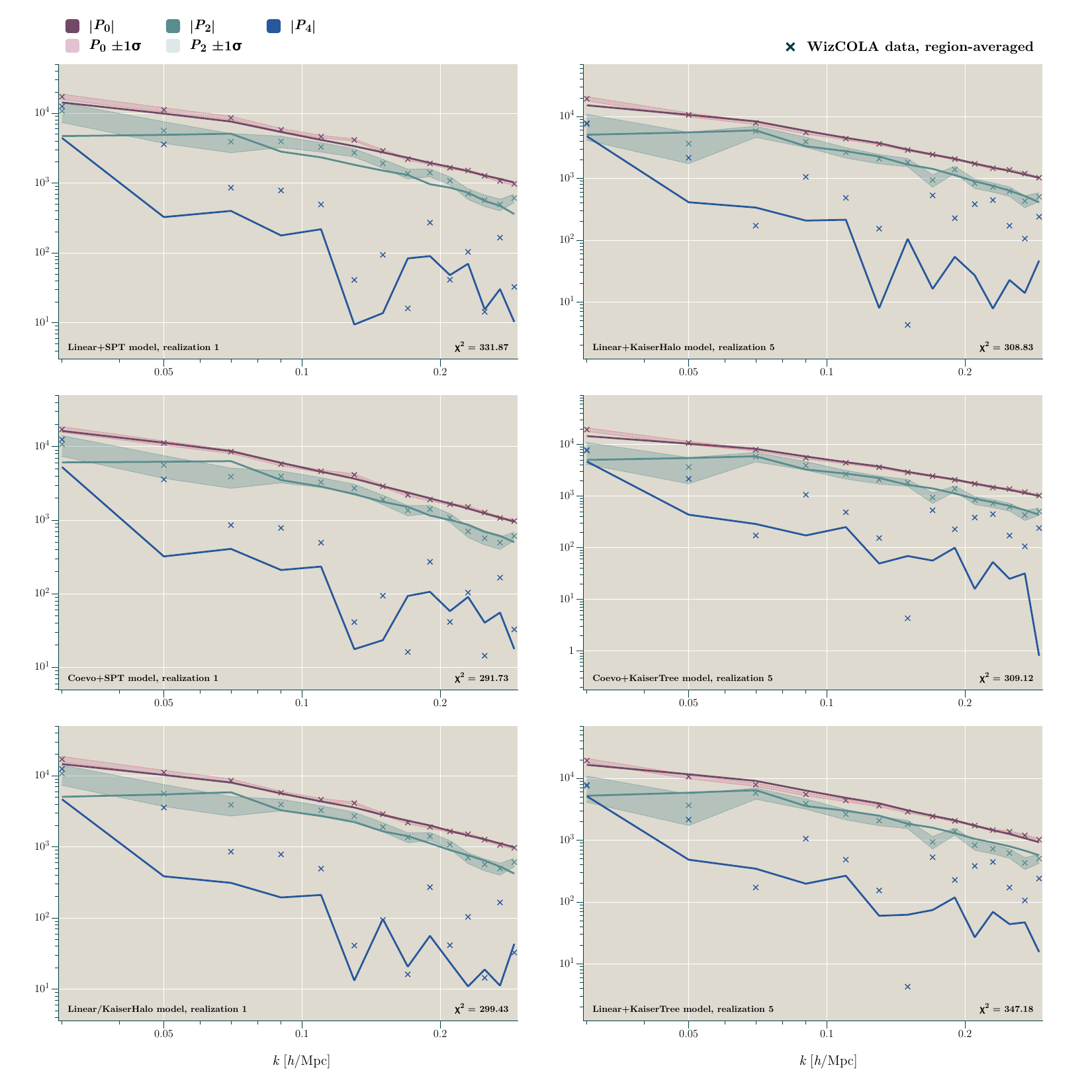}
    \caption{Specimen power spectrum fits.
    \semibold{Left side}: \textsl{Realization 1}.
    Top: {\Linear}+{\SPT}; middle: {\Coevo}+{\SPT}; bottom: {\Linear}+{\KaiserHalo}.
    In the top panel note the poor fit to $P_0$ at low $k$, and for $P_2$
    generally. In the middle panel, the more general bias model allows a good
    fit to the low-$k$ slope. In the bottom panel the {\KaiserHalo} model matches the
    spectral slope even in combination with {\Linear}.
    \semibold{Right side}: \textsl{Realization 5}.
    Top: {\Linear}+{\KaiserHalo}; middle: {\Coevo}+{\KaiserTree};
    bottom {\Linear}+{\KaiserTree}.
    The same general features are present.\label{fig:specimen-Pk}
    }
\end{figure}

The most striking features of Fig.~\ref{figure:rsd-improvement}
are the large negative shifts for the {\Linear} bias model (yellow bars)
in combination with the {\SPT} or {\KaiserTree} models.
Taken together with Fig.~\ref{figure:bias-improvement} these
show that the relatively rigid {\KaiserTree} and {\SPT} models do not have
the right shape
to match the spectral slope
of a typical {\WizCOLA} power spectrum at both small and large $k$, even
after accounting for fingers-of-God suppression.
(See Fig.~\ref{fig:specimen-Pk}
for an explicit demonstration of this in the {\KaiserTree}
and {\SPT} cases, respectively.)
This amounts to an adjustment
of the slope by a term of the form $k^2 P(k)$ at large $k$.%
	\footnote{It is well-known that the tree-level power spectrum significantly
	underpredicts the nonlinear power spectrum measured from
	simulations for quasi-linear wavenumbers, and that this
	underprediction is partially corrected by the 1-loop term.
	This underprediction is not directly the source of the mismatch under discussion.}

In the {\Linear} bias model
there is no option to change the shape of the underlying power spectrum;
only its normalization can be adjusted, which explains the poor performance
of {\Linear}
in combination
with the {\KaiserTree} and {\SPT} models.
The {\KaiserHalo} model is based on the {\Halofit} power spectrum which is
calibrated to match simulations, and therefore does not exhibit a mismatch in slope,
at least
on these scales. The {\EFT} model inherits its shape from {\SPT}, but the combination
of its additive and multiplicative counterterms can
account for some part of the mismatch.
Finally, even if we retain the rigid {\KaiserTree} or {\SPT} models,
any of the more complex bias prescriptions is apparently capable of
approximating the required change in slope.

\para{Typical improvements}
Figs.~\ref{figure:bias-improvement} and~\ref{figure:rsd-improvement}
show that, in typical circumstances, the improvement from better bias modelling
is comparable to, or perhaps
marginally greater than, than the improvement from better redshift-space
modelling.
If one is dealing with the {\Linear} bias model in conjunction with
{\KaiserTree} or {\SPT} then the gain from moving to \emph{any} other bias model
is very significant, as described above---more than $20$ units of $\chi^2$.
Otherwise, changing the bias model with
{\KaiserHalo} is worth perhaps $\sim 5$ units of $\chi^2$ (excluding {\Coevo}),
and with the {\EFT} model is worth in the range $10$--$15$ units of $\chi^2$.

Contrast this with the improvements from changing the redshift-space model, as in
Fig.~\ref{figure:rsd-improvement}.
Excluding the negative values associated with switching to
{\KaiserTree} or {\SPT} with the {\Linear} bias model,
these are mostly in the range $0$--$10$ units of $\chi^2$.
All of these numbers are comparable to the absolute improvement in best-fit
$\Delta \chi^2 = 14.5$ between the WiggleZ baseline and
{\Advective}+{\EFT}.

The discrepancy between the mean $\chi^2$ obtained from
{\Advective}+{\EFT} and the target mean $\chi^2 \sim 250$
implies that the {\WizCOLA} realizations contain further unmodelled effects.
We believe these relate to the variability of the BAO feature in
these realizations.
In each of our redshift-space models, the phase and amplitude of the BAO feature
is a rather rigid part of the template that cannot be adjusted independently---only
by adjusting the whole background cosmology.
The structure of the BAO feature \emph{can} be adjusted by nonlinear terms in
a bias model, but as described above these also change the spectral slope.
If the contribution from these nonlinear terms is too significant, it will degrade
the broadband fit well before it can be compensated by improvements to fitting the BAO
feature. Accordingly, variation in the BAO feature due to bias modelling is very modest,
giving limited scope to fit realization variance. In general, as we
now describe, this rigidity is
a positive feature that prevents some instances of overfitting.

\subsubsection{Bayesian Information Criterion analysis}

The analysis of {\S}\ref{sec:improvement}
demonstrates how well each combination of bias and redshift-space
model matches the ensemble of realizations at the level of raw $\chi^2$.
As expected, the outcome is that the most permissive model gives the best fit.
But this does not demonstrate that the large number of parameters required by
the model are all physically meaningful; some might match
features that vary randomly from realization to realization, like the BAO feature.
Otherwise might simply lack statistical value.
The first case is `overfitting', which we deal with below.
The second is `overparametrization'. As explained in {\S}\ref{app:bic},
we attempt to detect this using the Bayesian information criterion.

In this analysis, parameter degeneracies play a critical role.
The BIC formula~\eqref{eq:bic} depends strongly on the number of parameters
carried by the model. If degenerate parameters are included in the analysis then
they will unfairly downweight the BIC.
\begin{figure}
    \centering
        \includegraphics[width=.7\linewidth]{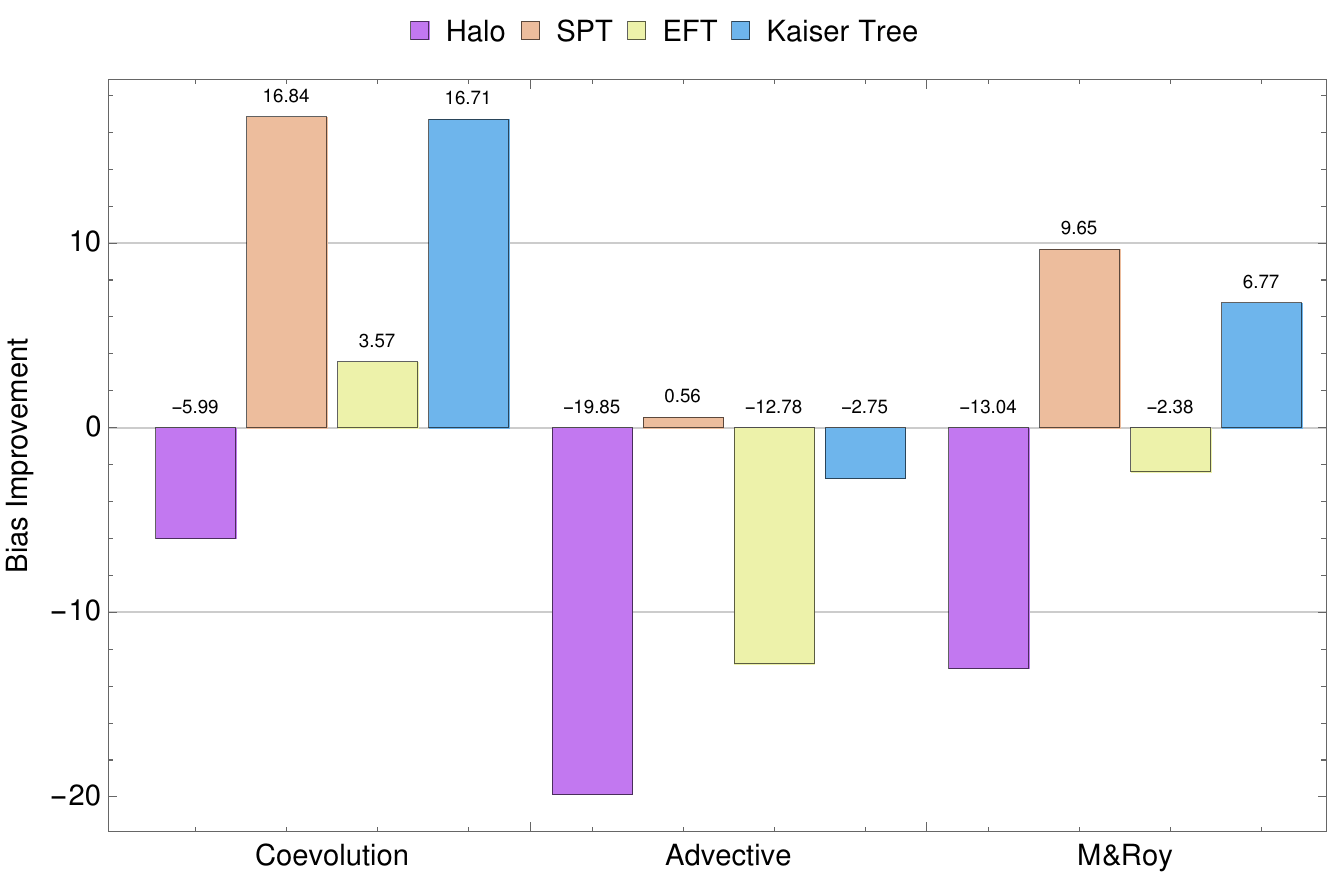}
\caption[Bias improvement]{Improvement in BIC for each bias model, measured relative
to the WiggleZ {\Linear} baseline with a fixed redshift-space model.}
\label{figure:bic_bias-improvement}
\end{figure}
\begin{figure}
    \centering
        \includegraphics[width=.7\linewidth]{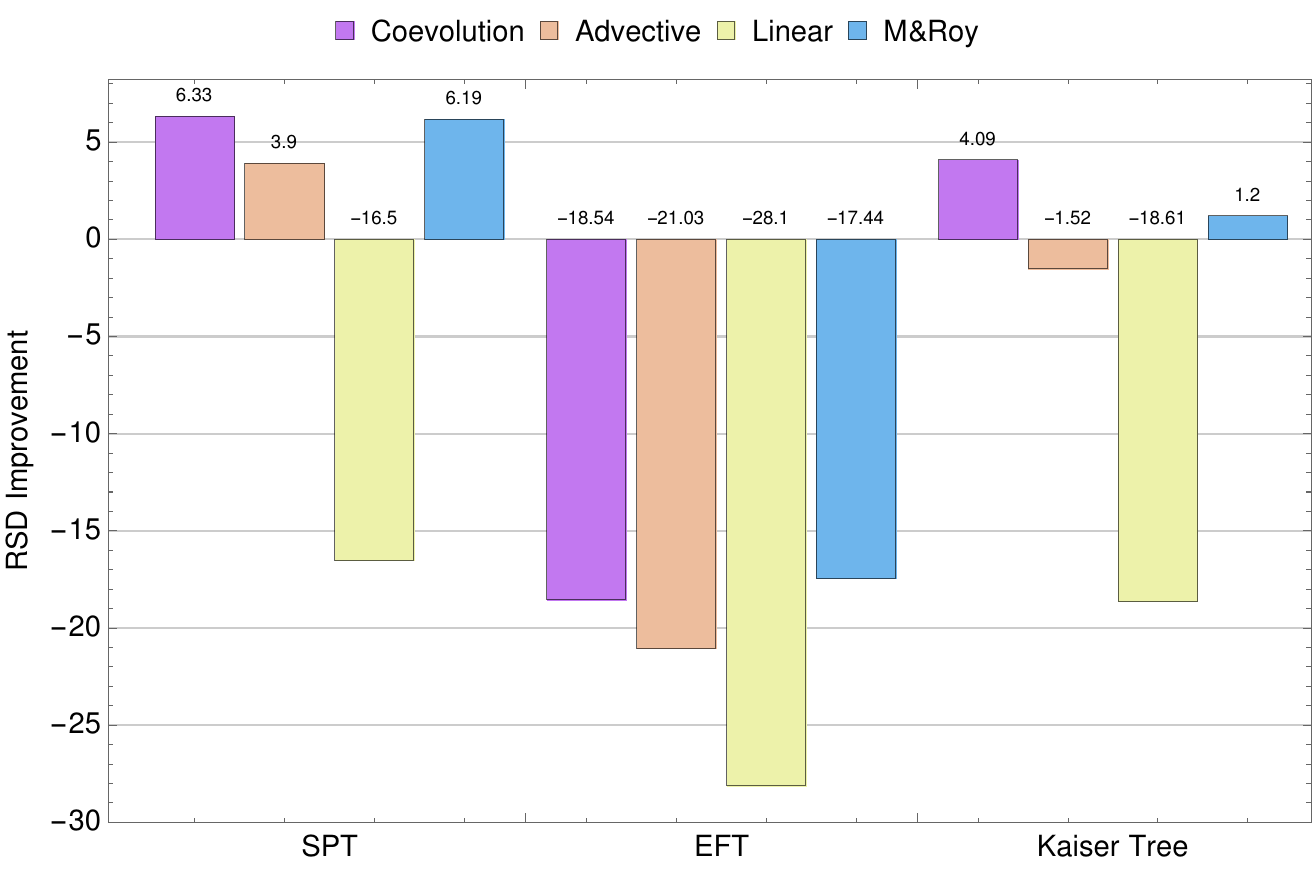}
    \caption{Improvement in BIC for each redshift-space model, measured relative
    to the WiggleZ {\KaiserHalo} baseline with a fixed bias model.}
    \label{figure:bic_rsd-improvement}
\end{figure}

After transformation from raw $\chi^2$ to BIC,
Figs.~\ref{figure:bias-improvement}--\ref{figure:rsd-improvement}
translate to
Figs.~\ref{figure:bic_bias-improvement}--\ref{figure:bic_rsd-improvement}.
Notice that many `improvements' have become negative, implying that the BIC ranks
the statistical power of these models lower than the WiggleZ baseline.

\para{Bias models}
Fig.~\ref{figure:bic_bias-improvement} suggests that there is little statistical value in
changing to the {\Advective} bias model, and only modest value in switching to the
McDonald \& Roy model {\MRoy}. On the other hand there is generally
clear value in
switching from the {\Linear} model to the {\Coevo} model.
We conclude that a WiggleZ-like survey with a generic redshift-space model
would typically benefit from
a bias model that is more permissive than {\Linear},
but there is not yet evidence that the most complex prescriptions are
required.

\para{Redshift-space models}
The BIC analysis strongly disfavours the {\EFT} model.
Reference to Table~\ref{table:results} shows that, with a sufficiently permissive
bias model, it produces typical $\chi^2$ that are close to the {\SPT}
model.
These models differ only in two respects:
(i) in the {\SPT} model, suppression of power at quasilinear $k$ is modelled
by a single fingers-of-God factor
$\exp(-k^2 \mu^2 f^2 \sigmav^2)$,
giving a common suppression for each $P_\ell$, whereas in {\EFT}
there are independent counterterms for each $P_\ell$;
and (ii) the {\EFT} model includes additive stochastic counterterms,
but {\SPT} does not.
The additive counterterms are significant in allowing {\EFT} to correct the spectral
slopes inherited from {\SPT},
but (as described above) this can equally by done by the bias model.

Meanwhile, the similar performance of {\EFT} and {\SPT} shows that there is no
significant
benefit from allowing different suppression scales for each of $P_0$, $P_2$ and $P_4$.
Under these circumstances there is no significant benefit from using {\EFT} in
preference to {\SPT}.

Of course, it is likely that this conclusion depends strongly on $\kmax$.
In a survey with large $\kmax$ the need for different suppression scales associated
with each $P_\ell$ may be more significant, in which case the value of the
{\EFT} model would need to be revisited.

\subsubsection{Overfitting: Comparison to ensemble average}

Finally we address the issue of overfitting.
For each model combination we compute the best-fit
parameter combination and its $\chi^2$.
We also compute the best-fit to the ensemble average of the full {\WizCOLA}
suite of over 600 realizations.
This enables us to assign a `shift' to each realization,
\begin{equation}
\label{eq:shift}
    \Delta \chi ^2 = \chi ^2_\text{ensemble} - \chi ^2_\text{bestfit}.
\end{equation}
The sign is chosen so that $\Delta\chi^2$ is typically positive.
A reasonable model for $\Delta\chi^2$ might be a $\chi^2$ distribution
with degrees of freedom equal to the number of parameters carried by the model,
given in Table~\ref{tab:bic}.

\begin{figure}
    \centering
    \includegraphics[scale=0.49]{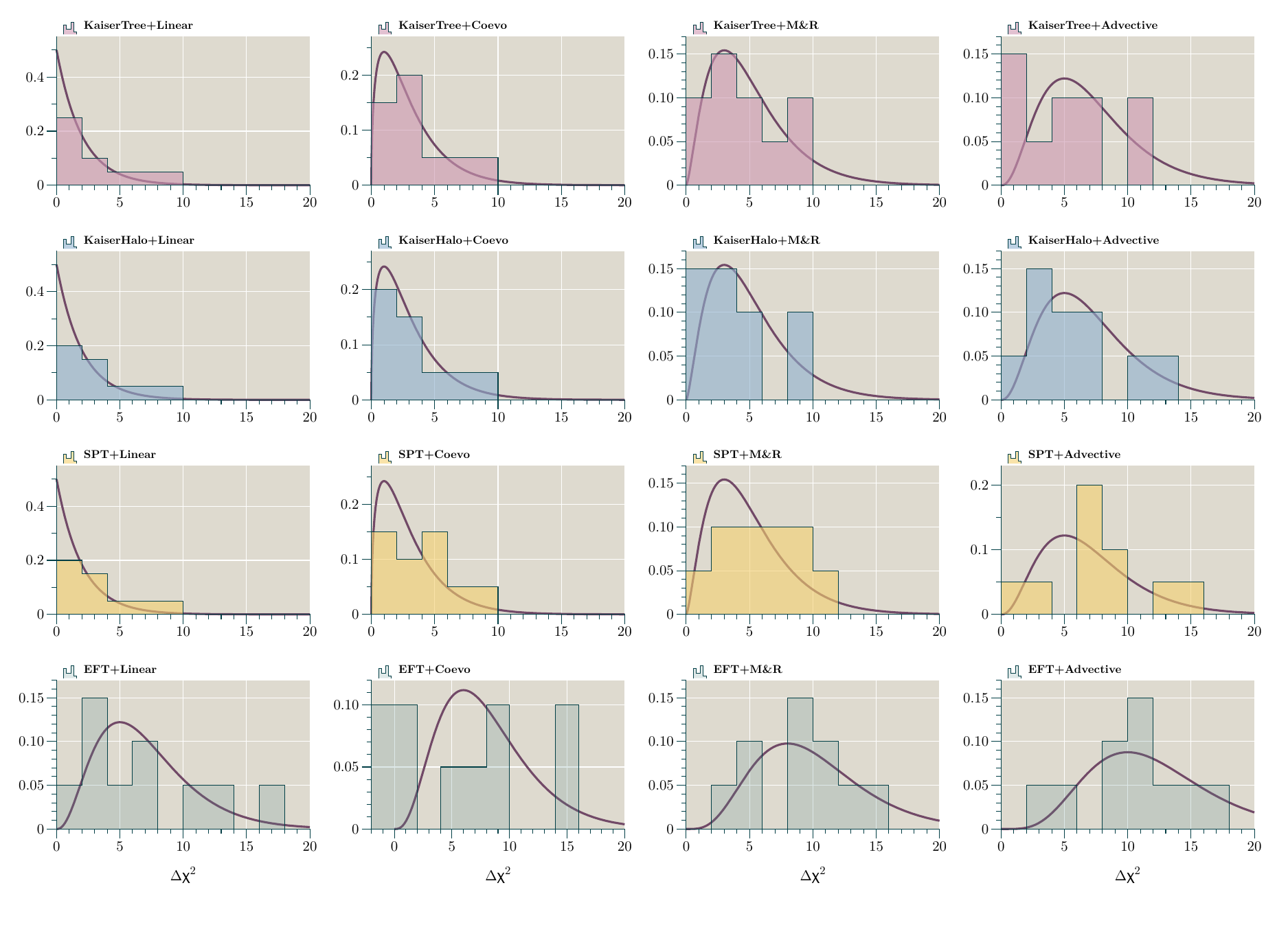}
    \caption{Distribution of $\Delta\chi^2$ for each combination of bias model
    (columns of the grid) and redshift-space model (rows of the grid).
    As a guide to the expected dispersion we
    overplot a $\chi^2$ distribution
    $\sim \chi^2_r$ with degrees-of-freedom $r$ equal to the number of
    parameters in the combined model; this is given in Table~\ref{tab:bic}.
    \label{fig:chisq-grid}}
\end{figure}

We interpret unusually large shifts as evidence that the model is adapting
to features present in the power spectrum of a given realization, but which
are \emph{not} present in the ensemble average.
In Fig.~\ref{fig:chisq-grid}
we show the distribution of $\Delta \chi^2$ for each combination of
bias and redshift-space model.
To give a sense of the expected dispersion
we overplot the corresponding $\chi^2$ distribution.
Since our distributions are poorly resolved we
limit ourselves to qualitative observations and do not attempt
a quantitative analysis.

The {\Linear} and {\Coevo} models contain fewest bias parameters.
Except in combination with {\EFT} we expect the distribution of
$\Delta\chi^2$ to show a sharp peak near $\Delta \chi^2 \approx 0$
and a strongly damped tail to larger values.
Our observations seem to reproduce this behaviour, and we conclude there
is no significant evidence for overfitting.

The {\MRoy} and {\Advective} models add more parameters and the
expected distribution of $\Delta\chi^2$ is broader,
without such a clear division into peak-and-tail.
With the small sample size we are using it is not possible to be
certain, but there are some suggestions from
the {\MRoy}+{\KaiserHalo}, {\MRoy}+{\SPT}
and {\Advective}+{\SPT}
distributions that the distribution of $\Delta\chi^2$ might show
more dispersion than expected.
This could be interpreted as weak evidence of overfitting
with these bias models.
To validate this conclusion would require a larger sample.

Finally, the {\EFT} model exhibits a broad distribution
of $\Delta\chi^2$ for all bias models.
At least for
{\Coevo}+{\EFT},
{\MRoy}+{\EFT}
and
{\Advective}+{\EFT}
the distribution appears to have too much weight at large
values of $\Delta\chi^2$.
We conclude that there is modest evidence
for overfitting using $\EFT$,
almost irrespective of the bias model chosen.
As before, this conclusion should be validated using a larger
sample.

\begin{figure}
    \centering
        \includegraphics[scale=0.58]{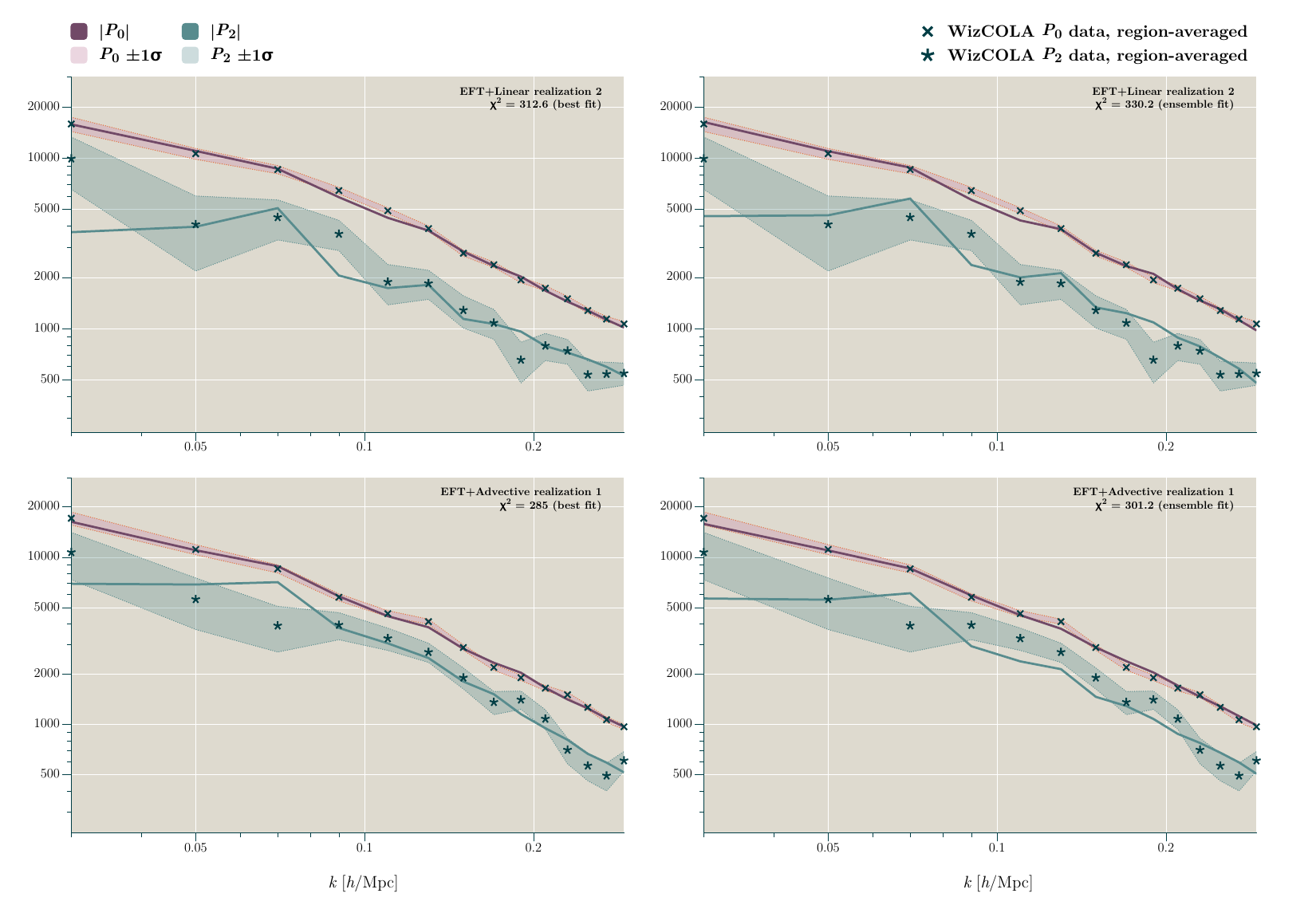}
    \caption[Comparison of overfitting]{Top panel: Realization 2/{\EFT}+{\Linear} with $\Delta \chi ^2  = 17.6$.
    Left plot is the best-fit to this realization, and right plot shows the fit to this realization
    using the ensemble-average best-fit parameters.
    Bottom panel: Realization 1/{\EFT}+{\Advective} with $\Delta \chi ^2 = 16.2$.
    The arrangement of left- and right-hand plots is the same.
    The pink shaded region shows the {\WizCOLA} $1\sigma$ confidence contour
    for $P_0$ and the green shaded region shows the same for $P_2$.}
    \label{figure:overfit-compare}
\end{figure}

\para{Origin of overfitting}
If overfitting is really occurring, where does it manifest itself?
As an example,
we explicitly compare
the ensemble-average best-fit with the individual best-fit
for Realization 1/{\Advective}+{\EFT}
and Realization 2/{\Linear}+{\EFT};
see Fig.~\ref{figure:overfit-compare}.
The $\chi^2$ shifts
for these combinations
are $\Delta\chi^2 = 17.6$ and $\Delta\chi^2 = 16.2$, respectively,
which are the largest in our grid of models.

First consider the top panel of Fig.~\ref{figure:overfit-compare},
showing Realization 2/{\Linear}+{\EFT}.
The difference between the left- and right-hand plots
appears to be driven by a response to the BAO feature.
In comparison with the ensemble average,
Realization 2 has a lower amplitude BAO feature
near $k = 0.1 h/\Mpc$ and $k = 0.2 h/\Mpc$ in $P_0$
and near $k = 0.07 h/\Mpc$ and $k = 0.2 h/\Mpc$ in $P_2$.
This could be regarded as evidence for the model adjusting itself
to realization variance in the BAO feature.

Now consider the bottom panel of Fig.~\ref{figure:overfit-compare}
for Realization 1/{\Advective}+{\EFT}.
Near $k = 0.13 h/\Mpc$
there is a relatively large excursion in the $P_0$ data to which
the individual best-fit responds but the ensemble best-fit does not.
More significantly, there is a weaker fit to $P_2$
over a relatively broad region $0.09 h/\Mpc \lesssim k \lesssim 0.21 h/\Mpc$.
While this is still evidence of adjustment to realization variance
it is not clear that the adjustment is driven by the BAO feature
rather than the  broadband effect in $P_2$.

\section{Conclusions}
\label{sec:Conclusions}

In this paper we have addressed the question: what is an appropriate level of modelling sophistication if we wish to extract cosmological information from present-day or next-generation galaxy surveys? A large selection of models are available to account for biasing or to predict the redshift-space power spectrum, with varying motivations---some empirical, and some motivated by theoretical considerations such as consistency of loop order in the description of clustering, bias, and redshift-space effects. Here we use two statistical diagnostics to test the rigour of the models available: the Bayesian Information Criteria, to evaluate the performance of the model given the data (i.e., model selection), and the distribution of $\chi^2$ values across realizations to test for over-fitting.

Our testing used the {\WizCOLA} simulation suite, originally developed to supply
realistic covariance matrices for the WiggleZ survey including details of
the survey geometry, incompleteness, selection function, and mask.
The power spectra and error bars derived from the suite therefore incorporate
a range of experimental effects that are relevant for state-of-the-art redshift
surveys.
Because our results already account for these effects, they provide
a picture of the performance of each model under real-world
circumstances
rather than a (possibly misleading) idealized case.

\para{Redshift-space power spectrum}
On the power spectrum side, the most complex model we consider is the
effective field-theory of large-scale structure.
Its construction is recounted in {\S}\ref{sec:modelling}.
In redshift space the basic power spectrum template matches the one-loop {\SPT} model,
modified by six subtractions known as `counterterms'.
Our testing shows that it can successfully match the broadband spectral slope
produced by the {\WizCOLA} realizations at both large and small $k$.
This cannot be done by the closely-related {\SPT} model, which often exhibits a mismatch
at small $k$. (The Monte Carlo fits prioritize high $k$, where the error bars are lower.
Although the small-$k$ region is likely to be most accurate in perturbation
theory this is not accounted for in the fitting, which explains why the region
we would na\"{\i}vely expect to be most accurate is in `tension' with the data.)
We ascribe this to a combination of the additive and multiplicative counterterms
that appear in the {\EFT} model.

Beside this,
the leading feature of the {\EFT} model is its ability to accommodate separate suppression
scales for each of the lowest-order multipoles $P_0$, $P_2$ and $P_4$.
However, our testing shows that the {\WizCOLA} realizations do not make significant
use of this freedom,
at least up to $\kmax = 0.29 h/\Mpc$;
clearly any such conclusion is
strongly $\kmax$-dependent. Since future surveys such as
Euclid, DESI and LSST will probe higher $\kmax$ it would be very interesting
to study the $\kmax$-dependence of the suppression scales.
If independent suppression is required
above $\kmax = 0.29 h/\Mpc$
then the {\EFT} model could become attractive.
With $\kmax = 0.29 h/\Mpc$, however, it does not provide enough statistical
return for its extra parameters and is strongly disfavoured by the Bayesian information criterion.
It would also be interesting to determine whether this
continues to be true when the BIC is replaced by a more
accurate measure of error-bar inflation due to marginalization over
the parameters of the model.

The {\SPT} model provides an underlying template that is very similar to the {\EFT} model.
Its subtler treatment of redshift-space effects compared to tree-level
enables it to improve its overall match to the structure of the $P_\ell$,
and the data do not penalize it for its single suppression scale
compared to the multiple suppression scales allowed by {\EFT}.
When paired with any sufficiently-permissive bias model,
to correct problems with the spectral slope,
it can nearly reproduce the $\chi^2$ values yielded by the much more permissive
{\EFT} model.
It is favoured by the BIC.
We conclude that a model of this type, including
\semibold{nonlinear redshift-space structure}
but only a \semibold{common suppression scale for $P_0$, $P_2$ and $P_4$}
seems likely to represent a good compromise for present-day or near-future
surveys with $\kmax$ not too dissimilar to $0.3 h/\Mpc$.
The `TNS' model of Taruya, Nishimichi \& Saito is of this type~\cite{Taruya:2010mx},
and we expect it would perform very similarly to our {\SPT} model.

\para{Bias models}
Our results do not show a preference for complex bias modelling.
Increasingly permissive models do give an improvement, but the effect is not large.
The critical feature is apparently the inclusion of \emph{some} nonlinearity,
which enables small changes to the slope of the power spectrum at large $k$.
Beyond this we do not see significant effects.
Accordingly, the BIC prioritizes the simplest nonlinear model we include---the
co-evolution model of Saito et al.~\cite{Saito:2014qha}.
As explained in~{\S}\ref{sec:Analysis} this is based on the 1-loop
McDonald \& Roy model with some nonlinear parameters constrained by matching to
analytic arguments and $N$-body simulations.

One might have imagined that the bias model could provide a means to shift
the amplitude and phase of the BAO feature.
It was explained in {\S}\ref{Results}
that this feature is quite variable in the {\WizCOLA} realizations.
This is simply a statistical effect: too few modes are measured to resolve it clearly.
In practice, however, it does not seem possible for nonlinear terms in the bias
to change the structure of the BAO feature significantly. This is
presumably
because such
changes would simultaneously change the broadband slope of the power spectrum,
at least in some regions, and would therefore significantly degrade the fit.
It is not likely that
this degradation could be compensated by the relatively small rewards on offer
for providing a better fit to the acoustic oscillation itself.

\para{Discussion}
A striking outcome of our analysis is that improvements in modelling
do not lead to a significant decrease in the overall $\chi^2$
when fitting to our {\WizCOLA} subsample;
we see only a $\Delta \chi^2 \approx 14.5$
decrease even from our best-performing model combination.
Since all our models fit the BAO variability equally badly,
the most significant
part of this improvement does not come from a better match to the BAO
feature, but rather from matching broadband features of the spectrum.
As we have argued, the rigidity of the BAO feature in our templates
is a good thing to the degree that it prevents overfitting.
In short, the tree-level Kaiser formula and the {\Halofit} power spectrum
do a surprisingly good job.
What could make a difference at larger $\kmax$ is the appearance of
different suppression scales for the low-order Legendre multipoles.
At present, however, there are no hints that this is required for the
{\WizCOLA} realizations.

We see some evidence for overfitting from the most permissive models---%
the {\EFT} model for the power spectrum,
and the {\MRoy} and {\Advective} models
for the bias,
although we require a larger sample size to resolve the shape of the
$\Delta\chi^2$ distributions with more certainty.
We do not see evidence for significant overfitting from the {\Linear},
{\KaiserTree} or {\KaiserHalo} models.
This risk of overfitting means that one should carefully characterize
whether the {\EFT} and {\Advective} models can really recover
unbiased estimates for the underlying cosmological parameters
which used in a parameter-estimation Monte Carlo.
This is an extremely interesting question to which we hope to return in the future.

\begin{acknowledgments}
The work reported in this paper
has been supported by the
European Research Council under the European Union's
Seventh Framework Programme (FP/2007--2013) and ERC Grant Agreement No. 308082
(DR, DS).
LFdlB acknowledges support from the UK Science and Technology
Facilities Council via Research Training Grant
ST/M503836/1.
The {\WizCOLA} simulations were performed
on the g2 supercomputer at Swinburne University of Technology.
\end{acknowledgments}

\newpage
\appendix

\section{Bias expansions: Dictionary between McDonald \& Roy and Chan et al.}
\label{app: Dictionary Perko-Assassi}

The relation between the basis of local, rotationally invariant operators
used by
Chan et al.~\cite{Chan:2012jj}
(also Assassi et al.~\cite{Assassi:2014fva})
and
McDonald \& Roy~\cite{McDonald:2009dh}
(also Senatore~\cite{Senatore:2014eva}, Angulo et al.~\cite{Angulo:2015eqa}
and Perko et al.~\cite{Perko:2016puo})
is given in Table \ref{tab: A vs P}.

\begingroup
\setlength{\tabcolsep}{10pt} 
\renewcommand{\arraystretch}{1.5} 
\begin{table}
\centering
\rowcolors{2}{gray!25}{white}
\begin{tabular}{lcc}
\toprule
& \semibold{McDonald \& Roy} & \semibold{Chan et al.} \\
\midrule
\multicolumn{1}{c}{\multirow{3}{*}{\semibold{Operators}}} & $\delta$ , $\theta$, $ \eta = \theta - \delta $                                                                                                                                                                                                                          & $\delta = \partial ^2 \Phi _g$ , $\theta =  \partial ^2 \Phi _v $                                                                                                        \\
& \begin{tabular}[c]{@{}c@{}}$\Psi = \eta -\eta ^{(2)}$\\ $s_{ij} = (\partial _i \partial _j \partial ^{-2} - \frac{1}{3}\delta _{ij}) \delta $\\ $t_{ij} = (\partial _i \partial _j \partial ^{-2} - \frac{1}{3}\delta _{ij}) \eta $, \end{tabular}                & \begin{tabular}[c]{@{}c@{}}$\mathcal{G}_2(\delta)$\\ $\Gamma _3(\delta, \theta)$\\ $\mathcal{G}_3(\delta)$\end{tabular}                                                                                                                                             \\
\midrule
\midrule
\multicolumn{1}{l}{$\Or(\delta)$} & $\delta ^{(1)}$                                                                                                                                                                                                                                                                                                    & $\delta ^{(1)}$                                                                                                                                                                                                                                                        \\
\multicolumn{1}{l}{$\Or(\delta^2)$} & $\delta ^{(2)}$ , $\delta ^{(1)\, 2}$, $s^2 = s_{ij} s^{ij}$                                                                                                                                                                                                                                                       & $\delta ^{(2)}$ , $\delta ^{(1)\, 2}$, $\mathcal{G}_2(\delta)$                                                                                                                                                                                                        \\
\multicolumn{1}{l}{$\Or(\delta^3)$} & \begin{tabular}[c]{@{}c@{}}$\delta ^{(3)}$, $\delta ^{(1)} \delta ^{(2)}$ , $\delta ^{(1)\, 3}$,\\  $\delta ^{(1)}s^2 $, $s_{ij} t^{ij}$, $\Psi $,\\ $s^ 3=s_{ij}s^{ik}s^j_k$,  $s_{ij}^{(1)} s^{ij \, (2)}$ \end{tabular} & \multicolumn{1}{l}{\begin{tabular}[c]{@{}l@{}}$\delta ^{(3)}$, $\delta ^{(1)} \delta ^{(2)}$ , $\delta ^{(1)\, 3}$,\\  $\mathcal{G}_2(\delta) \delta ^{(1)}$, $\mathcal{G}_2(\delta)^{(3)}$,\\ $\mathcal{G}_3(\delta)$, $\Gamma _3(\delta, \theta)$\end{tabular}} \\
\bottomrule
\end{tabular}
\caption{List of independent operators up to cubic order
used by McDonald \& Roy~\cite{McDonald:2009dh}
and Chan et al.~\cite{Chan:2012jj}, respectively,
as building blocks of local Eulerian bias.}
\label{tab: A vs P}
\end{table}
\endgroup

Bearing in mind that $\Phi _v^{(1)} = \Phi _g^{(1)}$,
the translation from the McDonald \& Roy basis
to the Chan et al. basis is
\begin{itemize}
\item Order $\Or(\delta^2)$
\begin{subequations}
\begin{align}
s^2 = & \, \frac{2}{3} \delta ^{(1)\,2} + \mathcal{G}_2(\delta)
\end{align}
\end{subequations}
\item Order $\Or(\delta^3)$
\begin{subequations}
\begin{align}
\delta ^{(1)} s^2 & = \frac{2}{3} \delta ^{(1)\,3} + \delta ^{(1)}\mathcal{G}_2(\delta)
\\
\Psi & = \frac{8}{21} \delta ^{(1)\,3} - \frac{4}{7} \delta ^{(1)} s^2 - \frac{2}{7} \mathcal{G}_2(\delta)
\\
s_{ij}t^{ij} & = \frac{1}{2} \Gamma _3(\delta, \theta) + \frac{1}{3} \delta ^2 - \frac{1}{3} \theta ^2
\\
s^3 & = \frac{5}{9} \delta ^{(1)\,3} - 2 \delta ^{(1)} \delta ^{(2)} - 2 \delta ^{(1)}\mathcal{G}_2(\delta)  - \mathcal{G}_3(\delta)
\\
2 s_{ij}^{(1)} s^{ij \, (2)} & = \frac{4}{3}   \delta ^{(1)} \delta ^{(2)}  + \mathcal{G}_2^{(3)}(\delta)
\end{align}
\end{subequations}
\end{itemize}
Additionally, Assassi et al. establish the relation
$\delta ^{(2)} - \theta ^{(2)} = - \frac{2}{7} \mathcal{G}_2^{(2)}$.

\section{Redshift-space halo power spectrum: Full calculation}
\label{app:p_calculations}

\subsection{Notation and operators in Fourier space}

Throughout this section we employ the following abbreviated notation
\begin{subequations}
\begin{align}
    \label{eq:not1}
    \int _{\vect{q},\vect{r}}^ {\vect{k}} & \equiv
        \int \frac{\d ^3 q }{(2 \pi)^3}\frac{\d ^3 r }{(2 \pi)^3} \, (2 \pi )^3 \delta(\vect{k} - \vect{q} - \vect{r}),
    \\
    \label{eq:not2}
    \int _{\vect{q},\vect{r},\vect{s}}^ {\vect{k}} & \equiv
        \int \frac{\d^3 q}{(2\pi)^3} \frac{\d^3r}{(2\pi)^3} \frac{\d^3 s}{(2\pi)^3} \, (2\pi)^3 \delta(\vect{k} - \vect{q} - \vect{r} - \vect{s}),
\end{align}
\end{subequations}
As in {\S}\ref{sec:Redshift-space distortions} we define
$\mu \equiv \hat{\vect{k}} \cdot \hat{\vect{x}}$
(see below Eq.~\eqref{eq:delta-s-oneloop}).
We also use the related variable
$\mu_q \equiv \hat{\vect{q}} \cdot \hat{\vect{x}}$.

The different operators appearing in the perturbative bias expansion
can all be defined in Fourier space.
Specifically, we have:
\begin{subequations}
\begin{align}
    \delta ^{(1)}_{\vect{k}} & = D(z) \delta ^{*}_{\vect{k}},
    \\
    \delta ^{(2)}_{\vect{k}} & = \int _{\vect{q},\vect{r}}^ {\vect{k}} F_{AB}(\vect{q},\vect{r};z) \delta_{\vect{q}}^{*} \delta_{\vect{r}}^{*} ,
    \\
    [\delta ^{(1) \,2}]_{\vect{k}} & = \int _{\vect{q},\vect{r}}^ {\vect{k}} D(z)^2 \delta_{\vect{q}}^{*} \delta_{\vect{r}}^{*},
    \\
    [\mathcal{G} _2^{(2)} (\delta)]_{\vect{k}} & = \int _{\vect{q},\vect{r}}^ {\vect{k}} G_2 (\vect{q},\vect{r}) D(z)^2 \delta_{\vect{q}}^{*} \delta_{\vect{r}}^{*} ,
    \\
    \delta ^{(3)}_{\vect{k}} & = \int _{\vect{q},\vect{s},\vect{r}}^ {\vect{k}}   T(\vect{q}, \vect{r}, \vect{r}, \vect{s} + \vect{r}; z) \delta_{\vect{q}}^{*} \delta_{\vect{s}}^{*}  \delta_{\vect{r}}^{*} ,
    \\
    [\delta ^{(1)} \delta ^{(2)}]_{\vect{k}} &= \int _{\vect{q},\vect{r}}^ {\vect{k}} \int _{\vect{p},\vect{s}}^ {\vect{r}}F_{AB}(\vect{p},\vect{s};z)\delta_{\vect{q}}^{*} \delta_{\vect{p}}^{*} \delta_{\vect{s}}^{*},
    \\
    [\delta ^{(1) \, 3}]_{\vect{k}} & = \int _{\vect{q},\vect{s},\vect{r}}^ {\vect{k}}  D(z)^3 \delta_{\vect{q}}^{*} \delta_{\vect{s}}^{*}  \delta_{\vect{r}}^{*},
    \\
    [\delta ^{(1)} \mathcal{G} _2 (\delta)]_{\vect{k}} & = \int _{\vect{q},\vect{s},\vect{r}}^ {\vect{k}}  G_2 (\vect{q},\vect{s}) D(z)^3 \delta_{\vect{q}}^{*} \delta_{\vect{s}}^{*} \delta_{\vect{r}}^{*},
    \\
    [\mathcal{G} _2 ^{(3)} (\delta)]_{\vect{k}} & = \int _{\vect{q},\vect{s},\vect{r}}^ {\vect{k}}  G_2 (\vect{q},\vect{s} + \vect{r}) F_{AB}(\vect{s},\vect{r};z) \delta_{\vect{q}}^{*} \delta_{\vect{s}}^{*}  \delta_{\vect{r}}^{*},
    \\
    [\mathcal{G} _3 (\delta)]_{\vect{k}} & = \int _{\vect{q},\vect{s},\vect{r}}^ {\vect{k}}  G_3 (\vect{q},\vect{s}, \vect{r}) D(z)^3 \delta_{\vect{q}}^{*} \delta_{\vect{s}}^{*}  \delta_{\vect{r}}^{*} ,
    \\
    [\Gamma _3 (\delta, \theta)]_{\vect{k}} & = \int _{\vect{q},\vect{r}}^ {\vect{k}}G_2 (\vect{q},\vect{r}) D(z)  \int _{\vect{p},\vect{s}}^ {\vect{r}}\left( F_{AB}(\vect{p},\vect{s};z) - \frac{1}{f}F_{KL}(\vect{p},\vect{s};z) \right) \delta_{\vect{q}}^{*} \delta_{\vect{p}}^{*}  \delta_{\vect{s}}^{*},
\end{align}
\end{subequations}
where $D(z)$ is the linear growth function whose growth factor is $f$.
The kernels appearing here are
\begin{subequations}
\begin{align}
    F_{AB}(\vect{q},\vect{r};z) & \equiv \DA(z) \bar{\alpha}(\vect{q}, \vect{r}) + \DB(z) \bar{\gamma}(\vect{q}, \vect{r}) ,
    \\
    F_{KL}(\vect{q},\vect{r};z) & \equiv \DK(z) \bar{\alpha}(\vect{q}, \vect{r}) + \DL(z) \bar{\gamma}(\vect{q}, \vect{r}) ,
    \\
    G_2 (\vect{q},\vect{r}) & \equiv -1 + \frac{(\vect{q} \cdot \vect{r})^2}{q^2 r^2} ,
    \\
    G_3 (\vect{q},\vect{r}, \vect{s}) & \equiv -\frac{1}{2} + \frac{3}{2} \frac{(\vect{q} \cdot \vect{r})^2}{q^2 r^2} -   \frac{(\vect{q} \cdot \vect{r})(\vect{q} \cdot \vect{s})(\vect{r} \cdot \vect{s}) }{q^2 r^2 s^2} ,
    \\
    T(\vect{q}, \vect{r}, \vect{r}, \vect{s} + \vect{r}; z) & \equiv
    \mbox{}
    \begin{aligned}[t]
        & 2  (\DD (z) - \DJ (z) ) \bar{\gamma}(\vect{s} + \vect{r}, \vect{q})  \bar{\alpha}(\vect{s}, \vect{r}) + 2 \DE (z) \bar{\gamma}(\vect{s} + \vect{r}, \vect{q})  \bar{\gamma}(\vect{s}, \vect{r})
        \\
        & \mbox{}
    	+ 2 (\DF (z) + \DJ (z) ) \bar{\alpha}(\vect{s} + \vect{r}, \vect{q})  \bar{\alpha}(\vect{s}, \vect{r}) + 2 \DG (z)  \bar{\alpha}(\vect{s} + \vect{r}, \vect{q})  \bar{\gamma}(\vect{s}, \vect{r})
        \\
        & \mbox{}
    	+ \DJ (z) \alpha(\vect{s} + \vect{r}, \vect{q})  \bar{\gamma}(\vect{s}, \vect{r}) - 2 \DJ (z) \alpha(\vect{s} + \vect{r}, \vect{q})  \bar{\alpha}(\vect{s}, \vect{r}) ,
    \end{aligned}
    \\
    \tilde{T}(\vect{q},\vect{s},\vect{r}, \vect{s} + \vect{r};z) & \equiv
    \mbox{}
    \begin{aligned}[t]
        & {-(1+z)} T'(\vect{q},\vect{s},\vect{r}, \vect{s} + \vect{r};z)
        \\
        & \mbox{}
    	- \alpha(\vect{q}, \vect{s} + \vect{r}) F_{AB}(\vect{s}, \vect{r}; z) f - \alpha( \vect{s} + \vect{r}, \vect{q}) F_{KL}(\vect{s}, \vect{r}; z) ,
    \end{aligned}
\end{align}
\end{subequations}
The functions $D_i$, where $i$ is one of $A$, $B$, $D$, $E$, $F$, $G$, $J$, $K$ or $L$, are the
one-loop growth functions defined by de la Bella et al.; see
Eqs.~(2.19a--b), (2.21a--e) and (2.32a--b) of Ref.~\cite{delaBella:2017qjy}.
Their corresponding growth factors
$f_i$ are defined by
$F_i = H^{-1} \d \ln D_i / \d t$.
The SPT kernels $\alpha(\vect{q},\vect{s})$ and $\gamma(\vect{q},\vect{s})$
are defined in Eqs.~(2.8a--b), (2.9) and (2.10) of Ref.~\cite{delaBella:2017qjy}
and satisfy
\begin{align}
    \alpha(\vect{q}, \vect{s}) & = \frac{\vect{q}\cdot(\vect{q}+\vect{s})}{q^2} ,
    \\
    \beta(\vect{q}, \vect{s}) & = \frac{\vect{q}\cdot\vect{s}}{2q^2 s^2}(\vect{q}+\vect{s})^2 ,
    \\
    \gamma(\vect{q}, \vect{s}) & = \alpha(\vect{q}, \vect{s}) + \beta(\vect{q}, \vect{s}) .
\end{align}
An overline denotes symmetrization with weight unity, so eg.
$\bar{\gamma}(\vect{q}, \vect{s}) = [ \gamma(\vect{q}, \vect{s}) + \gamma(\vect{s}, \vect{q}) ] / 2$.
A prime $'$ denotes a derivative with respect to redshift $z$.

\subsection{Halo density contrast in redshift space}

The linear, quadratic and cubic contributions to the redshift-space halo density contrast are
\begin{subequations}
\begin{align}
{\deltahalo_s}^{(1)} & = (b_1^{(1)} + f \mu ^2 ) \delta ^{(1)}_{\vect{k}},
\\
{\deltahalo_s}^{(2)} & = \mbox{}
    \begin{aligned}[t]
    & b_1^{(2)} \delta ^{(2)}_{\vect{k}} + \frac{b_{2}^{(2)}}{2!} [ \delta ^{(1)\,2} ]_{\vect{k}} + b_{\mathcal{G}_2}^{(2)} [\mathcal{G}_2^{(2)}]_{\vect{k}}
    \\
    & \mbox{}
    + \mu ^2 \int _{\vect{q},\vect{s}}^ {\vect{k}} F_{KL}(\vect{q},\vect{s};z) \delta_{\vect{q}}^{*} \delta_{\vect{s}}^{*}
    + \mu^2 k^2 \frac{( D f )^2}{2} \int _{\vect{q},\vect{s}}^ {\vect{k}} \frac{\mu _q \mu _s}{q s} \delta_{\vect{q}}^{*} \delta_{\vect{s}}^{*}
	+  b_1^{(1)} \mu k D^2 f \int _{\vect{q},\vect{s}}^ {\vect{k}} \frac{\mu _q}{q} \delta_{\vect{q}}^{*} \delta_{\vect{s}}^{*},
    \end{aligned}
\\
{\deltahalo_s}^{(3)} & = \mbox{}
    \begin{aligned}[t]
    & b_1^{(3)} \delta ^{(3)}_{\vect{k}}+ 2 \frac{b_{2}^{(3)}}{2!} [ \delta ^{(1)} \delta ^{(2)} ]_{\vect{k}}
    + b_{\mathcal{G}_2}^{(3)} [\mathcal{G}_2^{(3)}]_{\vect{k}}
    + \frac{b_{3}}{3!} [ \delta ^{(1)\,3} ]_{\vect{k}}
    + b_{1 \mathcal{G}_2} [\delta \mathcal{G}_2]_{\vect{k}}
    + b_{\mathcal{G}_3} [\mathcal{G}_3]_{\vect{k}}
    + b_{\Gamma _3} [ \Gamma _3]_{\vect{k}}
    \\
    & \mbox{}
    + \mu ^2 \int _{\vect{q},\vect{s},\vect{r}}^ {\vect{k}} \tilde{T}(\vect{q},\vect{s},\vect{r}, \vect{s}+\vect{r};z) \delta_{\vect{q}}^{*} \delta_{\vect{s}}^{*} \delta_{\vect{r}}^{*}
    + b_{1}^{(1)} \mu k D \int _{\vect{q},\vect{s}}^ {\vect{k}} \frac{\mu _q}{q} \int _{\vect{q}',\vect{s}'}^ {\vect{q}}F_{KL}(\vect{q}',\vect{s}';z) \delta_{\vect{q}'}^{*} \delta_{\vect{s}'}^{*} \delta_{\vect{s}}^{*}
    \\
    & \mbox{}
    + b_{1}^{(2)} \mu k f D \int _{\vect{q},\vect{s}}^ {\vect{k}} \frac{\mu _q}{q} \int _{\vect{q}',\vect{s}'}^ {\vect{s}} F_{AB}(\vect{q}',\vect{s}';z) \delta_{\vect{q}}^{*} \delta_{\vect{q}'}^{*} \delta_{\vect{s}'}^{*}
    + \frac{b_2^{(2)}}{2!} 2 \mu k f D^3  \int _{\vect{q},\vect{s},\vect{r}}^ {\vect{k}} \frac{\mu _q}{q} \delta_{\vect{q}}^{*} \delta_{\vect{s}}^{*} \delta_{\vect{r}}^{*}
    \\
    & \mbox{}
    + b_{\mathcal{G}_2}^{(2)}\mu k f D  \int _{\vect{q},\vect{s},\vect{r}}^ {\vect{k}} \frac{\mu _q}{q}  G_2(\vect{s},\vect{r}) \delta_{\vect{q}}^{*} \delta_{\vect{s}}^{*} \delta_{\vect{r}}^{*}
    + \mu^2 k^2 f D \int _{\vect{q},\vect{s}}^ {\vect{k}} \int _{\vect{q}',\vect{s}'}^ {\vect{s}} F_{KL}(\vect{q}',\vect{s}';z) \delta_{\vect{q}}^{*} \delta_{\vect{q}'}^{*} \delta_{\vect{s}'}^{*}
    \\
    & \mbox{}
    + b_1^{(1)} \frac{1}{2} \mu^2 k^2 f^2 D^3   \int _{\vect{q},\vect{s},\vect{r}}^ {\vect{k}} \frac{\mu _q \mu_s}{q s}   \delta_{\vect{q}}^{*} \delta_{\vect{s}}^{*} \delta_{\vect{r}}^{*}
    + \frac{1}{3} \mu^3 k^3 f^3 D^3   \int _{\vect{q},\vect{s},\vect{r}}^ {\vect{k}} \frac{\mu _q \mu_s \mu_r}{q s r}  \delta_{\vect{q}}^{*} \delta_{\vect{s}}^{*} \delta_{\vect{r}}^{*} .
    \end{aligned}
\end{align}
\end{subequations}

\subsection{Two-point statistics}
The one-loop halo--halo power spectrum breaks into
tree (`11') and one-loop (`13' and `22') pieces, defined by
\begin{subequations}
	\begin{align}
		(2 \pi)^3 \delta(\vect{k} + \vect{k}')P^{hh}_{s,11}(k) & =
		  \langle [{\deltahalo_s}^{(1)}]_{\vect{k}} [{\deltahalo_s}^{(1)}]_{\vect{k}'} \rangle ,
		\\
		(2 \pi)^3 \delta(\vect{k} + \vect{k}')P^{hh}_{s,13}(k) & =
		  \langle [{\deltahalo_s}^{(1)}]_{\vect{k}} [{\deltahalo_s}^{(3)}]_{\vect{k}'} \rangle
		  +
		  \langle [{\deltahalo_s}^{(3)}]_{\vect{k}} [{\deltahalo_s}^{(1)}]_{\vect{k}'} \rangle ,
		\\
		(2 \pi)^3 \delta(\vect{k} + \vect{k}')P^{hh}_{s,22}(k) & =
		\langle [{\deltahalo_s}^{(2)}]_{\vect{k}} [{\deltahalo_s}^{(1)}]_{\vect{k}'} \rangle
		.
	\end{align}
\end{subequations}

\subsubsection{Real space}
The real-space contributions
are those occurring at $\mu^0$ and yield
\begin{subequations}
\label{eq:phhreal}
\begin{align}
\label{eq:p11h}
P^{hh}_{11}(k,z) & = b_1^{(1)\,2} D(z)^2 P_{*}(k),
\\
\label{eq:p13h}
P^{hh}_{13}(k,z) & = \mbox{}
\begin{aligned}[t]
    & 4 b_1^{(1)}b_1^{(3)} D(z) P_{*}(k)  \int \frac{\d^3 q}{(2 \pi)^3} T(\vect{q}, \vect{k}, - \vect{q}, \vect{k} - \vect{q}; z) P_{*}(q) \\
    & \mbox{} + 8 b_1^{(1)} \frac{b_{2}^{(3)}}{2!} D(z)^2 P_{*}(k) \int \frac{\d^3 q}{(2 \pi)^3} P_{*}(q) F_{AB}(\vect{k},- \vect{q};z) \\
    & \mbox{} + 6 b_1^{(1)} \frac{b_{3}}{3!} D(z)^4 P_{*}(k)  \int \frac{\d^3 q}{(2 \pi)^3} P_{*}(q)\\
    & \mbox{} + 4 b_1^{(1)} b_{1\mathcal{G}_2} D(z)^4 P_{*}(k) \int \frac{\d^3 q}{(2 \pi)^3} P_{*}(q) G_2(\vect{k}, - \vect{q})\\
    & \mbox{} + 8 b_1^{(1)} b_{\mathcal{G}_2}^{(3)} D(z)^2 P_{*}(k) \int \frac{\d^3 q}{(2 \pi)^3} P_{*}(q)G_2(\vect{k}, \vect{k} - \vect{q}) F_{AB}(\vect{-q},\vect{k};z)\\
    & \mbox{} + 8 b_1^{(1)} b_{\Gamma _3} D(z)^2 P_{*}(k) \int \frac{\d^3 q}{(2 \pi)^3} P_{*}(q) G_2(\vect{k}, \vect{k}- \vect{q}) \Big( F_{AB}(\vect{k}, -\vect{q};z)- \frac{1}{f} F_{KL}(\vect{k}, -\vect{q};z)  \Big),
\end{aligned}
\\
\label{eq:p22h}
P^{hh}_{22}(k,z) & = \mbox{}
\begin{aligned}[t]
    & 2 b_1^{(2)\,2} \int \frac{\d^3 q}{(2 \pi)^3} P_{*}(q) P_{*}(|\vect{k}-\vect{q}|)F_{AB}(\vect{q}, \vect{k} - \vect{q}; z)^2 \\
    & + \mbox{} 2 \frac{b_{2}^{(2)\,2}}{2!^2} D(z)^4 \int \frac{\d^3 q}{(2 \pi)^3} P_{*}(q) P_{*}(|\vect{k}-\vect{q}|)\\
    & + \mbox{} 2 b_{\mathcal{G}_2}^{(2)\,2} D(z)^4 \int \frac{\d^3 q}{(2 \pi)^3} P_{*}(q) P_{*}(|\vect{k}-\vect{q}|)G_2(\vect{q}, \vect{k} - \vect{q})^2\\
    & + \mbox{} 4 b_{1}^{(2)} \frac{b_{2}^{(2)}}{2!} D(z)^2 \int \frac{\d^3 q}{(2 \pi)^3} P_{*}(q) P_{*}(|\vect{k}-\vect{q}|) F_{AB}(\vect{q},\vect{k}-\vect{q};z)\\
    & + \mbox{} 4 b_{1}^{(2)} b_{\mathcal{G}_2}^{(2)} D(z)^2 \int \frac{\d^3 q}{(2 \pi)^3} P_{*}(q) P_{*}(|\vect{k}-\vect{q}|) F_{AB}(\vect{q},\vect{k}-\vect{q};z) G_2(\vect{q}, \vect{k} - \vect{q})\\
    & + \mbox{} 4 \frac{b_{2}^{(2)}}{2!} b_{\mathcal{G}_2}^{(2)} D(z)^4 \int \frac{\d^3 q}{(2 \pi)^3} P_{*}(q) P_{*}(|\vect{k}-\vect{q}|)G_2(\vect{q}, \vect{k} - \vect{q}).
\end{aligned}
\end{align}
\end{subequations}

\subsubsection{Redshift space}
The corresponding redshift-space results are
\begin{center}
    \semibold{Tree level}
\end{center}
\begin{equation}
\label{eq:p11hs}
    P^{hh}_{s,\,11}(k,z) = \left( b_1^{(1)} + f \mu^2\right)^2 D(z)^2 P_{*}(k).
\end{equation}

\newpage
\begin{center}
    \semibold{13-type correlations}
\end{center}
\begin{equation}
    \label{eq:p13hs}
    P^{hh}_{s,\,13}(k,z)
    = \mbox{}
    \begin{aligned}[t]
        &
        \left( 1 + f \mu^2 /b_1^{(1)} \right) P^{hh}_{13}(k)
        \\
        & \mbox{}
        +
        4 D(z)^2 P_{*}(k)  \left( b_1^{(1)} + f \mu^2\right) \mu^2 \int  \frac{\d^3 q}{(2 \pi)^3} P_{*}(q) \tilde{T}(\vect{q}, \vect{k}, - \vect{q}, \vect{k} - \vect{q}; z)
        \\
        & \mbox{}
        +
        4 D(z)^2 P_{*}(k)  \left( b_1^{(1)} + f \mu^2\right) b_1^{(1)} \mu k \int  \frac{\d^3 q}{(2 \pi)^3} P_{*}(q)F_{KL}(\vect{k},- \vect{q};z)\frac{k\mu - q \mu_q}{|\vect{k} - \vect{q}|^2}
        \\
        & \mbox{}
        +
        4 D(z)^2 P_{*}(k)  \left( b_1^{(1)} + f \mu^2\right) b_1^{(2)} f \mu k \int  \frac{\d^3 q}{(2 \pi)^3} P_{*}(q)F_{AB}(\vect{k},- \vect{q};z)\frac{\mu_q}{q}
        \\
        & \mbox{}
        +
        4 D(z)^2 P_{*}(k)  \left( b_1^{(1)} + f \mu^2\right)  f \mu^2 k^2 \int  \frac{\d^3 q}{(2 \pi)^3} P_{*}(q)F_{KL}(\vect{k},- \vect{q};z)\frac{\mu_q}{q}\frac{k\mu - q \mu_q}{|\vect{k} - \vect{q}|^2}
        \\
        & \mbox{}
        +
        2 D(z)^4 P_{*}(k)  \left( b_1^{(1)} + f \mu^2\right) \frac{b_2^{(2)}}{2!} f\mu ^2 \int  \frac{\d^3 q}{(2 \pi)^3} P_{*}(q)
        \\
        & \mbox{}
        +
        D(z)^4 P_{*}(k)  \left( b_1^{(1)} + f \mu^2\right)^2 (f\mu k) ^2 \int  \frac{\d^3 q}{(2 \pi)^3} P_{*}(q)\left( \frac{\mu_q}{q}\right)^2
        .
    \end{aligned}
\end{equation}

\newpage
\begin{center}
    \semibold{22-type correlations}
\end{center}
\begin{equation}
\label{eq:p22hs}
    P^{hh}_{s,\,22}(k,z)
    = \mbox{}
    \begin{aligned}[t]
        &
        P^{hh}_{22}(k)
        \\
        & \mbox{}
        +
        4 b_1^{(2)} \mu^2 \int  \frac{\d^3 q}{(2 \pi)^3} P_{*}(q)P_{*}(|\vect{k} - \vect{q}|) F_{AB}(\vect{q}, \vect{k} - \vect{q}; z)F_{KL}(\vect{q}, \vect{k} - \vect{q}; z)
        \\
        & \mbox{}
        +
        2 \mu^4 \int  \frac{\d^3 q}{(2 \pi)^3} P_{*}(q)P_{*}(|\vect{k} - \vect{q}|) F_{KL}(\vect{q}, \vect{k} - \vect{q}; z)^2
        \\
        & \mbox{}
        +
        4 D(z)^2 \frac{b_2^{(2)}}{2!} \mu^2 \int  \frac{\d^3 q}{(2 \pi)^3} P_{*}(q)P_{*}(|\vect{k} - \vect{q}|) F_{KL}(\vect{q}, \vect{k} - \vect{q}; z)
        \\
        & \mbox{}
        +
        4 D(z)^2 b_{\mathcal{G}_2}^{(2)} \mu^2 \int  \frac{\d^3 q}{(2 \pi)^3} P_{*}(q)P_{*}(|\vect{k} - \vect{q}|) F_{KL}(\vect{q}, \vect{k} - \vect{q}; z)G_2(\vect{q}, \vect{k} - \vect{q})
        \\
        & \mbox{}
        +
        4 D(z)^2 b_{1}^{(1)}b_{1}^{(2)} f \mu k \int  \frac{\d^3 q}{(2 \pi)^3} P_{*}(q)P_{*}(|\vect{k} - \vect{q}|) F_{AB}(\vect{q}, \vect{k} - \vect{q};z) \frac{\mu_q}{q}
        \\
        & \mbox{}
        +
        4 D(z)^2 b_{1}^{(1)} \mu^2 f \mu k \int  \frac{\d^3 q}{(2 \pi)^3} P_{*}(q)P_{*}(|\vect{k} - \vect{q}|) F_{KL}(\vect{q}, \vect{k} - \vect{q};z) \frac{\mu_q}{q}
        \\
        & \mbox{}
        +
        2 D(z)^2 b_{1}^{(2)} (f \mu k)^2 \int  \frac{\d^3 q}{(2 \pi)^3} P_{*}(q)P_{*}(|\vect{k} - \vect{q}|) F_{AB}(\vect{q}, \vect{k} - \vect{q};z) \frac{\mu_q}{q} \frac{\mu_{\vect{k}-\vect{q}}}{|\vect{k}-\vect{q}|}
        \\
        & \mbox{}
        +
        2 D(z)^2 \mu^2 (f \mu k)^2 \int  \frac{\d^3 q}{(2 \pi)^3} P_{*}(q)P_{*}(|\vect{k} - \vect{q}|) F_{KL}(\vect{q}, \vect{k} - \vect{q};z) \frac{\mu_q}{q} \frac{\mu_{\vect{k}-\vect{q}}}{|\vect{k}-\vect{q}|}
        \\
        & \mbox{}
        +
        4 D(z)^4 b_{1}^{(1)}\frac{b_{2}^{(2)}}{2!} f \mu k \int  \frac{\d^3 q}{(2 \pi)^3} P_{*}(q)P_{*}(|\vect{k} - \vect{q}|) \frac{\mu_q}{q}
        \\
        & \mbox{}
        +
        4 D(z)^4 b_{1}^{(1)}b_{\mathcal{G}_2}^{(2)} f \mu k \int  \frac{\d^3 q}{(2 \pi)^3} P_{*}(q)P_{*}(|\vect{k} - \vect{q}|) G_2(\vect{q}, \vect{k} - \vect{q}) \frac{\mu_q}{q}
        \\
        & \mbox{}
        +
        2 D(z)^4 \frac{b_{2}^{(2)}}{2!} (f \mu k)^2 \int  \frac{\d^3 q}{(2 \pi)^3} P_{*}(q)P_{*}(|\vect{k} - \vect{q}|) \frac{\mu_q}{q}\frac{\mu_{\vect{k}-\vect{q}}}{|\vect{k}-\vect{q}|}
        \\
        & \mbox{}
        +
        2 D(z)^4 b_{\mathcal{G}_2}^{(2)} (f \mu k)^2 \int  \frac{\d^3 q}{(2 \pi)^3} P_{*}(q)P_{*}(|\vect{k} - \vect{q}|) G_2(\vect{q}, \vect{k} - \vect{q}) \frac{\mu_q}{q}\frac{\mu_{\vect{k}-\vect{q}}}{|\vect{k}-\vect{q}|}
        \\
        & \mbox{}
        +
        2 D(z)^4 b_{1}^{(1)\,2} (f \mu k)^2 \int  \frac{\d^3 q}{(2 \pi)^3} P_{*}(q)P_{*}(|\vect{k} - \vect{q}|)\frac{\mu_q}{q}\left( \frac{\mu_q}{q} + \frac{\mu_{\vect{k}-\vect{q}}}{|\vect{k}-\vect{q}|}\right)
        \\
        & \mbox{}
        +
        2 D(z)^4 b_1^{(1)} (f \mu k)^3 \int  \frac{\d^3 q}{(2 \pi)^3} P_{*}(q)P_{*}(|\vect{k} - \vect{q}|) \left( \frac{\mu_q}{q}\right)^2\frac{\mu_{\vect{k}-\vect{q}}}{|\vect{k}-\vect{q}|}
        \\
        & \mbox{}
        +
        \frac{1}{2} D(z)^4 (f \mu k)^4 \int  \frac{\d^3 q}{(2 \pi)^3} P_{*}(q)P_{*}(|\vect{k} - \vect{q}|)\left(\frac{\mu_q}{q}\frac{\mu_{\vect{k}-\vect{q}}}{|\vect{k}-\vect{q}|}\right)^2
    \end{aligned}
\end{equation}

\para{Advective terms} The advective contributions are
\begin{subequations}
	\begin{align}
		(2 \pi)^3 \delta(\vect{k} + \vect{k}')P^{\text{Adv}}_{s,\,13}(k)
		& =
		2 \langle [\deltahalo_s]^{(1)}_{\vect{k}} [\deltaadvect]^{(3)}_{\vect{k}'} \rangle
		+
		2 \langle [\deltahalo_s]^{(1)}_{\vect{k}} [\deltaadvects]^{(3)}_{\vect{k}'} \rangle
		,
		\\
		(2 \pi)^3 \delta(\vect{k} + \vect{k}')P^{\text{Adv}}_{s,\,22}(k)
		& =
		\langle [\deltahalo_s]^{(2)}_{\vect{k}} [\deltaadvect]^{(2)}_{\vect{k}'} \rangle
		+
		\langle [\deltaadvect]^{(2)}_{\vect{k}} [\deltahalo_s]^{(2)}_{\vect{k}'} \rangle
		+
		\langle [\deltaadvect]^{(2)}_{\vect{k}} [\deltaadvect]^{(2)}_{\vect{k}'} \rangle
		.
	\end{align}
\end{subequations}
Working out the $13$-type correlation functions, we find
\begin{equation}\small
P^{\text{Adv}}_{s,\,13}(k)
=
\begin{aligned}[t]
&
D^2 P_{*}(k) \, 4 \left( b_1^{(1)} + f \mu^2 \right) \left( b_1^{(2)} - b_1^{(3)} \right) \int \frac{\d ^3q}{(2 \pi)^3} P_{*}(q) \frac{\vect{q} \cdot (\vect{k} - \vect{q}) }{q^2} F_{AB}(\vect{k},\vect{q};z)\\
& +
D^2 P_{*}(k)  \frac{1}{f}   \left( b_1^{(1)} + f \mu^2 \right) \left( b_1^{(1)} - b_1^{(3)} \right)  \int \frac{\d ^3q}{(2 \pi)^3} P_{*}(q) \frac{\vect{q} \cdot (\vect{k} - \vect{q}) }{|\vect{k} - \vect{q}|^2} F_{KL}(\vect{k},\vect{q};z)\\
& +
D^4 P_{*}(k) \, 2  \left( b_1^{(1)}  + f \mu^2 \right) \left( b_1^{(1)} - b_1^{(2)} \right) \mu k \int \frac{\d ^3q}{(2 \pi)^3} P_{*}(q) \frac{\vect{q}\cdot \vect{k}}{q^2}  \frac{q^2 + k^2}{k^2} \frac{\mu_q}{q} \\
& +
D^4 P_{*}(k) \,2 \left( b_1^{(1)}  + f \mu^2 \right) \left( \frac{b_1^{(1)} +  b_1^{(3)}}{2} -  b_1^{(2)}\right) \int \frac{\d ^3q}{(2 \pi)^3} P_{*}(q) \frac{\vect{q} \cdot (\vect{k} - \vect{q})}{q^2} \left(\frac{k}{q} + \frac{q}{k} \right)\\
& +
D^4 P_{*}(k) \, 4 \left( b_1^{(1)} + f \mu^2  \right) \left( \frac{b_2^{(2)}}{2!} - \frac{b_2^{(3)}}{2!} \right) \int \frac{\d ^3q}{(2 \pi)^3} P_{*}(q) \frac{\vect{q} \cdot (\vect{k} - \vect{q}) }{q^2} \\
& +
D^4 P_{*}(k) \, 4 \left( b_1^{(1)} + f \mu^2  \right) \left( b_{\mathcal{G}_2}^{(2)} - b_{\mathcal{G}_2}^{(3)} \right) \int \frac{\d ^3q}{(2 \pi)^3} P_{*}(q) \frac{\vect{q} \cdot (\vect{k} - \vect{q}) }{q^2} G_2(\vect{k}, \vect{q}).
\end{aligned}
\end{equation}
The $22$-type contributions read
\begin{equation}\small
P^{\text{Adv}}_{s,\,22}(k)
=
\begin{aligned}[t]
&
4 D^2   \left( b_1^{(1)} - b_1^{(2)} \right) b_1^{(2)}  \int \frac{\d ^3q}{(2 \pi)^3} P_{*}(q) P_{*}(|\vect{k} - \vect{q}|)\frac{\vect{q} \cdot (\vect{k} - \vect{q}) }{q^2} F_{AB}(\vect{q},\vect{k}- \vect{q};z)\\
& +
4 D^2 \left( b_1^{(1)} - b_1^{(2)} \right) \mu^2 \int \frac{\d ^3q}{(2 \pi)^3} P_{*}(q)P_{*}(|\vect{k} - \vect{q}|) \frac{\vect{q} \cdot (\vect{k} - \vect{q}) }{q^2} F_{KL}(\vect{q},\vect{k}- \vect{q};z)\\
& +
4 D^4   \left( b_1^{(1)} - b_1^{(2)} \right) \frac{b_2^{(2)}}{2!}  \int \frac{\d ^3q}{(2 \pi)^3} P_{*}(q) P_{*}(|\vect{k} - \vect{q}|)\frac{\vect{q} \cdot (\vect{k} - \vect{q}) }{q^2}\\
& +
4 D^4   \left( b_1^{(1)} - b_1^{(2)} \right) b_{\mathcal{G}_2}^{(2)}  \int \frac{\d ^3q}{(2 \pi)^3} P_{*}(q) P_{*}(|\vect{k} - \vect{q}|)\frac{\vect{q} \cdot (\vect{k} - \vect{q}) }{q^2} G_{2}(\vect{q},\vect{k}- \vect{q})\\
& +
D^4   \left( b_1^{(1)} - b_1^{(2)} \right)^2  \int \frac{\d ^3q}{(2 \pi)^3} P_{*}(q) P_{*}(|\vect{k} - \vect{q}|)\left( \frac{\vect{q} \cdot (\vect{k} - \vect{q}) }{q^2}\right)^2  \frac{q^2 + |\vect{k} - \vect{q}|^2}{|\vect{k} - \vect{q}|^2}\\
& +
2 D^4   \left( b_1^{(1)} - b_1^{(2)} \right) b_1^{(1)} f \mu k \int \frac{\d ^3q}{(2 \pi)^3} P_{*}(q) P_{*}(|\vect{k} - \vect{q}|)  \frac{\vect{q} \cdot (\vect{k} - \vect{q})}{q^2}\left( \frac{\mu_q}{q} +  \frac{\mu_{\vect{k}- \vect{q}}}{|\vect{k} - \vect{q}|} \right)\\
& +
2 D^4   \left( b_1^{(1)} - b_1^{(2)} \right) (f \mu k)^2 \int \frac{\d ^3q}{(2 \pi)^3} P_{*}(q) P_{*}(|\vect{k} - \vect{q}|)  \frac{\vect{q} \cdot (\vect{k} - \vect{q}) }{q^2} \frac{\mu_q \mu_{\vect{k} - \vect{q}}}{q|\vect{k} - \vect{q}|}.
\end{aligned}
\end{equation}

\subsection{Algorithm for evaluating the loop integrals}
\label{app:evaluate-rsd-twopf}

Evaluation of the loop integrals above appears to be
a daunting calculation. In this section, we summarize the algorithm
presented in an earlier paper, Ref.~\cite{delaBella:2017qjy},
in order to deal with the tensorial parts of the one-loop correlation functions
in redshift space.

\begin{enumerate}
\item In integrals of the form
\begin{equation}
    I = \int \frac{\d^3 q}{(2\pi)^3} \;
    P_\ast(q) P_\ast(|\vect{k}-\vect{q}|) \times\cdots ,
\end{equation}
introduce a variable $\vect{s} = \vect{k} - \vect{q}$ and a $\delta$-function constraint:
\begin{equation}
    I = \int \frac{\d^3 q \; \d^3 s}{(2\pi)^6} \;
    (2\pi)^3 \delta(\vect{k} - \vect{q} - \vect{s})
    P_\ast(q) P_\ast(r) \times \cdots .
\end{equation}

\item Replace the Dirac $\delta$-function by its Fourier representation, and
expand the resulting exponential using the Rayleigh plane wave formula,
\begin{equation}
    \e{\im \vect{k}\cdot\vect{x}}
    =
    \sum_{\ell=0}^\infty
    (2\ell+1)
    \im^\ell
    j_\ell(kx)
    \Legendre{\ell}{\hat{\vect{k}}\cdot\hat{\vect{x}}}
    \label{eq:rayleigh-formula}
\end{equation}
where $j_\ell$ is the spherical Bessel function of order $\ell$
and $\Legendre{\ell}{x}$ is the $\ell^{\text{th}}$ Legendre polynomial.

\item The angular part of the $\vect{q}$, $\vect{s}$ and $\vect{x}$
integrations can be done using
the generalized orthogonality relation
\begin{equation}
    \int \d^2 \hat{\vect{x}} \;
    \Legendre{\ell}{\hat{\vect{a}}\cdot\hat{\vect{x}}}
    \Legendre{\ell'}{\hat{\vect{b}}\cdot\hat{\vect{x}}}
    =
    \frac{4\pi}{2\ell+1}
    \delta_{\ell\ell'}
    \Legendre{\ell}{\hat{\vect{a}}\cdot\hat{\vect{b}}} .
    \label{eq:generalized-Legendre-orthogonality}
\end{equation}

\item For $P_{22}$ integrals, involving $P_{*}(q)P_{*}(|\vect{k}-\vect{q}|)$, we use the  3-Bessel integral $\FabrikantThree{\mu}{\nu}{\sigma}$
\begin{equation}
    \FabrikantThree{\mu}{\nu}{\sigma}
    \equiv
    \int_0^\infty \d x \; x^2 j_\mu(k x) j_\nu(q x) j_\sigma(s x) ,
    \label{eq:fabrikant-3J}
\end{equation}
where $s = (q^2 + k^2 - 2 k q \cos\theta)^{1/2}$ and the different subscripts are associated to the different wavenumbers: $\mu \mapsto k$, $\nu \mapsto q$ and $\sigma \mapsto s$\footnote{The analytical solution to these integrals for general $k$, $q$ and $s$
and arbitrary orders $\mu$, $\nu$ and $\sigma$ was solved by Gervois \& Navelet~\cite{doi:10.1137/0520067} and Fabrikant~\cite{fabrikant2013elementary}.}. In general,  $\FabrikantThree{\mu}{\nu}{\sigma} \neq 0$ where $k$, $q$ and $s$ satisfy the triangle condition
$|k-q| < s < |k+q|$. The result is a scalar integral over $q$ and $\theta$.
We collect the results needed for the computation of the power spectrum:
\begin{subequations}
\begin{align}
    \FabrikantThree{0}{0}{0} & = \frac{\pi}{4 k q s} \\
    \FabrikantThree{1}{1}{0} & = \frac{\pi}{8} \frac{k^2 + q^2 - s^2}{k^2 q^2 s} \\
    \FabrikantThree{2}{2}{0} & = \frac{\pi}{32} \frac{3k^4 + 2k^2(q^2 - 3s^2) + 3(q^2 - s^2)^2}{k^3 q^3 s} \\
    \FabrikantThree{2}{2}{2} & = \frac{\pi}{64} \frac{(3k^4 + 2k^2 q^2 + 3 q^4) s^2 + 3(k^2 + q^2)s^4 -
        3 (k^2 - q^2)^2(k^2 + q^2) - 3 s^6}{k^3 q^3 s^3}
    \label{eq:fabrikant231}
\end{align}
\end{subequations}
\begin{subequations}
\begin{align}
\FabrikantThree{2}{3}{1} & = \frac{\pi}{64} \frac{3k^4(q^2 + 5s^2) + (q^2 -s^2)^2(q^2 + 5s^2) + k^2(q^4 + 6 q^2 s^2 - 15 s^4) - 5k^6}
        {k^3 q^4 s^2} \\
    \nonumber
    \FabrikantThree{2}{4}{2} & = \frac{\pi}{512} \frac{1}{k^3 q^5 s^3} \Big(
        35 k^8 - 20 k^6(3q^2 + 7s^2) + 6k^4(3q^4 + 10 q^2 s^2 + 35s^2)
    \\ & \qquad\qquad \mbox{} + (q^2 - s^2)^2(3q^4 + 10 q^2 s^2 + 35 s^4)
        + 4 k^2 (q^6 + 3q^4 s^2 + 15 q^2 s^4 - 35 s^6)
    \Big)
    \label{eq:fabrikant330}
\end{align}
\end{subequations}
\begin{subequations}
\begin{align}
    \FabrikantThree{3}{3}{0} & = \frac{\pi}{64} \frac{(k^2 + q^2 - s^2)\big[5k^4 + 5(q^2 - s^2)^2 - 2k^2(q^2 + 5s^2)\big]}
        {k^4 q^4 s} \\
    \nonumber
    \FabrikantThree{4}{4}{0} & = \frac{\pi}{512} \frac{1}{k^4 q^5 s} \Big(
        35 k^8 + 20 k^2(q^2 - 7s^2)\big[ k^4 + (q^2 - s^2)^2\big] + 35(q^2 - s^2)^4
    \\ & \qquad\qquad \mbox{} + 6k^4(3q^4 - 30q^2 s^2 + 35 s^4)
    \Big)
\end{align}
\end{subequations}
Index permutations can be obtained by making suitable exchanges of $k$, $q$ and $s$;
for example, $\FabrikantThree{2}{1}{3}$ can be obtained from~\eqref{eq:fabrikant231}
by exchanging $q$ and $s$,
and $\FabrikantThree{0}{3}{3}$ can be obtained from~\eqref{eq:fabrikant330}
by exchanging $k$ and $s$.

\item For $P_{13}$ integrals
the procedure is very similar.  These are typically simpler because they involve integration only
over $P_{*}(q)$ and therefore  can be performed analytically using the Fourier transform
\begin{equation}
\int \d^3 s \, s^{-2} \e{\im \vect{s}\cdot\vect{x}} = 2\pi^2/x .
\end{equation}
Consequently, 13-type integrals require only 2-Bessel integrals of the form
\begin{equation}
   \FabrikantOne{\mu}
    \equiv
    \int_0^\infty \d x \; x j_\mu(k x) j_\mu(q x).
    \label{eq:fabrikant-2J}
\end{equation}
The Bessel functions are easily computable using (for example) \sansbold{Mathematica}.
\end{enumerate}

\newpage
\subsection{Final results}

\subsubsection{Power spectrum}

\para{13-type integrals}
\begin{itemize}
\item $\mathbf{\mu^0}$ \textbf{terms}
\begin{subequations}
\begin{align}
\begin{split}
b_1^{(1)}b_1^{(3)} \quad \rightarrow & \quad \quad  - D k P(k) \int_{q_{IR}}^{q_{UV}}\d q \int_{-1}^1 \d x \,\frac{ P(q)}{2 \pi ^2 \left(k^2-2 k q x+q^2\right)} \times \\
&\times \Big[ x\Big(2 D_D
		\left(k^3 x-k^2 q 			\left(x^2+2\right)		+3 k q^2 x-q^3 x^2 \right) \\
		&\quad  +4 D_E  (k x-q) (k-q x)^2
		-D_J		\left(k^3 x-3 k^2 q+3 k q^2 x-2 q^3 x^2+q^3\right)
	\Big)\\
&+ D_F \left(k^3 x^2-k^2 q	 \left(2  x^3+x\right)+k q^2 \left(5 x^2-2\right)+q^3 x \left(1-2 x^2\right)
		  \right)\\
   &	+2 D_G	 \left(k^3 x^2-3 k^2 q x^3 + k q^2 \left(2   x^4+2 x^2-1\right)+q^3 x \left(1-2 x^2\right)
			 \right)	 \Big]
\end{split}
\\
\begin{split}
b_1^{(1)}b_2^{(3)} \quad \rightarrow & \quad - D^2 P(k) \int_{q_{IR}}^{q_{UV}}\d q \int_{-1}^1 \d x \, \frac{q P(q)}{2 \pi ^2 k} \times \\
& \quad \times \left[ D_A \left(k^2 x-2 k q+q^2 x\right)+2 D_B \left(k^2 x-k q \left(x^2+1\right)+q^2
   x\right)\right]
   \end{split}
   \\
b_1^{(1)}b_3 \quad \rightarrow & \quad \quad  D^4 P(k) \int_{q_{IR}}^{q_{UV}}\d q \int_{-1}^1 \d x \,\frac{ q^2  P(q)}{4 \pi ^2} \\
b_1^{(1)}b_{1\mathcal{G}_2}  \quad \rightarrow  & \quad \quad D^4 P(k) \int_{q_{IR}}^{q_{UV}}\d q \int_{-1}^1 \d x \, \frac{ q^2 \left(x^2-1\right) P(q)}{\pi ^2} \\
\begin{split}
b_1^{(1)}b_{\mathcal{G}_2}^{(3)}  \quad \rightarrow & \quad - D^2 k P(k) \int_{q_{IR}}^{q_{UV}}\d q \int_{-1}^1 \d x \,\frac{q \left(x^2-1\right) P(q)}{\pi ^2 \left(k^2-2 k q x+q^2\right)} \times \\
& \quad \times \left(D_A \left(k^2 x-2 k q+q^2 x\right)+2 D_B \left(k^2 x-k q
   \left(x^2+1\right)+q^2 x\right)\right)
\end{split}
\\
\begin{split}
b_1^{(1)}b_{\Gamma_3}  \quad \rightarrow & \quad - D^2 k P(k) \int_{q_{IR}}^{q_{UV}}\d q \int_{-1}^1 \d x \,\frac{q \left(x^2-1\right)  P(q)}{\pi ^2 f \left(k^2-2 k q
   x+q^2\right)} \times \\
   & \quad \times \Big[ D_A (f-f_A) \left(k^2 x-2 k q+q^2 x\right) +2 D_B (f-f_B) \left(k^2 x-k q \left(x^2+1\right)+q^2 x\right)\\
   & \quad \quad +f g^2 \left(k^2 x-2 k q+q^2 x\right)\Big]
\end{split}
\end{align}
\end{subequations}

\item $\mathbf{\mu^2}$ \textbf{terms}
\begin{subequations}
\begin{align}
\begin{split}
b_1^{(1)} \quad \rightarrow & \quad - D^2  P(k) \int_{q_{IR}}^{q_{UV}}\d q \int_{-1}^1 \d x \, -\frac{P(q)}{96 \pi^2 k^3 q^3 \left|k^2-2 k q x+q^2\right|}\times \\
   & \quad \times
   \Big[
   48 k^3 q^3 \Big( \Big(
   			-k q x \Big(k^2 (D_A f_A D + 2 D_B f_B D +  4 D_D f_D + 4  D_E f_E + F f_F - 3  J f_J) \\
   			& \quad \quad \quad \quad
   			+ q^2 (3 A f_A D + 4 B f_B D - F f_F - 2 D_G f_G + D_J f_J) \Big)\\
   			& \quad \quad \quad
   			x^2 \Big(k^2 q^2 (D_A f_A D + 4 D_B f_B D +  6 D_D f_D + 8  D_E f_E + 5  F f_F + 4 D_G f_G- 3  J f_J) \\
   			& \quad \quad \quad \quad
   					+ D q^4 (D_A f_A D + 2 D_B f_B D + k^4\left( 2 D_D f_D + 4  D_E f_E +  F f_F + 2 D_G f_G-   J f_J) \right) \Big)
   			  	\Big)\\
   			  & \quad \quad \quad
			 + 2 k^2 q^2 (D_A f_A D + 2 D_B f_B D  -	 F f_F - D_G f_G	)
   			 \Big)
   \\
   &\quad \quad
   2 f k q \Big(2 k^5 q^3 x \left(12 D \left(2 x^2 (D_A + 3 D_B) + 2 D_A + 2 D_B + D^2 \right) - 11(D_K + D_L) \right)\\
   				& \quad \quad \quad
   				 - 2 k^4 q^4 \left(12 D \left(x^2 \left(6 D_A + 8 D_B + D^2 \right) + 4 D_B x^4 + 2 D^2 \right) - 11(D_K + D_L) \right)\\
   				 & \quad \quad \quad
   				  + 2 k^3 q^5 x \left(12 D \left(2 x^2 (D_A + 3 D_B) + 2 D_A + 2 D_B + 3 D^2 \right) - 11 (D_K + D_L) \right)\\
   				 & \quad \quad \quad
   				 + 8 k^2 q^6 \left(-3 D x^2 \left(D_A + 2 D_B + D^2 \right) + D_K + D_L \right)\\
   				 & \quad \quad \quad
   				 +  8 k^6 q^2 \left(-3 D x^2 (D_A + 2 D_B) + D_K + D_L \right)\\
   				 & \quad \quad \quad
   				 -  3 k^8   (D_K + D_L) + 6 k^7 q x (D_K + D_L) +  6 k q^7 x (D_K + D_L) - 3 q^8 (D_K + D_L)
   	   		\Big)\\
   & \quad \quad
   + 3 f \left(k^2 - q^2 \right)^4   (D_K + D_L) \left|k^2 -2 k q x +  q^2 \right| \ln\left| \frac {k + q}{k - q}  \right|
   \Big]
\end{split}
\\
\begin{split}
b_1^{(3)} \quad \rightarrow & \quad - f D k P(k) \int_{q_{IR}}^{q_{UV}}\d q \int_{-1}^1 \d x \,\frac{ P(q)}{2 \pi ^2 \left(k^2-2 k q x+q^2\right)} \times \\
&\times \Big[ x\Big(2 D_D
		\left(k^3 x-k^2 q 			\left(x^2+2\right)		+3 k q^2 x-q^3 x^2 \right) \\
		&\quad  +4 D_E  (k x-q) (k-q x)^2
		-D_J		\left(k^3 x-3 k^2 q+3 k q^2 x-2 q^3 x^2+q^3\right)
	\Big)\\
&+ D_F \left(k^3 x^2-k^2 q	 \left(2  x^3+x\right)+k q^2 \left(5 x^2-2\right)+q^3 x \left(1-2 x^2\right)
		  \right)\\
   &	+2 D_G	 \left(k^3 x^2-3 k^2 q x^3 + k q^2 \left(2   x^4+2 x^2-1\right)+q^3 x \left(1-2 x^2\right)
			 \right)	 \Big]
\end{split}
\\
\begin{split}
b_2^{(3)} \quad \rightarrow & \quad - f D^2 P(k) \int_{q_{IR}}^{q_{UV}}\d q \int_{-1}^1 \d x \, \frac{q P(q)}{2 \pi ^2 k} \times \\
& \quad \times \left[ D_A \left(k^2 x-2 k q+q^2 x\right)+2 D_B \left(k^2 x-k q \left(x^2+1\right)+q^2
   x\right)\right]
   \end{split}
\end{align}
\end{subequations}
\begin{subequations}
\begin{align}
b_3 \quad \rightarrow & \quad \quad  f D^4 P(k) \int_{q_{IR}}^{q_{UV}}\d q \int_{-1}^1 \d x \,\frac{ q^2  P(q)}{4 \pi ^2} \\
b_{1\mathcal{G}_2}  \quad \rightarrow & \quad \quad f D^4 P(k) \int_{q_{IR}}^{q_{UV}}\d q \int_{-1}^1 \d x \, \frac{ q^2 \left(x^2-1\right) P(q)}{\pi ^2} \\
\begin{split}
b_{\mathcal{G}_2}^{(3)} \quad \rightarrow & \quad - f D^2 k P(k) \int_{q_{IR}}^{q_{UV}}\d q \int_{-1}^1 \d x \,\frac{q \left(x^2-1\right) P(q)}{\pi ^2 \left(k^2-2 k q x+q^2\right)} \times \\
& \quad \times \left(D_A \left(k^2 x-2 k q+q^2 x\right)+2 D_B \left(k^2 x-k q
   \left(x^2+1\right)+q^2 x\right)\right)
\end{split}
\\
\begin{split}
b_{\Gamma_3}  \quad \rightarrow & \quad -f D^2 k P(k) \int_{q_{IR}}^{q_{UV}}\d q \int_{-1}^1 \d x \,\frac{q \left(x^2-1\right)  P(q)}{\pi ^2 f \left(k^2-2 k q
   x+q^2\right)} \times \\
   & \quad \times \Big[ D_A (f-f_A) \left(k^2 x-2 k q+q^2 x\right) +2 D_B (f-f_B) \left(k^2 x-k q \left(x^2+1\right)+q^2 x\right)\\
   & \quad \quad +f g^2 \left(k^2 x-2 k q+q^2 x\right)\Big]
\end{split}
\end{align}
\end{subequations}
\begin{subequations}
\begin{align}
\begin{split}
b_1^{(1)}b_1^{(1)} \quad \rightarrow & \quad \quad D^2  P(k) \int_{q_{IR}}^{q_{UV}}\d q \int_{-1}^1 \d x \, \frac{P(q)}{24 \pi ^2 k^3} \times \\
& \quad \times
\Big[
\frac{3 }{q}\left(k^2+q^2\right) \left(k^2-q^2\right)^2 (D_K+D_L) \ln \left| \frac{k+q}{k-q}\right| \\
& \quad \quad +
\frac{2 k}{\left|k^2-2 k q x+q^2\right|}
	\Big(6 k^5 q x \left(-D_A f_A-2 D_B f_B+f g^2 \left(fx^2+1\right)+D_K+D_L\right)\\
	& \quad \quad \quad
	+ k^4 q^2 \left(12 D_A f_A+12 D_B f_B \left(x^2+1\right)-3 f g^2 \left(f x^2+4\right)+3 D_K+7 D_L\right)\\
	& \quad \quad \quad
	-2 k^3 q^3 x \left(3 D_A f_A+6 D_B f_B-3 f g^2+6 D_K+10 D_L\right)\\
	& \quad \quad \quad
	-3  k^6 \left(f^2 g^2 x^2+D_K+D_L\right)+k^2 q^4 (3 D_K+7 D_L)\\
	& \quad \quad \quad
	+6 k q^5 x (D_K+D_L)-3 q^6
   (D_K+L)
   \Big)
\Big]
\end{split}
\\
\begin{split}
b_1^{(1)}b_1^{(2)} \quad \rightarrow & \quad - f D^2   P(k) \int_{q_{IR}}^{q_{UV}}\d q \int_{-1}^1 \d x \, \frac{P(q) }{3 \pi ^2} (A+2 B)  \left(k^2+q^2\right)
\end{split}
\\
\begin{split}
b_1^{(1)}b_2^{(2)} \quad \rightarrow & \quad \quad f D^4  P(k) \int_{q_{IR}}^{q_{UV}}\d q \int_{-1}^1 \d x \, \frac{ q^2 P(q)}{2 \pi ^2}
\end{split}
\end{align}
\end{subequations}
\item $\mathbf{\mu^4}$ \textbf{terms}
\begin{subequations}
\begin{align}
\begin{split}
1 \quad \rightarrow & \quad \quad   - f D^2  P(k) \int_{q_{IR}}^{q_{UV}}\d q \int_{-1}^1 \d x \, -\frac{P(q)}{96 \pi^2 k^3 q^3 \left|k^2-2 k q x+q^2\right|}\times \\
   & \quad \times
   \Big[
   48 k^3 q^3 \Big( \Big(
   			-k q x \Big(k^2 (D_A f_A D + 2 D_B f_B D +  4 D_D f_D + 4  D_E f_E + F f_F - 3  J f_J) \\
   			& \quad \quad \quad \quad
   			+ q^2 (3 A f_A D + 4 B f_B D - F f_F - 2 D_G f_G + D_J f_J) \Big)\\
   			& \quad \quad \quad
   			x^2 \Big(k^2 q^2 (D_A f_A D + 4 D_B f_B D +  6 D_D f_D + 8  D_E f_E + 5  F f_F + 4 D_G f_G- 3  J f_J) \\
   			& \quad \quad \quad \quad
   					+ D q^4 (D_A f_A D + 2 D_B f_B D + k^4\left( 2 D_D f_D + 4  D_E f_E +  F f_F + 2 D_G f_G-   J f_J) \right) \Big)
   			  	\Big)\\
   			  & \quad \quad \quad
			 + 2 k^2 q^2 (D_A f_A D + 2 D_B f_B D  -	 F f_F - D_G f_G	)
   			 \Big)
   \\
   &\quad \quad
   2 f k q \Big(2 k^5 q^3 x \left(12 D \left(2 x^2 (D_A + 3 D_B) + 2 D_A + 2 D_B + D^2 \right) - 11(D_K + D_L) \right)\\
   				& \quad \quad \quad
   				 - 2 k^4 q^4 \left(12 D \left(x^2 \left(6 D_A + 8 D_B + D^2 \right) + 4 D_B x^4 + 2 D^2 \right) - 11(D_K + D_L) \right)\\
   				 & \quad \quad \quad
   				  + 2 k^3 q^5 x \left(12 D \left(2 x^2 (D_A + 3 D_B) + 2 D_A + 2 D_B + 3 D^2 \right) - 11 (D_K + D_L) \right)\\
   				 & \quad \quad \quad
   				 + 8 k^2 q^6 \left(-3 D x^2 \left(D_A + 2 D_B + D^2 \right) + D_K + D_L \right)\\
   				 & \quad \quad \quad
   				 +  8 k^6 q^2 \left(-3 D x^2 (D_A + 2 D_B) + D_K + D_L \right)\\
   				 & \quad \quad \quad
   				 -  3 k^8   (D_K + D_L) + 6 k^7 q x (D_K + D_L) +  6 k q^7 x (D_K + D_L) - 3 q^8 (D_K + D_L)
   	   		\Big)\\
   & \quad \quad
   + 3 f \left(k^2 - q^2 \right)^4   (D_K + D_L) \left|k^2 -2 k q x +  q^2 \right| \ln\left| \frac {k + q}{k - q}  \right|
   \Big]
\end{split}
\\
\begin{split}
b_1^{(1)} \quad \rightarrow & \quad  f D^2 P(k) \int_{q_{IR}}^{q_{UV}}\d q \int_{-1}^1 \d x \, \frac {P(q)} {96 \pi ^2 k^3 q^3} \times \\
& \quad \times
\Big[
- 3 \left(k^2-q^2\right)^3 \left(k^2+3 q^2\right) (D_K+D_L) \ln \left|\frac{k+q}{k-q}\right|\\
   & \quad \quad
   + 6 k^7 q (D_K+D_L)+2 k^5 q^3 (15 D_K+31 D_L)\\
   & \quad \quad
   + 42 k^3 q^5 (D_K+D_L)+18 k q^7 (D_K+D_L) \\
   & \quad
   +
   \frac{3 }{q}\left(k^2+q^2\right) \left(k^2-q^2\right)^2 (D_K+D_L) \ln \left| \frac{k+q}{k-q}\right| \\
& \quad \quad +
\frac{2 k}{\left|k^2-2 k q x+q^2\right|}
	\Big(6 k^5 q x \left(-D_A f_A-2 D_B f_B+f g^2 \left(fx^2+1\right)+D_K+D_L\right)\\
	& \quad \quad \quad
	+ k^4 q^2 \left(12 D_A f_A+12 D_B f_B \left(x^2+1\right)-3 f g^2 \left(f x^2+4\right)+3 D_K+7 D_L\right)\\
	& \quad \quad \quad
	-2 k^3 q^3 x \left(3 D_A f_A+6 D_B f_B-3 f g^2+6 D_K+10 D_L\right)\\
	& \quad \quad \quad
	-3  k^6 \left(f^2 g^2 x^2+D_K+D_L\right)+k^2 q^4 (3 D_K+7 D_L)\\
	& \quad \quad \quad
	+6 k q^5 x (D_K+D_L)-3 q^6
   (D_K+L)
\Big]\\
\end{split}
\\
\begin{split}
b_1^{(2)} \quad \rightarrow & \quad   - f D^2   P(k) \int_{q_{IR}}^{q_{UV}}\d q \int_{-1}^1 \d x \, \frac{P(q) }{3 \pi ^2} (A+2 B)  \left(k^2+q^2\right)
\end{split}
\\
\begin{split}
b_2^{(2)} \quad \rightarrow & \quad \quad    f D^4  P(k) \int_{q_{IR}}^{q_{UV}}\d q \int_{-1}^1 \d x \, \frac{ q^2 P(q)}{2 \pi ^2}
\end{split}
\end{align}    \\
\end{subequations}
\item $\mathbf{\mu^6}$ \textbf{terms}
\begin{equation}
\begin{split}
1 \quad \rightarrow & \quad - f^2 D^2   P(k) \int_{q_{IR}}^{q_{UV}}\d q \int_{-1}^1 \d x \, \frac{P(q)}{96 \pi ^2 k^3 q^3} \times \\
& \quad \times \Big[ 3 \left(k^2-q^2\right)^3 \left(k^2+3 q^2\right) (D_K+L) \log \left(\left|
   \frac{k+q}{k-q}\right| \right) \\
  \quad & +2 k^5 q^3 \left(12 f^2 D^2 x^2+15 D_K+31 D_L\right) \\
  \quad & -6 k^7 q (D_K+D_L)-42
   k^3 q^5 (D_k+D_L)+18 k q^7 (D_K+D_L)\Big]
\end{split}
\end{equation}
\end{itemize}

\para{22-type integrals}
\begin{itemize}
\item $\mathbf{\mu^0}$ \textbf{terms}
\begin{subequations}
\begin{align}
\begin{split}
b_1^{(2)}b_1^{(2)} \quad \rightarrow & \quad \quad   \int_{q_{IR}}^{q_{UV}}\d q \int_{-1}^1 \d x \, P(q)P\left(\left|\vect{k} - \vect{q}\right|\right)\frac{k^4 }{8 \pi ^2 \left(k^2-2 k q x+q^2\right)^2}  \times \\
& \quad \times
\left(D_A \left(k x-2 q x^2+q\right)+2 D_B x (k-q x)\right)^2
\end{split}
\\
\begin{split}
b_1^{(2)}b_2^{(2)} \quad \rightarrow & \quad \quad    \int_{q_{IR}}^{q_{UV}}\d q \int_{-1}^1 \d x \, P(q)P\left(\left|\vect{k} - \vect{q}\right|\right)\frac{D^2 k^2 q }{4 \pi ^2 \left(k^2-2 k q x+q^2\right)} \times \\
& \quad \times \left(D_A \left(k x-2 q x^2+q\right)+2 D_B x (k-q x)\right)
\end{split}
\\
\begin{split}
b_1^{(2)}b_{\mathcal{G}_2}^{(2)} \quad \rightarrow & \quad \quad    \int_{q_{IR}}^{q_{UV}}\d q \int_{-1}^1 \d x \, P(q)P\left(\left|\vect{k} - \vect{q}\right|\right)\frac{g^2 k^4 q \left(x^2-1\right) }{2 \pi ^2 \left(k^2-2 k q x+q^2\right)^2} \times \\
& \quad \times  \left(D_A \left(k x-2 q x^2+q\right)+2 D_B x (k-q x)\right)
\end{split}
\\
\begin{split}
b_2^{(2)}b_2^{(2)} \quad \rightarrow & \quad \quad    \int_{q_{IR}}^{q_{UV}}\d q \int_{-1}^1 \d x \, P(q)P\left(\left|\vect{k} - \vect{q}\right|\right) \frac{D^4 q^2 }{8 \pi ^2}
\end{split}
\\
\begin{split}
b_2^{(2)}b_{\mathcal{G}_2}^{(2)} \quad \rightarrow & \quad \quad    \int_{q_{IR}}^{q_{UV}}\d q \int_{-1}^1 \d x \, P(q)P\left(\left|\vect{k} - \vect{q}\right|\right)\frac{D^4 k^2 q^2 \left(x^2-1\right) }{2 \pi ^2
   \left(k^2-2 k q x+q^2\right)}
\end{split}
\\
\begin{split}
b_{\mathcal{G}_2}^{(2)}b_{\mathcal{G}_2}^{(2)} \quad \rightarrow & \quad \quad    \int_{q_{IR}}^{q_{UV}}\d q \int_{-1}^1 \d x \, P(q)P\left(\left|\vect{k} - \vect{q}\right|\right) \frac{D^4 k^4 q^2 \left(x^2-1\right)^2 }{2 \pi ^2
   \left(k^2-2 k q x+q^2\right)^2}
\end{split}
\end{align}
\end{subequations}
\item $\mathbf{\mu^2}$ \textbf{terms}
\begin{subequations}
\begin{align}
\begin{split}
b_1^{(2)} \quad \rightarrow & \quad \quad    \int_{q_{IR}}^{q_{UV}}\d q \int_{-1}^1 \d x \, P(q)P\left(\left|\vect{k} - \vect{q}\right|\right)\frac{1}{64 \pi ^2 q^2 \left| k^2-2 q x k+q^2\right|^3 }\times \\
& \quad \times
\Big[
f^2 g^2 (-(D_A+D_B))  \left| k^2-2 q x k+q^2\right| ^5+\\
& \quad \quad
f^2 g^2  \left(3 D_A k^2+4 D_A q^2+2   D_B k^2+4 D_B q^2\right) \left| k^2-2 q x k+q^2\right| ^4\\
& \quad \quad
-f^2 g^2   \left(3 D_A \left(k^4+k^2 q^2+2 q^4\right)+2 D_B q^2 \left(k^2+3 q^2\right)\right) \left| k^2-2 q x k+q^2\right|
   ^3\\
   &\quad \quad
   + \Big(D_A
   \Big(f^2 g^2 \left(k^6-2 k^4 q^2-3 k^2 q^4+4 q^6\right)  \left| k^2-2 q x k+q^2\right| ^2\\
  & \quad \quad \quad
  +16 k^4 q^2  \left(D_K \left|k x-2 q x^2+q\right|^2+2 D_L x  (k-q x)\right)  \Big)\\
  & \quad \quad \quad
	+ D_B  \Big(  -2 f^2 g^2 \left(k^6+k^2 q^4-2 q^6\right) \left|k^2-2 k q x+q^2\right|\\
	& \quad \quad \quad
	+
   32  k^4 q^2 x (k-q x)  \left(D_K \left|k x-2 q x^2+q\right|+2 D_L x (k-q x)\right)
   \Big)\\
   &\quad \quad
   +f^2 g^2 \left(k^2-q^2\right)^3 \left|k^2-2 k q x+q^2\right| \left(D_A q^2+D_B \left(k^2+q^2\right)\right)
\Big]
\end{split}
\\
\begin{split}
b_2^{(2)} \quad \rightarrow & \quad \quad  D^2 \int_{q_{IR}}^{q_{UV}}\d q \int_{-1}^1 \d x \, P(q)P\left(\left|\vect{k} - \vect{q}\right|\right) \frac{1}{32 \pi ^2  \left|k^2-2 k q x+q^2\right|} \times \\
   & \quad \times
   \Big[
   f^2 g^2 \left| k^2-2 q x   k+q^2\right|  \left(\left| k^2-2 q x k+q^2\right| -2 \left(k^2+q^2\right)\right)\\
   & \quad
   +f^2 g^2
   \left(k^2-q^2\right)^2+8 k^2 q \left(k x (D_K+2 D_L)-2 q x^2 (D_K+D_L)+D_K q\right)
   \Big]
\end{split}
\\
\begin{split}
b_{\mathcal{G}_2}^{(2)} \quad \rightarrow & \quad \quad  \int_{q_{IR}}^{q_{UV}}\d q \int_{-1}^1 \d x \, P(q)P\left(\left|\vect{k} - \vect{q}\right|\right) \frac{1}{64 \pi ^2 q^2  \left|k^2-2 k q x+q^2\right|^3} \times \\
& \quad
\times
\Big[
f^2 D^4  \left| k^2-2 q x k+q^2\right| ^5\\
&\quad \quad
-4 f^2 D^4 \left(k^2+q^2\right) \left|k^2-2 k q x+q^2\right|^4\\
&\quad \quad
+2 f^2 D^4 \left(3 k^4+2 k^2 q^2+3 q^4\right) \left| k^2-2 q x k+q^2\right| ^3 \\
&\quad \quad
-4 D^2 \left| k^2-2 q x k+q^2\right|
\Big(   f^2 D^2 \left(k^2-q^2\right)^2 \left(k^2+q^2\right) \left(k^2-2 k q x+q^2\right)\\
	& \quad \quad \quad
         -8 k^4 q^3 \left(x^2-1\right)
          \left(D_K \left(k x-2 q x^2+q\right) +2 D_L x (k-q x)\right)
\Big)\\
&\quad \quad
+f^2 g^4 \left(k^2-q^2\right)^4 \left|k^2-2 k q x+q^2\right|
\Big]
\end{split}
\end{align}
\end{subequations}
\begin{subequations}
\begin{align}
\begin{split}
b_1^{(1)}b_1^{(1)} \quad \rightarrow & \quad - f^2 D^4   \int_{q_{IR}}^{q_{UV}}\d q \int_{-1}^1 \d x \, P(q)P\left(\left|\vect{k} - \vect{q}\right|\right)\frac{1}{96 \pi ^2} \left(\frac{1}{q^2}-\frac{1}{k^2-2 k q x+q^2}\right) \times \\
& \quad
\times
  \left(2 k^2 \left(q^2-3 \left| k^2-2 q x   k+q^2\right| \right)
+3 \left(q^2-\left| k^2-2 q x k+q^2\right| \right)^2+3 k^4\right)
\end{split}
\\
\begin{split}
b_1^{(1)}b_1^{(2)} \quad \rightarrow & \quad \quad  f D^2   \int_{q_{IR}}^{q_{UV}}\d q \int_{-1}^1 \d x \, P(q)P\left(\left|\vect{k} - \vect{q}\right|\right) \frac{\left(-\left| k^2-2 q x k+q^2\right| +k^2+q^2\right)}{8 \pi ^2 q^2 \left| k^2-2 q x  k+q^2\right| } \times \\
& \quad
 \Big[
 \left| k^2-2 q x k+q^2\right|  \left(-(D_A+D_B) \left| k^2-2 q x k+q^2\right|
  +2 q^2 (D_A+D_B)+D_A k^2\right)\\
 & \quad \quad
 -q^4 (D_A+D_B)+D_A k^2 q^2+D_B k^4
 \Big]
\end{split}
\\
\begin{split}
b_1^{(1)}b_2^{(2)} \quad \rightarrow & \quad \quad f D^4   \int_{q_{IR}}^{q_{UV}}\d q \int_{-1}^1 \d x \, P(q)P\left(\left|\vect{k} - \vect{q}\right|\right) \frac{-\left| k^2-2 q x k+q^2\right| +k^2+q^2 }{4 \pi ^2}
\end{split}
\\
\begin{split}
b_1^{(1)}b_{\mathcal{G}_2}^{(2)} \quad \rightarrow & \quad \quad  f D^4 \int_{q_{IR}}^{q_{UV}}\d q \int_{-1}^1 \d x \, P(q)P\left(\left|\vect{k} - \vect{q}\right|\right) \frac{ \left((k-q)^2-\left| k^2-2 q x k+q^2\right| \right)}{8 \pi ^2 q^2 \left| k^2-2 q x k+q^2\right|} \times \\
& \quad
\times
 \left(-\left| k^2-2 q x k+q^2\right| +k^2+q^2\right) \left((k+q)^2-\left| k^2-2 q x k+q^2\right| \right)
\end{split}
\end{align}
\end{subequations}

\item $\mathbf{\mu^4}$ \textbf{terms}
\begin{subequations}
\begin{align}
\begin{split}
1 \quad \rightarrow & \quad \quad    \int_{q_{IR}}^{q_{UV}}\d q \int_{-1}^1 \d x \, P(q)P\left(\left|\vect{k} - \vect{q}\right|\right) \frac{1}{1024 \pi
   ^2 q^2 \left| k^2 - 2 k q x + q^2 \right| ^2}\times \\
   & \quad \times
   \Big[
		3 f^4 g^4 \left(	\left|k^2 - 2 q  x k + q^2 \right|^2 - 2 \left(k^2 + q^2 \right) \left| k^2 - 2 q x k + q^2 \right|
        +\left(k^2 - q^2 \right)^2 \right)^2 \\
        & \quad \quad
        +
        16 f^2 g^2 \left(
        		\left| k^2 - 2 q x k + q^2 \right|^2 - 2 \left(k^2 + q^2 \right)\left| k^2 - 2 q x k + q^2 \right|
       		+\left(k^2 - q^2 \right)^2
       			   \right) \times \\
       			   & \quad \quad \quad
       			   			\times \Big(\left|k^2 - 2 q x k + q^2 \right|  \Big(-(D_K + D_L) \left|k^2 - 2 q x k + q^2 \right|\\
       			   			& \quad \quad \quad \quad +k^2 D_K + 2 q^2 (D_K + D_L) \Big) + k^4 D_L + k^2  q^2 D_K - q^4 (D_K + D_L) \Big)\\
       	& \quad \quad
       	+
       	128 k^4 q^2   \left(k x (D_K + 2 D_L) - 2 q x^2 (D_K + D_L) + D_K q \right)^2
   \Big]
\end{split}
\\
\begin{split}
b_1^{(1)} \quad \rightarrow & \quad \quad  f D^2 \int_{q_{IR}}^{q_{UV}}\d q \int_{-1}^1 \d x \, P(q)P\left(\left|\vect{k} - \vect{q}\right|\right)\frac{1
}{32 \pi ^2 q^2 \left| k^2-2 q x k+q^2\right|^2 }\times \\
& \quad \times
\Big(\left| k^2-2 q x k+q^2\right| ^3 \left(f^2 g^2 \left(7 k^2+9 q^2\right)+4 (D_K+L) \left(k^2-2 k q x+q^2\right)\right)\\
& \quad \quad -3 f^2
   g^2 \left| k^2-2 q x k+q^2\right| ^4\ -\left| k^2-2 q x k+q^2\right| ^2  \Big(f^2 g^2 \left(5k^4+2 k^2 q^2+9 q^4\right)\\
    & \quad \quad \quad +4 \left(k^2-2 k q x+q^2\right) \left(k^2 (2 D_K+L)+3 q^2  (D_K+L)\right)\Big)\\
    & \quad \quad \quad +\left| k^2-2 q x k+q^2\right|  \Big(f^2 g^2 \left(k^2+3 q^2\right)
   \left(k^2-q^2\right)^2\\
    & \quad \quad \quad
   +4 \left|k^2-2 k q x+q^2\right| \left(k^4 (D_K-L)+2 k^2 q^2 (D_K+L)+3 q^4
   (D_K+L)\right)\Big)\\
   & \quad \quad \quad +4 \left(k^4-q^4\right) \left(k^2-2 k q x+q^2\right) \left(k^2 L+q^2  (D_K+L)\right)\Big)
\end{split}
\\
\begin{split}
b_1^{(2)} \quad \rightarrow & \quad \quad  f^2 D^2 \int_{q_{IR}}^{q_{UV}}\d q \int_{-1}^1 \d x \, P(q)P\left(\left|\vect{k} - \vect{q}\right|\right)
\frac {1}{64 \pi ^2 q^2 \left| k^2-2 q x  k+q^2 \right|^2 }\times
\\
& \quad \times \left (\left (-3 \left|    k^2 - 2 q x k + q^2 \right| +2 k^2 + 6 q^2 \right) \left| k^2 - 2 q k + q^2 \right| +k^4 + 2 k^2 q^2 - 3 q^4 \right) \times \\
& \quad \quad \times
  \Big[\left| k^2-2 q x k+q^2 \right|  \left(-(D_A+D_B) \left| k^2-2 q x k+q^2 \right| +2 q^2 (D_A+D_B)+D_A
   k^2\right) \\
   & \quad \quad \ -q^4 (D_A+D_B)+D_A k^2 q^2+D_B k^4 \Big]
\end{split}
\\
\begin{split}
b_2^{(2)} \quad \rightarrow & \quad \quad f^2 D^4   \int_{q_{IR}}^{q_{UV}}\d q \int_{-1}^1 \d x \, P(q)P\left(\left|\vect{k} - \vect{q}\right|\right)\frac{1}{32 \pi   ^2 \left|k^2-2 k q x+q^2\right|} \times \\
& \quad
\times
\left(\left(-3 \left| k^2-2 q x k+q^2\right| +2 k^2+6 q^2\right) \left| k^2-2 q x k+q^2\right| +k^4+2 k^2 q^2-3 q^4\right)
\end{split}
\\
\begin{split}
b_{\mathcal{G}_2}^{(2)} \quad \rightarrow & \quad \quad  f^2 D^4 \int_{q_{IR}}^{q_{UV}}\d q \int_{-1}^1 \d x \, P(q)P\left(\left|\vect{k} - \vect{q}\right|\right)\frac{ \left((k-q)^2- \left| k^2-2 q x k+q^2 \right| \right)}{ 64 \pi ^2 q^2 \left| k^2-2 q x k+q^2 \right|^2 } \times \\
& \quad \Big[
 \left((k + q)^2 - \left|k^2 - 2 q x k +   q^2 \right| \right)\times \\
 & \quad \quad \times  \left(\left(-3  \left|   k^2 - 2 q x k + q^2 \right| +2 k^2 + 6 q^2 \right)
           \left| k^2 - 2 q x k + q^2 \right| +k^4 + 2 k^2 q^2 - 3 q^4 \right)
           \Big]
\end{split}
\end{align}
\end{subequations}
\begin{subequations}
\begin{align}
\begin{split}
b_1^{(1)}b_1^{(1)} \quad \rightarrow & \quad \quad  f^2 D^4 k   \int_{q_{IR}}^{q_{UV}}\d q \int_{-1}^1 \d x \, P(q)P\left(\left|\vect{k} - \vect{q}\right|\right) \frac{1 }{32 \pi ^2 q^2 \left|k^2-2   k q x+q^2\right|} \times \\
& \quad
\times
\Big[
\left| k^2-2 q x k+q^2\right|  \Big(3 (k-2 q x) \left| k^2-2 q x k+q^2\right| +12 q x \left(k^2+q^2\right)\\
&\quad \quad -2 k \left(3 k^2+5 q^2\right)\Big)+3 k^5-6 k^4 q x+6 k^3 q^2-4 k^2 q^3 x+7 k q^4-6 q^5 x
\Big]
\end{split}
\end{align}
\end{subequations}

\item $\mathbf{\mu^6}$ \textbf{terms}
\begin{subequations}
\begin{align}
\begin{split}
1 \quad \rightarrow & \quad \quad f^2 D^2   \int_{q_{IR}}^{q_{UV}}\d q \int_{-1}^1 \d x \, P(q)P\left(\left|\vect{k} - \vect{q}\right|\right)\frac{1}{512 \pi ^2 q^2\left| k^2-2 q x k+q^2\right|^2}\times \\
	& \quad \times \Big[f^2 g^2
	\Big(-15 \left| k^2-2 q x k+q^2\right| ^4+12 \left(3 k^2+5 q^2\right) \left|  k^2 -2 q x k+q^2\right| ^3\\
	& \quad \quad
	-2 \left(13 k^4+18 k^2 q^2+45 q^4\right) \left| k^2-2 q x k+q^2\right| ^2\\
	& \quad \quad
	+4 \left(k^6+k^4 q^2-9 k^2 q^4+15 q^6\right) \left| k^2-2 q x k+q^2\right| +\left(k^2-q^2\right)^2 \left(k^4+6
   k^2 q^2-15 q^4\right)\Big)
   \\
   & \quad
   + 8  \left(-3 \left| k^2-2 q x k+q^2\right| ^2+2 \left(k^2+3 q^2\right) \left| k^2-2 q x k+q^2\right| +k^4+2 k^2 q^2-3 q^4\right)
   \\
   & \quad \quad \times \Big(-(D_K+D_L) \left| k^2-2 q x
   k+q^2\right| ^2+\left| k^2-2 q x k+q^2\right|  \left(k^2 D_K+2 q^2
   (D_K+D_L)\right)\\
   & \quad \quad
   +\left(k^2-q^2\right) \left(k^2 D_L+q^2 (D_K+D_L)\right)\Big)\Big]
\end{split}
\\
\begin{split}
b_1^{(1)} \quad \rightarrow & \quad \quad  f^3 D^4   \int_{q_{IR}}^{q_{UV}}\d q \int_{-1}^1 \d x \, P(q)P\left(\left|\vect{k} - \vect{q}\right|\right) \frac{1 }{32 \pi ^2 q^2 \left(k^2-2 k q
   x+q^2\right)}\times \\
   & \quad \times \Big[5 \left| k^2-2 q x k+q^2\right| ^3-3 \left(3 k^2+5 q^2\right) \left| k^2-2 q x
   k+q^2\right| ^2\\
   & \quad +3 \left(k^4+2 k^2 q^2+5 q^4\right) \left| k^2-2 q x k+q^2\right| +k^6+k^4 q^2+3 k^2
   q^4-5 q^6\Big]
\end{split}
\end{align}
\end{subequations}
\item $\mathbf{\mu^8}$ \textbf{terms}
\begin{subequations}
\begin{align}
\begin{split}
1 \quad \rightarrow & \quad \quad f^4 D^4  \int_{q_{IR}}^{q_{UV}}\d q \int_{-1}^1 \d x \, P(q)P\left(\left|\vect{k} - \vect{q}\right|\right) \frac{1}{1024 \pi ^2 q^2 \left|
   k^2-2 q x k+q^2\right| ^2}\times\\
& \quad \times \Big[ 35 \left| k^2-2 q x k+q^2\right| ^4-20 \left(3 k^2+7 q^2\right) \left| k^2-2 q x
   k+q^2\right| ^3\\
   & \quad +6 \left(3 k^4+10 k^2 q^2+35 q^4\right) \left| k^2-2 q x k+q^2\right| ^2\\
   & \quad +4 \left(k^6+3
   k^4 q^2+15 k^2 q^4-35 q^6\right) \left| k^2-2 q x k+q^2\right| \\
   & \quad +\left(k^2-q^2\right)^2 \left(3 k^4+10 k^2
   q^2+35 q^4\right)\Big]
\end{split}
\end{align}
\end{subequations}
\end{itemize}

\subsubsection{Advective terms}
\para{13-type integrals}
We would need the following integrands
\begin{subequations}\small
\begin{align}
\begin{split}
\Iint{1}_{13}(k,q,x) = & \quad D^2 P(k) q^2 P(q) \frac{k q x-q^2}{4 \pi ^2 \left( k^2-2 k q x+q^2 \right) }
 \Big[  \left(1-\frac{1}{2} x   \left(\frac{k}{q}+\frac{q}{k}\right)\right)   \left(D_A f_A-f g^2\right)\\
  & \quad \quad \quad \quad +  D_B f_B   \left(-\frac{x \left(k^2-2 k q x+q^2\right)}{2 k   q}-\frac{1}{2} x \left(\frac{k}{q}+\frac{q}{k}\right) +1 \right) \Big],
\end{split}
\\
\begin{split}
\Iint{2}_{13}(k,q,x) = &  \quad D^4 x P(k)  \left(\frac{k}{q}+\frac{q}{k}\right) P(q) \frac{q^2-k q x}{4 \pi ^2},
\end{split}
\\
\begin{split}
\Iint{3}_{13}(k,q,x) = & \quad D^2 P(k) P(q)  \frac{k q x-q^2}{\pi ^2} \Big[ D_A
   \left(1-\frac{1}{2} x  \left(\frac{k}{q}+\frac{q}{k}\right)\right) \\
    & \quad \quad \quad \quad + D_B \left(-\frac{x \left(k^2-2 k q x+q^2\right)}{2 k  q}-\frac{1}{2} x  \left(\frac{k}{q}+\frac{q}{k}\right)+1\right)\Big],
\end{split}
\\
\begin{split}
\Iint{4}_{13}(k,q,x) = &  \quad D^4 P(k) P(q) \frac{k q x-q^2}{2 \pi ^2},
\end{split}
\\
\begin{split}
\Iint{5}_{13}(k,q,x) = & \quad D^4 \left(x^2-1\right) P(k) P(q)\frac{ k q x-q^2 }{\pi ^2},
\end{split}
\\
\begin{split}
\Iint{6}_{13}(k,q,x) = &  \quad f D^4  P(k) \left[ -\frac{k}{3q} \left(\frac{k}{q}+\frac{q}{k}\right)-1\right] P(q)\frac{q^2}{2 \pi ^2}.
\end{split}
\end{align}
\end{subequations}
Then we list the different contributions according to the order in power of $\mu$ as well as to the combination of the different bias parameters
\begin{itemize}
\item $\mathbf{\mu^0}$ \textbf{terms}
\begin{subequations}
\begin{align}
b_1^{(1)}b_1^{(1)} \quad \rightarrow &\quad \quad \int_{q_{IR}}^{q_{UV}}\d q \int_{-1}^1 \d x \,2\, \left( \frac{1}{f}\Iint{1}_{13}(k,q,x) + \Iint{2}_{13}(k,q,x)\right),
\\
b_1^{(1)}b_1^{(2)} \quad \rightarrow &\quad \quad \int_{q_{IR}}^{q_{UV}}\d q \int_{-1}^1 \d x \, \left( -2 \Iint{2}_{13}(k,q,x) + \Iint{3}_{13}(k,q,x)\right),
\\
b_1^{(1)}b_1^{(3)} \quad \rightarrow &\quad \quad \int_{q_{IR}}^{q_{UV}}\d q \int_{-1}^1 \d x \, \left( \frac{2}{f}\Iint{1}_{13}(k,q,x) + 2 \Iint{2}_{13}(k,q,x) -\Iint{3}_{13}(k,q,x)  \right),
\\
b_1^{(1)}b_2^{(2)} \quad \rightarrow &\quad \quad \int_{q_{IR}}^{q_{UV}}\d q \int_{-1}^1 \d x  \,\Iint{4}_{13}(k,q,x),
\\
b_1^{(1)}b_2^{(3)} \quad \rightarrow &\quad - \int_{q_{IR}}^{q_{UV}}\d q \int_{-1}^1 \d x  \, \Iint{4}_{13}(k,q,x),
\\
b_1^{(1)}b_{\mathcal{G}_2}^{(2)} \quad \rightarrow &\quad \quad \int_{q_{IR}}^{q_{UV}}\d q \int_{-1}^1 \d x  \,\Iint{5}_{13}(k,q,x),
\\
b_1^{(1)}b_{\mathcal{G}_2}^{(3)} \quad \rightarrow &\quad - \int_{q_{IR}}^{q_{UV}}\d q \int_{-1}^1 \d x \, \Iint{5}_{13}(k,q,x).
\end{align}
\end{subequations}
\item $\mathbf{\mu^2}$ \textbf{terms}
\begin{subequations}
\begin{align}
b_1^{(1)} \quad \rightarrow &\quad \quad \int_{q_{IR}}^{q_{UV}}\d q \int_{-1}^1 \d x \,2\, \left( \Iint{1}_{13}(k,q,x) + f \Iint{2}_{13}(k,q,x)\right),
\\
b_1^{(2)} \quad  \rightarrow &\quad \quad \int_{q_{IR}}^{q_{UV}}\d q \int_{-1}^1 \d x \, \left( -4 f \Iint{2}_{13}(k,q,x) + f \Iint{3}_{13}(k,q,x)\right),
\\
b_1^{(3)} \quad \rightarrow &\quad \quad \int_{q_{IR}}^{q_{UV}}\d q \int_{-1}^1 \d x \, \left(- 2 \Iint{1}_{13}(k,q,x) + 2 f \Iint{2}_{13}(k,q,x) - f \Iint{3}_{13}(k,q,x)  \right),
\\
b_2^{(2)} \quad \rightarrow &\quad \quad \int_{q_{IR}}^{q_{UV}}\d q \int_{-1}^1 \d x \, f \Iint{4}_{13}(k,q,x),
\\
b_2^{(3)} \quad \rightarrow &\quad -\int_{q_{IR}}^{q_{UV}}\d q \int_{-1}^1 \d x \, f \Iint{4}_{13}(k,q,x),
\\
b_{\mathcal{G}_2}^{(2)} \quad \rightarrow &\quad \quad \int_{q_{IR}}^{q_{UV}}\d q \int_{-1}^1 \d x  \,f \Iint{5}_{13}(k,q,x),
\\
b_{\mathcal{G}_2}^{(3)} \quad \rightarrow &\quad - \int_{q_{IR}}^{q_{UV}}\d q \int_{-1}^1 \d x  \,f \Iint{5}_{13}(k,q,x),
\\
b_1^{(1)}b_1^{(1)} \quad \rightarrow &\quad \quad \int_{q_{IR}}^{q_{UV}}\d q \int_{-1}^1 \d x  \,\Iint{6}_{13}(k,q,x),
\\
b_1^{(1)}b_1^{(2)} \quad \rightarrow &\quad - \int_{q_{IR}}^{q_{UV}}\d q \int_{-1}^1 \d x  \,\Iint{6}_{13}(k,q,x).
\end{align}
\end{subequations}
\item $\mathbf{\mu^4}$ \textbf{terms}
\begin{subequations}
\begin{align}
b_1^{(1)} \quad \rightarrow &\quad \quad \int_{q_{IR}}^{q_{UV}}\d q \int_{-1}^1 \d x  \, f \Iint{6}_{13}(k,q,x),
\\
b_1^{(2)} \quad \rightarrow &\quad - \int_{q_{IR}}^{q_{UV}}\d q \int_{-1}^1 \d x f \,\Iint{6}_{13}(k,q,x).
\end{align}
\end{subequations}
\end{itemize}

\para{22-type integrals}
We define the following integrands
\begin{subequations}\small
\begin{align}
\Iint{1}_{22}(k,q,x) = & \quad  D^4 P(q)P\left(\left|\vect{k} - \vect{q}\right|\right)  \frac{\left(\left| k^2-2 q x k+q^2\right| +q^2\right) \left(\left| k^2-2 q x k+q^2\right| -k^2+q^2\right)^2 }{16 \pi ^2 q^2 \left| k^2-2 q x
   k+q^2\right| }\\
\Iint{2}_{22}(k,q,x) = & \quad - D^2 k^2 P(q)P\left(\left|\vect{k} - \vect{q}\right|\right) \frac{ (q-k x) \left(D_A \left(k x-2 q x^2+q\right)+2 D_B x (k-q x)\right) }{2 \pi ^2 \left(k^2-2 k q x+q^2\right)}\\
\Iint{3}_{22}(k,q,x) = & \quad   -D^4 q P(q)P\left(\left|\vect{k} - \vect{q}\right|\right) \frac{(q-k x) }{2 \pi ^2}\\
\Iint{4}_{22}(k,q,x) = & \quad D^4 k^2P(q)P\left(\left|\vect{k} - \vect{q}\right|\right) \frac{  q \left(x^2-1\right) (k x-q) }{\pi ^2
   \left(k^2-2 k q x+q^2\right)}\\
\Iint{5}_{22}(k,q,x) = & \quad   -D^2 k^2 P(q) P\left(\left|\vect{k} - \vect{q}\right|\right) \frac{ (q-k x) \left(D_K \left(k x-2 q
   x^2+q\right)+2 D_L x (k-q x)\right)}{2 \pi ^2 \left(k^2-2 k q x+q^2\right)}\\
   \begin{split}
\Iint{6}_{22}(k,q,x) = & \quad f^2 D^4 P(q) P\left(\left|\vect{k} - \vect{q}\right|\right) \Big[ \frac{\left(k^2-q^2\right)^3}{32 \pi ^2 q^2\left| k^2-2 q x k+q^2\right| } + \\
& + \frac{-\left| k^2-2 q x k+q^2\right| ^2+\left(3 k^2+q^2\right) \left| k^2-2 q x
   k+q^2\right| +\left(-3 k^4+2 k^2 q^2+q^4\right)} {32 \pi ^2 q^2} \Big]
   \end{split}
   \\
\begin{split}
\Iint{7}_{22}(k,q,x) = & \quad f^2 D^4 P(q) P\left(\left|\vect{k} - \vect{q}\right|\right) \Big[\frac{\left(k^2-q^2\right)^2\left(k^2+3 q^2\right)}{32 \pi ^2 q^2\left| k^2-2 q x k+q^2\right| } +\\
& + \frac{3 \left| k^2-2 q x k+q^2\right| ^2-\left(5 k^2+3 q^2\right) \left| k^2-2 q x k+q^2\right|+\left(k^4+2 k^2 q^2-3 q^4\right) }{32 \pi ^2 q^2 }
\end{split}
\\
\begin{split}
\Iint{8}_{22}(k,q,x) = & \quad f D^4 k P(q)  P\left(\left|\vect{k} - \vect{q}\right|\right)   \frac{\left(-\left| k^2-2 q x k+q^2\right| +k^2-q^2\right)}{8 \pi ^2 q^2}\times \\
& \times \Big[ -(k- 2qx) + \frac{(k^2 + q^2)(k- 2qx) + 2k q^2 }{\left| k^2-2 q x k+q^2\right|} \Big]
 \end{split}
\end{align}
\end{subequations}
Then we list the different contributions according to the order in power of $\mu$ as well as to the combination of the different bias parameters
\begin{itemize}
\item $\mathbf{\mu^0}$ \textbf{terms}
\begin{subequations}
\begin{align}
b_1^{(1)}b_1^{(1)} \quad \rightarrow &\quad \quad \int_{q_{IR}}^{q_{UV}}\d q \int_{-1}^1 \d x \, \Iint{1}_{22}(k,q,x),
\\
b_1^{(1)}b_1^{(2)} \quad \rightarrow &\quad \quad \int_{q_{IR}}^{q_{UV}}\d q \int_{-1}^1 \d x \left( - \,2 \Iint{1}_{22}(k,q,x) + \Iint{2}_{22}(k,q,x)\right),
\\
b_1^{(1)}b_2^{(2)} \quad \rightarrow &\quad \quad \int_{q_{IR}}^{q_{UV}}\d q \int_{-1}^1 \d x \,\Iint{3}_{22}(k,q,x),
\\
b_1^{(1)}b_{\mathcal{G}_2}^{(2)} \quad \rightarrow &\quad \quad \int_{q_{IR}}^{q_{UV}}\d q \int_{-1}^1 \d x  \,\Iint{4}_{22}(k,q,x),
\\
b_1^{(2)}b_1^{(2)} \quad \rightarrow &\quad \quad \int_{q_{IR}}^{q_{UV}}\d q \int_{-1}^1 \d x \,\left( \Iint{1}_{22}(k,q,x) -  \Iint{2}_{22}(k,q,x)\right)  ,
\\
b_1^{(1)}b_2^{(2)} \quad \rightarrow &\quad - \int_{q_{IR}}^{q_{UV}}\d q \int_{-1}^1 \d x \, \Iint{3}_{22}(k,q,x),
\\
b_1^{(2)}b_{\mathcal{G}_2}^{(2)} \quad \rightarrow &\quad - \int_{q_{IR}}^{q_{UV}}\d q \int_{-1}^1 \d x \, \Iint{4}_{22}(k,q,x).
\end{align}
\end{subequations}
\item $\mathbf{\mu^2}$ \textbf{terms}
\begin{subequations}
\begin{align}
b_1^{(1)} \quad \rightarrow &\quad \quad \int_{q_{IR}}^{q_{UV}}\d q \int_{-1}^1 \d x \,\left(\Iint{5}_{22}(k,q,x) + \Iint{6}_{22}(k,q,x)\right),
\\
b_1^{(2)} \quad \rightarrow &\quad - \int_{q_{IR}}^{q_{UV}}\d q \int_{-1}^1 \d x \,\left(\Iint{5}_{22}(k,q,x) + \Iint{6}_{22}(k,q,x)\right),
\\
b_1^{(1)}b_1^{(1)} \quad \rightarrow &\quad \quad \int_{q_{IR}}^{q_{UV}}\d q \int_{-1}^1 \d x \,\Iint{8}_{22}(k,q,x),
\\
b_1^{(1)}b_1^{(2)} \quad \rightarrow &\quad - \int_{q_{IR}}^{q_{UV}}\d q \int_{-1}^1 \d x  \,\Iint{8}_{22}(k,q,x).
\end{align}
\end{subequations}
\item $\mathbf{\mu^4}$ \textbf{terms}
\begin{subequations}
\begin{align}
b_1^{(1)} \quad \rightarrow &\quad \quad \int_{q_{IR}}^{q_{UV}}\d q \int_{-1}^1 \d x  \, \Iint{7}_{22}(k,q,x),
\\
b_1^{(2)} \quad \rightarrow &\quad - \int_{q_{IR}}^{q_{UV}}\d q \int_{-1}^1 \d x f \,\Iint{7}_{22}(k,q,x).
\end{align}
\end{subequations}
\end{itemize}

\clearpage
\phantomsection
\renewcommand{\bibname}{References}
\bibliographystyle{JHEP}
\bibliography{article}

\end{document}